\documentclass[10pt]{iopart}

\usepackage{amssymb}
\usepackage{bm}
\usepackage{color}
\usepackage{wasysym} %for hexagon (in Eq environment or in text)
\usepackage{hyperref} %for links in pdf

\newcommand{\eqref}{\eref}

\usepackage{tikz}

\def\plaquette{\tikz[baseline=.05ex]{
\fill (0,0) circle (1pt) coordinate (A);
\fill (1.5ex,0) circle (1pt) coordinate (B);
\fill (2.25ex,1.3ex) circle (1pt) coordinate (C);
\fill (0.75ex,1.3ex) circle (1pt) coordinate (D);
\draw (A)--(B);
\draw (B)--(C);
\draw (C)--(D);
\draw (D)--(A);
\draw (B)--(D);}
}

\def\plaquettev{\tikz[baseline=.1ex]{
\fill (0,0) circle (1pt) coordinate (A);
\fill (1.5ex,0) circle (1pt) coordinate (B);
\fill (2.25ex,1.3ex) circle (1pt) coordinate (C);
\fill (0.75ex,1.3ex) circle (1pt) coordinate (D);
\draw (A)--(B);
\draw [ultra thick] (B)--(C);
\draw (C)--(D);
\draw [ultra thick] (D)--(A);
\draw (B)--(D);}
}

\def\plaquetteh{\tikz[baseline=.1ex]{
\fill (0,0) circle (1pt) coordinate (A);
\fill (1.5ex,0) circle (1pt) coordinate (B);
\fill (2.25ex,1.3ex) circle (1pt) coordinate (C);
\fill (0.75ex,1.3ex) circle (1pt) coordinate (D);
\draw [ultra thick] (A)--(B);
\draw (B)--(C);
\draw [ultra thick] (C)--(D);
\draw (D)--(A);
\draw (B)--(D);}
}
\begin{document}

\title{Quantum Spin Liquids }

\author{Lucile Savary$^1$, Leon Balents$^2$}

\address{$^1$Department of Physics, Massachusetts Institute of Technology, Cambridge, MA 02139, U.S.A.}

\address{$^2$Kavli Institute for Theoretical Physics, University of
  California, Santa Barbara, CA 93106-4030, U.S.A.}

\date{\today}
\begin{abstract}
  Quantum spin liquids may be considered ``quantum disordered'' ground states of spin systems, in which zero point fluctuations are so strong that they prevent conventional magnetic long range order.   More interestingly, quantum spin liquids are prototypical examples of ground states with massive many-body entanglement, of a degree sufficient to render these states distinct {\em phases} of matter.   Their highly entangled nature imbues quantum spin liquids with unique physical aspects, such as non-local excitations, topological properties, and more.  In this review, we discuss the nature of such phases and their properties based on paradigmatic models and general arguments, and introduce theoretical technology such as gauge theory and partons that are conveniently used in the study of quantum spin liquids. An overview is given of the different types of quantum spin liquids and the models and theories used to describe them.  We also provide a guide to the current status of experiments to study quantum spin liquids, and to the diverse probes used therein.  
 \end{abstract}


%\submitto{\RPP}

\maketitle
\ioptwocol

\section{Introduction}
\label{sec:introduction}

What is a ``Quantum Spin Liquid'' (QSL)?  Probably the most common perception is that a QSL is a system of quantum spins which are highly correlated with one another due to their mutual interactions, yet do not order even at very low or zero temperature.  This is a rather loose idea and also an unsatisfying one insofar as it is mainly a statement of what these spins {\em do not} do.  Despite the relatively long history of this field, however, it is actually hard to agree on a {\em positive} definition of a QSL.  Here, we will take the point of view that the essential ingredient of a QSL is not the lack of order but the presence of an anomalously high degree of entanglement, or massive quantum superposition.  QSLs are exemplars of highly entangled quantum matter in the context of magnetism.

We will not attempt to make this notion truly precise, but roughly by a highly entangled state we mean one which is not smoothly connected to a product state over any finite spatial blocks (see Sec.~\ref{sec:highly-entangl-quant}).  The looseness of this definition reflects that fact that entanglement of many body systems is still an evolving subject.  It is being actively explored not only throughout diverse parts of condensed matter physics, but also in the context of quantum information science and even in the study of quantum gravity.  However, many aspects of highly entangled quantum phases are understood, and we will review some key points of their structure, and the application of these ideas to quantum spins.  

Probably the most interesting {\em implication} of this kind of many-body entanglement is the ability of such states to support non-local excitations.  That is, excitations which individually cannot be created by any local operator, but only by an infinite product of local operators, can have finite energy and even behave as sharp quasiparticles.  In QSLs, such excitations may exist as ``spinons'' -- strange spin excitations that behave as fractions of ordinary magnons or spin waves---as well as other forms.  Highly entangled QSLs include so-called topological phases, but are not limited to them.  The possibilities are very rich, and we will try to give a sense of this abundance here.

The focus of this review is primary theoretical, but it is driven by the goal of bringing the theory of highly entangled matter into confrontation with experiment. We take a syncretic approach, bringing together what we feel are the key threads of the very vast body of work on QSLs.  At the same time, we try to present the material using the minimal amount of formalism possible.  For example, we will try to avoid excessive use of field theory, even though this might seem natural to some readers.  Similarly, we will refrain from entering into any even semi-complete {\em classification} of phases, which has become very popular in recent years, as we feel this introduces unnecessary formal mathematics.    We will describe and review some mathematical and computational formalisms, which we feel are most relevant to understanding the current state of and making future progress on QSL theory.  In the process, we will highlight what we view as the most important QSL states: their universal properties, some of their mathematical descriptions, and the models and physical systems in which they are being contemplated.   A rather abbreviated summary of this can be found in Table~\ref{tab:QSLs}.

The structure of the article is as follows.  Sec.~\ref{sec:highly-entangl-quant} takes Kitaev's toric code model as an illustrative example of highly entangled quantum matter, and identifies some general features from it.  We also briefly review the basic notions of intrinsic topological order.  In Sec.~\ref{sec:gauge-theory}, we introduce gauge theory as a tool for describing highly entangled states, and show how non-local excitations are naturally captured by it.   Next, we turn to the connection of these phenomena to more physical microscopic spin models, through a widely used technique of ``partons,'' through which many QSL phases can be described.  This is discussed in Sec.~\ref{sec:partons}.  Up to this point, the focus is on entanglement and emergent non-local excitations, but for QSLs, it is also important to understand symmetry and quantum numbers, to which we turn in Sec.~\ref{sec:symm-fract}.   This concludes the general discussion of QSLs, and the remainder of the review is devoted to a pr\'{e}cis of models for QSLs, and one of the relation to experiment, in Sections \ref{sec:models-methods} and \ref{sec:mater-exper}, respectively, and a Summary and Discussion of open issues in Sec.~\ref{sec:future-directions}.

\section{Highly entangled quantum matter}
\label{sec:highly-entangl-quant}

In this section, we discuss highly entangled quantum phases, with a focus on their essential universal properties, without regard to their realization in quantum magnets.  What is quantum entanglement?  It is a property of certain quantum states, that in those states, the result of a measurement of one observable affects the outcome of the measurement of others, even when all observables themselves are undetermined, and when they are {\em a priori} independent.  We add to this notion the important ingredient of locality, and consider entanglement between observables that are spatially separated.  A state is entangled if it is a superposition state: it cannot be written as a product state even under an arbitrary {\em local} change of basis.  States of two observables can range from completely disentangled product states, to maximally entangled states, with a ``generic'', randomly chosen state somewhere in between.  

This idea is adequate for discussing entanglement for two observables.  But to think about entanglement in many-particle systems in the thermodynamic limit requires more care.  Some degree of entanglement is inevitable in any generic state.  Here we seek to describe states in which the entanglement is {\em essential} to the properties of the system.  The qualitative properties of a system define a {\em phase} of matter.  Hence we really are trying to characterize the entanglement of phases and not so much individual quantum states.  We are interested in phases of matter for which there is {\em no} representative member of the phase which is disentangled.  So we look for ground states that cannot be continuously deformed into a product state while staying within the phase.  

There is some subtlety in defining what an allowed ``continuous deformation'' is.  One possible but restrictive definition is to say that deformations correspond to continuous changes of a ground state wavefunction in response to variation of parameters of a {\em local Hamiltonian}, such that throughout the variation a non-zero gap is maintained above the ground state.  A state which under these conditions cannot be deformed to a product state represents the simplest type of highly entangled phase.   We will discuss a storied example in the next subsection.  We will also later encounter other phases which are gapless, so this definition is inadequate.  Rigorous definitions are for mathematicians, not physicists, so we do not worry about this further.

\subsection{Example: the toric code}
\label{sec:example:-toric-code}

Topological phases are by far the most studied and best understood examples of highly entangled quantum phases, and the subject is quite evolved.  It is not our goal to elaborate the full theory of topological phases, but to use them as illustrative examples for the effects of massive entanglement on QSLs.  So we proceed by discussing the canonical example of a topological phase, the toric code model of Kitaev.  This model has a ready generalization to three dimensions, but for simplicity, we focus here on Kitaev's original case of two dimensions.  Consider a set of spin-1/2 ``spins'' on the middle of the links of a square lattice, with the Hamiltonian
\begin{equation}
  \label{eq:1}
  H_{\rm tc} = - K \sum_p P_p - K' \sum_s S_s,
\end{equation}
where the sums are over plaquettes $p$ and sites $s$, the plaquette operator $P_p = \prod_{i \in p} \sigma_i^z$ is a product over the spins on the bonds surrounding the plaquette $p$, and the ``star'' operator $S_s = \prod_{i \in s} \sigma_i^x$ is a product over the spins on bonds neighboring the site $s$.  Obviously all the star operators commute as do the plaquette operators, and one can easily verify that the stars and plaquettes do as well, $[S_s, P_p]=0$ for all $p,s$.  This makes the toric code model especially easy to solve: ground states are simply those states for which $S_s=P_p=+1$ for every star and plaquette.  

\begin{figure}[htbp]
  \centering
  \includegraphics[width=3.0in]{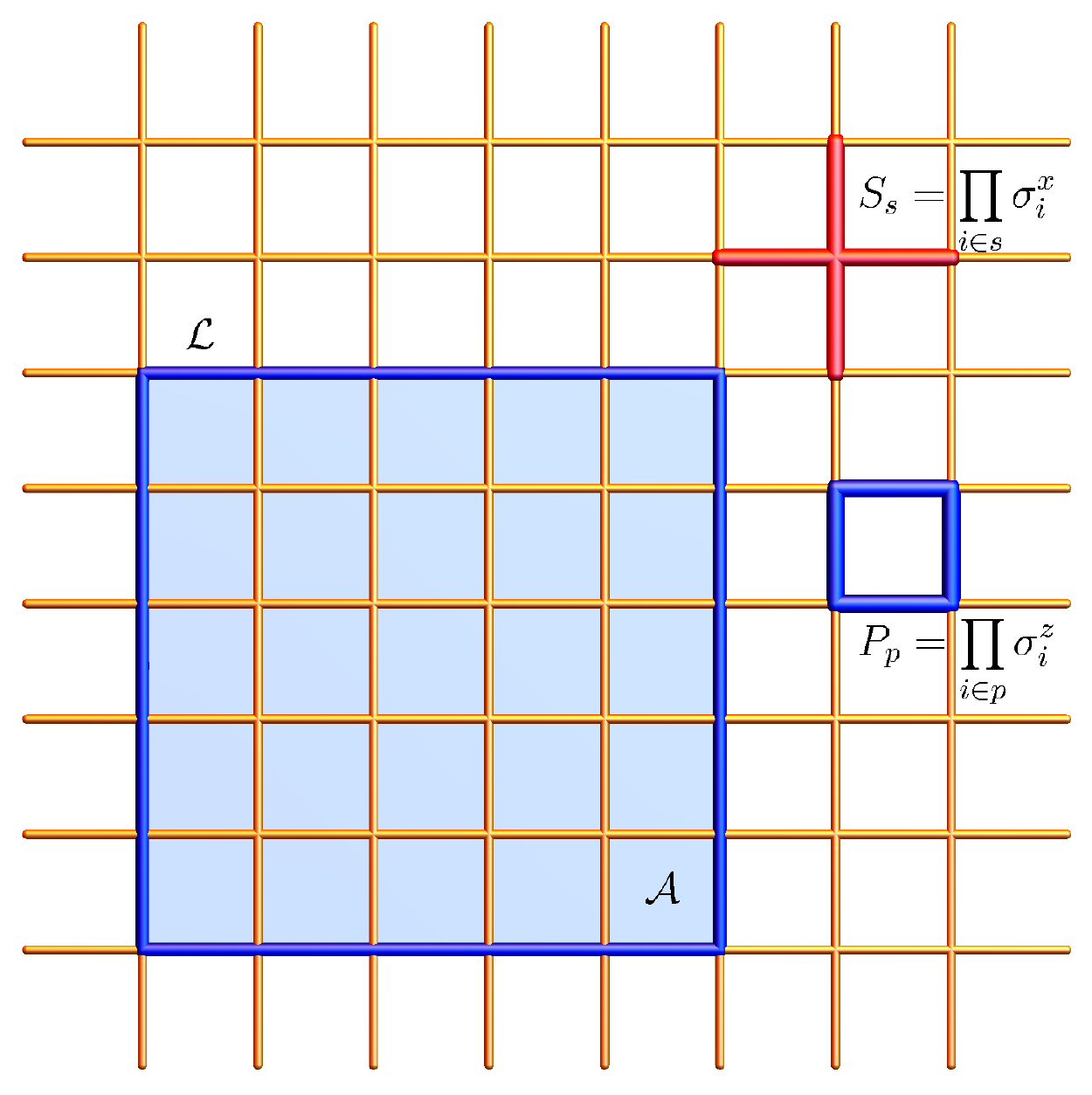}
  \caption{Operators in the toric code on the square lattice.  Blue and red links denote $\sigma_i^x$ and $\sigma_i^z$ operators, respectively.  A star $S_s$ and plaquette $P_p$ are shown, as is a loop operator on the contour $\mathcal{L}$ enclosing the area $\mathcal{A}$. }
  \label{fig:toric}
\end{figure}

While this looks fairly trivial in these variables, the state itself in any {\em local} basis is highly entangled.  Consider for instance the $\sigma_i^x$ basis, which is complete and local.  States with $S_s=1$ are those in which an even number of spins composing each star have eigenvalues $\sigma_i^x=-1$.  One may represent these states by coloring the links with negative spins, and according to $S_s=1$ the links form closed loops.  Since the operator $P_p$ is off-diagonal in this basis, it {\em requires} superposition of the loop states.  One can construct a state with $P_p=+1$ everywhere by taking as a ``base'' state a direct product state (eigenstate of $\sigma_i^x$) and acting on it with the projectors 
\begin{equation}
  \label{eq:3}
  Q_p = \frac{1+P_p}{2},
\end{equation}
which projects onto states with $P_p=1$.  Using this,
\begin{equation}
  \label{eq:2}
  |0\rangle = \prod_p Q_p\left( \otimes_i |\sigma_i^x=s_i\rangle \right),
\end{equation}
where the $s_i=\pm 1$ are chosen to satisfy the star rule $S_s=1$.  For example, if we take $s_i=1$ everywhere, then $|0\rangle$ appears to be a massive superposition of loop states, in which two loop configurations related to one another by the flip of a minimal square plaquette appear with equal weight.  This looks highly entangled indeed.  

It might seem there are many such states, depending upon the choice of $s_i$.   However, using $Q_p = Q_p P_p$, we can see that states in which the $s_i$ are related by flipping spins around a minimal square plaquette are the same (after projection).  In fact, there are very few such states, and the number depends upon the boundary conditions.  With periodic boundary conditions, i.e.\ on the torus, one can only make 4 = 2$\times$2 inequivalent choices.  In the loop picture, we may consider a base state which has a fixed {\em parity} of the number of loops winding in the $x$ or $y$ direction of the torus.  Mathematically, the parity of loops winding in the $x$ direction is given by a product over all the $\sigma_i^x=s_i$ on a full column of horizontal bonds.  The parity is unchanged by (commutes with) the action of $P_p$, and so states with different parity are orthogonal, even after projection.  This is the famous ``topological degeneracy'' of the toric code.  The interesting thing about this degeneracy is that the different degenerate states cannot be distinguished locally.   More precisely,  if we take any local operator $\mathcal{O}$ (i.e.\ a product over $\sigma_i^\mu$ operators over some set of nearby sites smaller than the size of the system), then 
\begin{equation}
  \label{eq:4}
  \langle m| \mathcal{O} |n\rangle = \overline{\mathcal{O}} \delta_{mn},
\end{equation}
if $|m\rangle$ with $m=1,2,3,4$ are the orthonormal degenerate ground states.  This can be shown formally, but the physics is simply that the ground states differ only by the global parity of the loops, and this does not affect the local configuration space.  Essentially because of the above property, one can show that the topological degeneracy is very robust: it is maintained even when the toric code Hamiltonian is perturbed by arbitrary perturbations, provided these are below some threshold, which defines the stability region for the toric code phase. 

% \oim{Trying something. Maybe wrong:}  In that sense is the phase ``topological.'' A topological set (or ``topology'') $\mathcal{T}$ is a set made of a collection of open sets, including the empty set and [muni de] the union and intersection operations. $\mathcal{T}$ should be stable under the union of an arbitrary number of its constituent, as well as under the intersection of a finite number of them. The notion of a open sets allows for / is tightly connected to  the notion of continuity. A topology has very little structure as compared to, e.g.\ a Hilbert space (a Hilbert space is a topology, with ... the set). In a sense, topological phases, to exist, need only {\em some} of the structure of a Hilbert space. This is a way to see that such phases should be stable. Upon providing the topology with some extra features, one can introduce the notion of homotopy etc..

\subsubsection{Anyons}
\label{sec:anyons}

The topological degeneracy is useful for formal theory and also numerical calculations, but is not very relevant to real systems.  However, it is actually symptomatic of a more fundamental property of the topologically ordered state, which is that it supports exotic excitations known as {\em anyons}.  The elementary excitations of the toric code appear rather simple: the minimal way to raise the energy of the Hamiltonian in Eq.~\eqref{eq:1} is to choose a {\em single} star or plaquette to be negative, i.e.\ $S_{s_0}=-1$ or $P_{p_0}=-1$, respectively.  This clearly costs an excitation energy of $2K'$ or $2K$, respectively.  We refer to these excitations as ``electric'' ({\sf e}) or ``magnetic'' ({\sf m}) particles, respectively.  The interesting thing about these particles is that they cannot be created locally as individuals.  For example, if one acts on the ground state with $\sigma_i^z$, this creates {\em two} {\sf e} particles because the spin $i$ is shared by two stars.  One may also see this by noting that if one multiplies all the star operators, $\prod_s S_s=1$ identically, as each $\sigma_i^x$ operator appears twice in this product.  To actually create an isolated {\sf e} particle one needs to act on the ground state with a ``string'' of adjacent $\sigma_i^z$ operators, which results in an {\sf e} particle at each end of the string.  Similarly, {\sf m} particles are always created in pairs at the ends of analogous strings.  

\begin{figure}[htbp]
  \centering
  \includegraphics[width=3.0in]{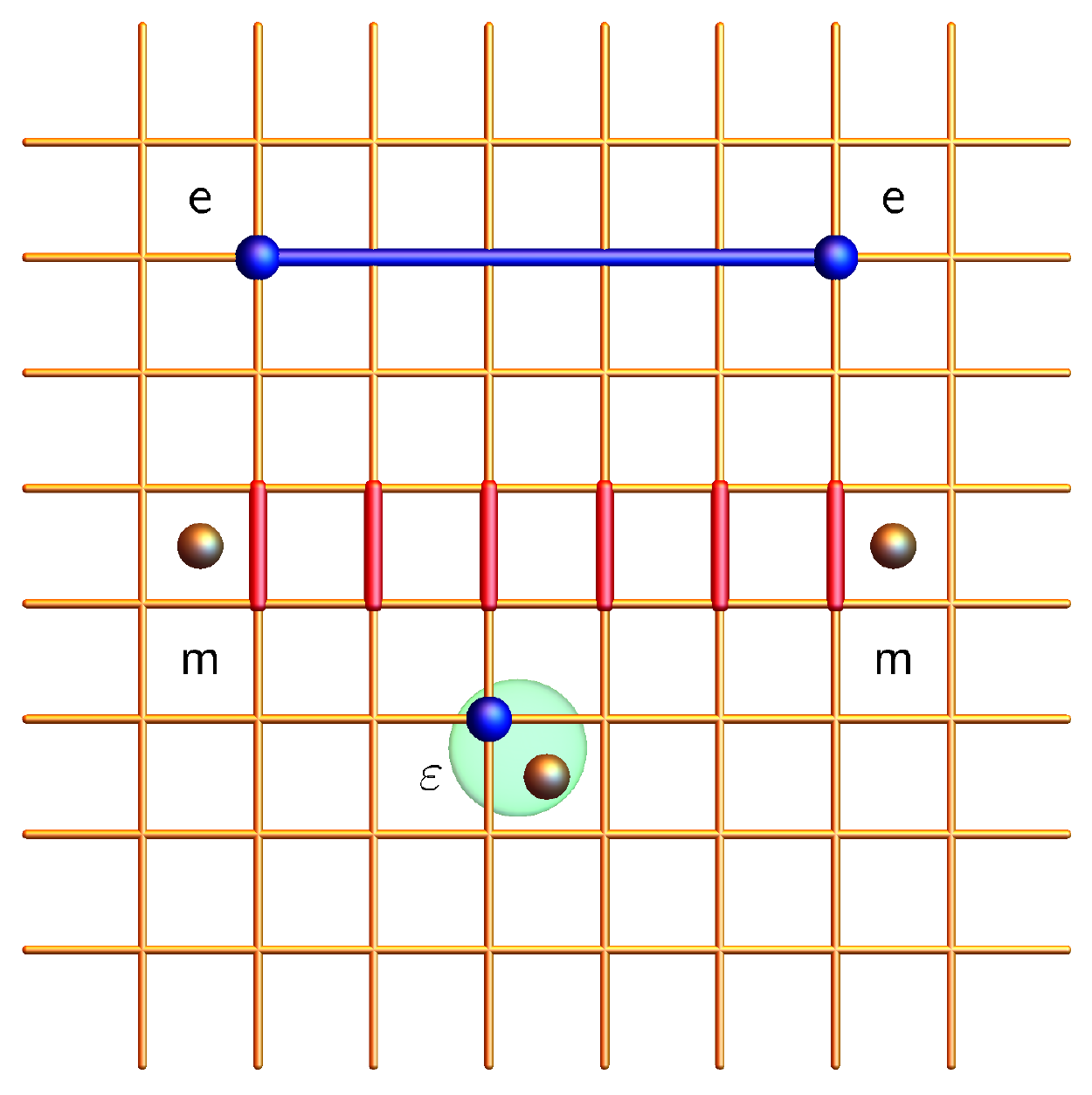}
  \caption{Anyons in the toric code.  A pair of {\sf e} anyons is created at the ends of a string of $\sigma_i^x$ operators as shown.  A pair of {\sf m} anyons is likewise created at the ends of a string of $\sigma_i^z$ operators, as indicated.  The $\varepsilon$ anyon is simply an {\sf e} and {\sf m} in close proximity. }
  \label{fig:toric}
\end{figure}

The non-locality of the {\sf e} and {\sf m} particles imbues them with unusual ``mutual statistics''.  Consider a state in which an {\sf m} particle is at the origin, and an {\sf e} particle is far away from it (see Fig.~\ref{fig:toric}).  This is defined simply by the state in which $S_s=1$ and $P_p=1$ for all stars and plaquettes {\em except} those where the particles are located, where they take a negative sign.  Now let us move the {\sf e} particle in a closed path around the {\sf m} particle.  We move the {\sf e} particle from one site to the next by acting with $\sigma_i^z$ where $i$ labels the link connecting the two sites.  By moving the particle one link at a time, we find that the final state is connected to the initial state by the action of a product of spins around this closed loop $\mathcal{L}$,
\begin{equation}
  \label{eq:5}
  |\psi_{\rm fin}\rangle = \prod_{i \in \mathcal{L}} \sigma_i^z |\psi_{\rm init}\rangle.
\end{equation}
Now we can use the discrete equivalent of Stokes' theorem, to rewrite
\begin{equation}
  \label{eq:6}
  \prod_{i \in \mathcal{L}} \sigma_i^z = \prod_{p \in \mathcal{A}} P_p,
\end{equation}
where $\mathcal{L} = \partial \mathcal{A}$, i.e.\ $\mathcal{A}$ is the area bounded by $\mathcal{L}$.  Since the initial state has a magnetic particle inside $\mathcal{L}$, then exactly one $P_p$ in this product is negative, and we find that
\begin{equation}
  \label{eq:7}
  |\psi_{\rm fin}\rangle = -|\psi_{\rm init}\rangle.
\end{equation}
Thus the act of bringing the {\sf e} particle around the {\sf m} one has induced a $\pi$ phase shift.  We say that the {\sf e} and {\sf m} particles have non-trivial {\em mutual statistics} (specifically they are mutual ``semions'').  In addition to the {\sf e} and {\sf m} particles, one can also consider a composite $\varepsilon$ particle which is just an {\sf e} and {\sf m} particle sitting next to one another.  Unlike the {\sf e} and {\sf m} particles, which have trivial bosonic self-statistics, the $\varepsilon$ particle is actually a {\em fermion} under self-exchange.  This can be seen by similar considerations to above.  
\begin{figure}[htbp]
  \centering
  \includegraphics[width=3.3in]{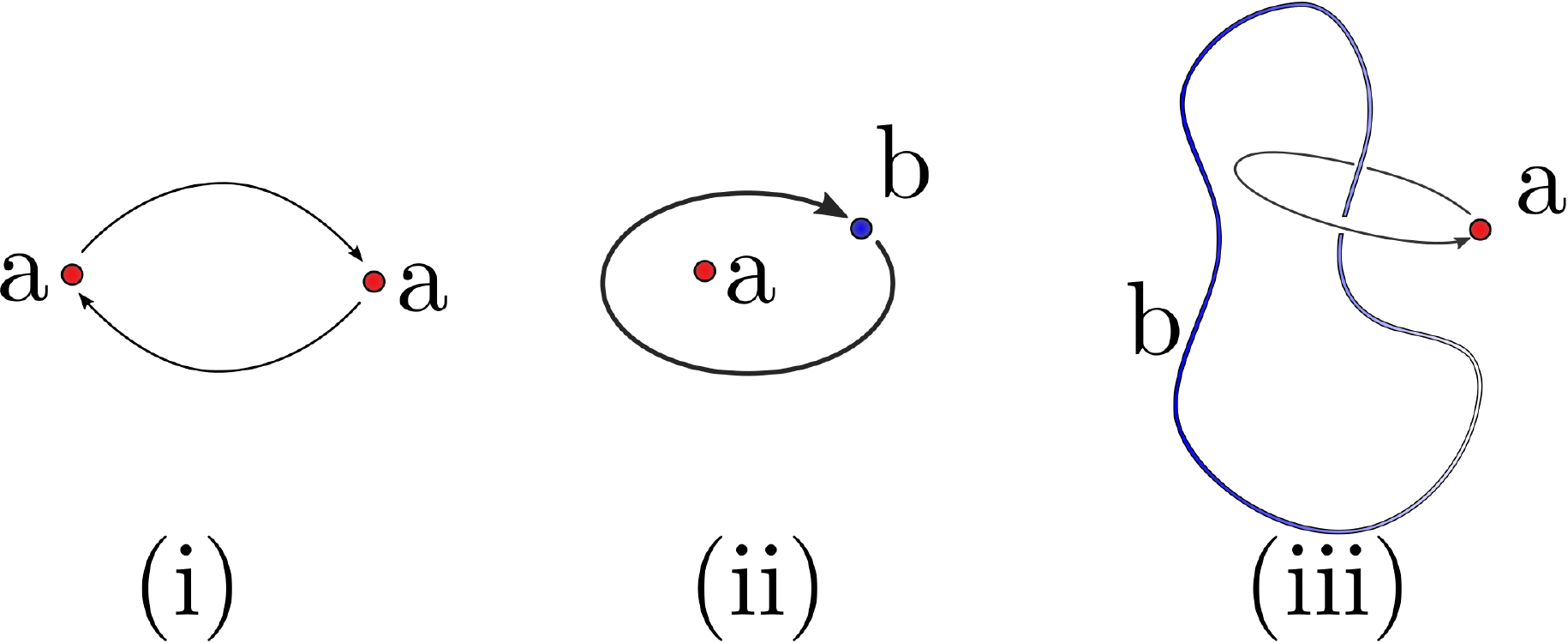}
  \caption{Statistics and mutual statistics in topological phases.  In (i), two identical quasiparticles of type a are exchanged.  This exchange may be accompanied by an arbitrary phase in two dimensions, which defines the self-statistics of an anyon.  In (ii) a quasiparticle of type b is moved around a quasiparticle of type a, and the wavefunction accumulates a phase which defines the mutual statistics of the two anyons in two dimensions.  In three dimensions, as shown in (iii), the mutual statistics of particles is replaced by the phase accumulated on moving a pointlike quasiparticle b around a closed loop which links with a line-like defect b.  The pointlike particles, however, must be bosons or fermions.  }
  \label{fig:braiding}
\end{figure}

The existence of these emergent anyons can be considered the defining feature of the toric code phase.  Although we have discussed them based on the exactly soluble Hamiltonian in Eq.~\eqref{eq:1}, they persist as excitations in the presence of arbitrary modifications to the Hamiltonian, provided those changes are no too large in magnitude.   All the other properties of the phase can be understood based on these quasiparticles (see e.g.\ Ref.\cite{kitaev2009}).   For example, the topological degeneracy can be derived by considering the process of locally creating a quasiparticle pair, transporting one of the quasiparticles around a cycle of the torus, and then annihilating it with its partner.  

\subsection{Anyons need entanglement}
\label{sec:entanglement}

How are these anyons connected to entanglement?  From several perspectives, it is clear that they cannot exist without massive entanglement.  If it were possible to approximate the ground state by a product form, then we would expect that excitations can be built from excitations of a single block.  But the excitations of a finite block must always have quantum numbers that can be obtained by adding those of the microscopic components of the system, e.g.\ electrons or spins.  Such excitations can always be created ``locally'', at least by operators defined within a single block.  These contradict the emergent electric and magnetic quantum numbers of the toric code anyons, and the fact that they can be created only in pairs.

\begin{figure}[htbp]
  \centering
  \includegraphics[width=3.3in]{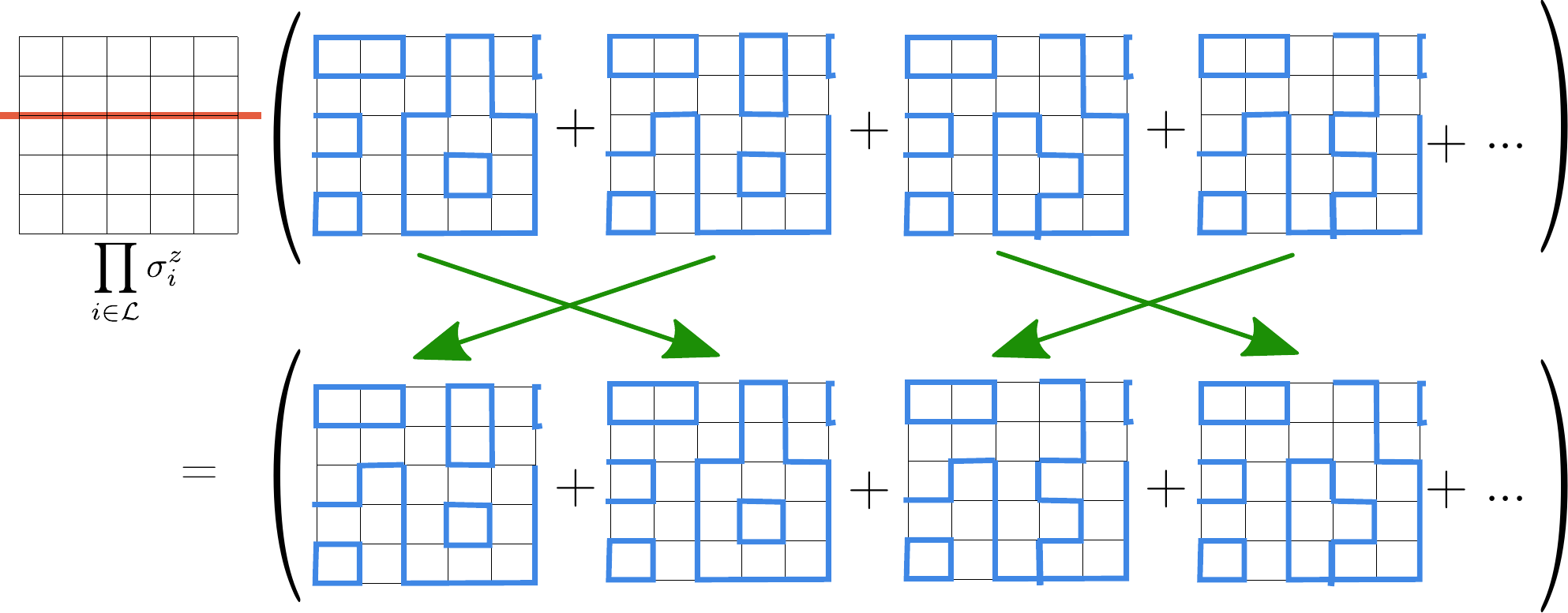}
  \caption{Illustration of how entanglement supports anyons in the toric code.  A pair of {\sf e} anyons at the ends of a line $\mathcal{L}$ is created by the action of a string of $\sigma_i^x$ along $\mathcal{L}$.  It is crucial that this string does not modify the ground state away from the ends, otherwise the state created would have an energy proportional to its length.  Here we show the action of this string, shown in red, away from its ends, on the toric code state, which is a superposition of loops.  Each component in the second line corresponds to the result of the action of the string operator on the component directly above it, in the first line. We can see that, while the action of the string modifies each component of the wavefunction all along the line, the result is simply another component of the original state, as shown by the arrows.  Consequently, the highly entangled superposition state is {\em not} modified by the string (except at its ends, which are not shown here). }
  \label{fig:superpose}
\end{figure}

Another symptom of the key role of entanglement in the topological phase is the non-trivial mutual statistics itself.  This implies a kind of ``action at a distance'': two anyons sense each other even when they are arbitrarily far away.  There must be some structure in the background of the wavefunction that allows them to maintain that information.  A more direct connection to entanglement can be made by considering the string operators that create pairs of anyons.  Consider a long string operator as in Eq.~\eqref{eq:6}.  When acting on the ground state it creates a pair of anyons which are well-separated, at an energy cost which is non-zero but finite (equal to $4K'$ in the ideal model).  It is actually surprising that this operator, which acts on a number of spins equal to its length, only increases the energy by a finite amount, and not by an amount proportional to its length!  If the ground state had a product form, we would expect that the portion of the string intersecting a particular block would disrupt the state on that block, leading to a finite increase of the energy of that block.  As a result, the total energy of the state acted upon by the string would be raised by an amount proportional to the length of the string.  The toric code model evades this energy cost by its massive entanglement:  because the ground state is itself a massive superposition, the string manages not to disrupt that ground state but instead {\em it simply reshuffles the components of the superposition}.  In this way, the anyons, which are created by a non-local process, can be supported by the entanglement of the ground state.  

% \begin{figure}[htbp]
%   \centering
%   \includegraphics[width=3.3in]{figures/cylinder.png}
%   \caption{Cylinder.}
%   \label{fig:cylinder}
% \end{figure}

\begin{figure}[htbp]
  \centering
  \includegraphics[width=3.3in]{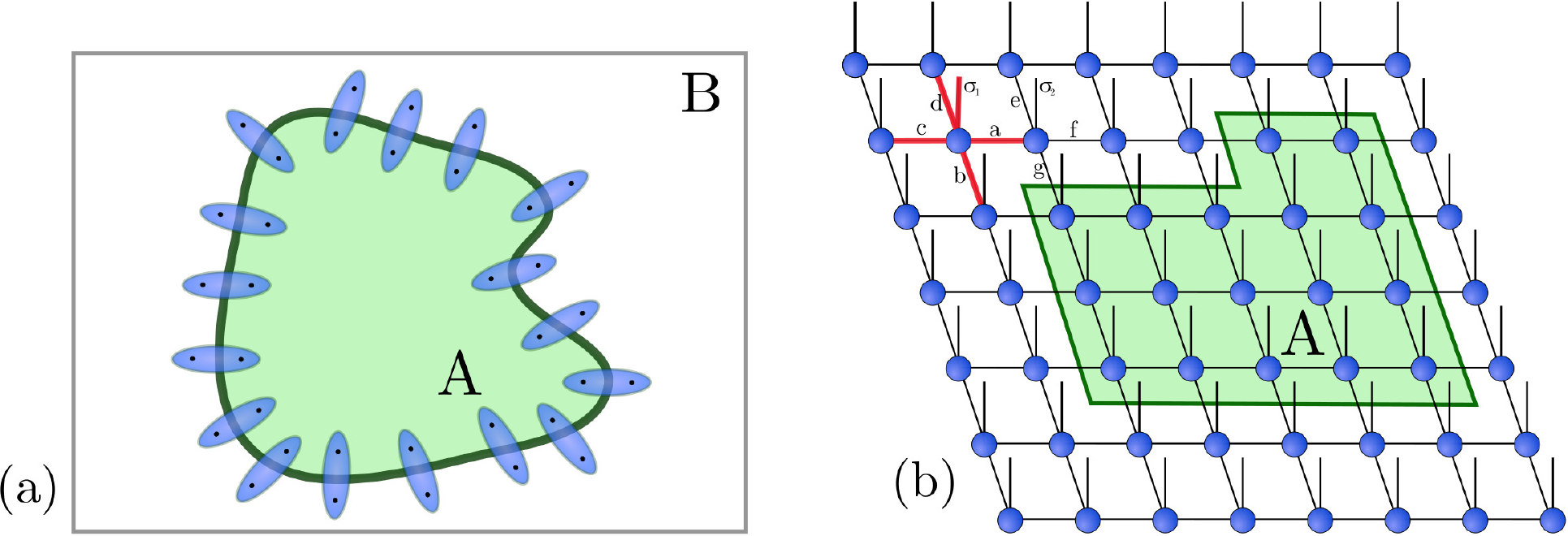}
  \caption{Area law and tensor network states.  In (a), we schematically show the division of space into a region A and its complement B.  Nearly all ground states of local Hamiltonians obey the ``area law'', where the leading term in the entanglement entropy for a large region A is proportional to the $d-1$-dimensional area of the boundary in $d$ dimensions, i.e.\ the length $L$ of the boundary in two dimensions, shown.  One can visualize the area law term as arising from the sum of local contributions due to entangled pairs (e.g.\ singlets) spanning the boundary, as shown in (a).   In a gapped system the next correction in two dimensions is a constant, which is known as the topological entanglement entropy, see Eq.~(\ref{eq:10}).  In (b) we show the representation of a quantum state by a Tensor Network State (TNS), specifically a Projected Entangled Pair State (PEPS), as in Eq.~(\ref{eq:75}).  One elementary tensor is shown in red, and contains here four internal indices and one physical one.  By construction, the entanglement between a region A (in green) and its complement is communicated only through the tensor legs crossing the boundary, so that the TNS construction automatically satisfies the area law.}
  \label{fig:peps}
\end{figure}

An even more direct connection to entanglement has been made, which proves very useful both conceptually and for numerical calculations.  This is made through the {\em entanglement entropy}, which is a general tool for studying entanglement in many body systems. Consider a system described by a wavefunction $|\psi\rangle$, and divide that system {\em in real space} into a region A and its complement B.  We form the reduced density matrix 
\begin{equation}
  \label{eq:8}
  \rho_A = {\rm Tr}_B |\psi\rangle \langle \psi|,
\end{equation}
which contains the information on the region A.  The (von Neumann) entanglement entropy which quantifies the entanglement between regions A and B is
\begin{equation}
  \label{eq:9}
  S(A) =- {\rm Tr}_A \, \rho_A \log\rho_A.
\end{equation}
This vanishes if the ground state is a product, i.e.\ $|\psi\rangle = |\psi_A\rangle \otimes |\psi_B\rangle$.  For a typical ground state, however, it is expected to behave according to
\begin{equation}
  \label{eq:10}
  S(A) \sim s_0 L - \gamma,
\end{equation}
where $L$ is the length of the boundary between regions A and B, and $s_0$ and $\gamma$ are constants, and we neglect terms that vanish as $L \rightarrow \infty$.  This form is generic for ground states with a gap to all excitations, and often also applies even to gapless systems.  The leading $s_0$ term describes local entanglement occurring near the boundary of the two regions, and is non-universal and depends upon details of the Hamiltonian.  This is known as the ``area law'' term, since it is proportional to the size of the boundary between A and B (which three dimensions would be an area). If there is an excitation gap, however, the correction $\gamma$ is universal if the boundary between A and B is smooth (the smoothness condition can be somewhat subtle\cite{jiang2012identifying,jiang2013accuracy}, and the correction may also be defined by a subtraction procedure\cite{kitaev2006topological,levin2006detecting}), and is topological in origin.  It is known as the {\em topological entanglement entropy}, and takes special quantized values.  For the toric code model, it is given by $\gamma = \ln 2$.  

The presence of a {\em negative} correction to the entanglement entropy may be surprising and not intuitive for a highly entangled ground state.  However its significance is understood as representing an obstruction to finding a disentangled wavefunction.  In particular, by its definition, the entanglement entropy itself is manifestly positive semidefinite, $S(A) \geq 0$.  The presence of the non-zero entanglement negativity $\gamma>0$ means that it is impossible to deform the ground state to obtain $s_0=0$, i.e.\ to remove the area law contribution.  When $\gamma=0$, which is the case in a non-topological gapped phase, the wavefunction can be deformed into a pure product state with $s_0=0$.  Thus a non-zero topological entanglement entropy is a direct indication of the ``long range entanglement'' in the ground state.   Numerical calculation of the topological entanglement entropy is difficult but beginning to be possible for non-trivial models of interest\cite{jiang2012identifying,isakov2011topological}.  

\subsection{Other topological phases}
\label{sec:other-topol-phas}

The toric code provides just one example of a topological phase.  Amongst highly entangled phases, topological phases are the simplest examples: there are no gapless excitations, and all local operators have exponentially decaying correlations.  Because of these facts, we expect low-lying quasiparticles to be sharply defined: they cannot decay, and they are as local as they can be (while still maintaining their anyonic nature).  Consequently, at least in two dimensions, the essential data characterizing topological phases is very minimal: a list of the quasiparticles, and rules for how they behave if two are brought together (``fusion rules'') and how they accumulate Berry phases when one is wound about another (``mutual statistics'').  With this in mind, one can contemplate a ``complete'' classification of topological phases in two dimensions, or at least a very thorough one.  This may already have been achieved.  In three dimensions, both quasiparticles and line-like defects may exist, and the interplay between these is much more complex, so that a full description of all possible topological phases seems more challenging.  An exposition of the classification of even two-dimensional topological phases is outside the scope of this paper, and probably not necessary or even useful for understanding QSLs.  However, it is instructive to understand, in addition to the two dimensional toric code, a few other cases of topological phases which can appear at least in relatively reasonable models of quantum magnetism.  

\subsubsection{Fractional quantum Hall states}
\label{sec:fract-quant-hall}

We begin with the first topological phases discovered: the Laughlin states in the fractional quantum Hall effect.  Such fractional quantum Hall (FQH) states exist for both fermions (as in the much studied physical situation) and for bosons or spins.  There is a family of such states, characterized by an integer $k$.  In the $k^{\rm th}$ state, there are $k$ anyons, which can be thought of as carrying a ``charge'' $n/k$ and a ``flux'' of $n$, where $n=1,\cdots, k$.   The charge and flux are emergent, not real charge and flux, but act similarly.    Here, in the FQH context, $k$ is related to the ``filling fraction'' by $\nu=1/k$, and to the Hall conductance by $\sigma_{xy} = \nu e^2/h$.  However, the fractional statistics of the anyons is in general more fundamental: one can start with a Laughlin state and add perturbations to the Hamiltonian which violate all the symmetries, including charge conservation, so that neither the filling fraction nor Hall conductance are well-defined.  Nevertheless, the topological order of the state and the mutual statistics of the anyons are preserved.  Thus these Laughlin states can appear in much more general situations than in the FQH effect, including quantum spin systems in which there is no magnetic field, no charges, and there may be only discrete symmetries.

Specifically, bringing a charge $q$ clockwise around a flux $f$ results in a phase factor of $e^{i2\pi q f}$, so that taking an $n$ anyon clockwise around an $n'$ anyon results in a phase factor of $e^{i2\pi n n'/k}$.  Interchanging two identical anyons of type $n$ is equivalent to half the rotation of one around the other, and so corresponds to a Berry phase of $e^{i\pi n^2/k}$.  These unimodular numbers $e^{i2\pi n n'/k}$ and  $e^{i\pi n^2/k}$ describe the mutual and exchange statistics of the anyons.  The fusion rule is simply that combining an $n$ and an $n'$ anyon results in an $n+n'$ anyon (mod $k$).  In the Laughlin sequence for fermions, $k$ is an odd integer, and so we can see that fusing $k$ primitive anyons with $n=1$ gives the $n=k$ anyon whose exchange results in the factor angle $e^{i\pi k^2/k} =-1$. This has Fermi statistics, and can be identified with the bare electron.  For bosons, $k$ is even, and the same procedure of fusing $k$ anyons gives a boson, corresponding to the bare boson.  Indeed this correspondence tells us that when the microscopic constituents of the system are bosonic, as is the case for spin systems and hence QSLs, only the Laughlin states with $k$ even can exist.

The Laughlin states are ``chiral'': the definition of the anyonic statistics requires us to specify the clockwise sense of rotation of one anyon about another, and if we reversed this sense (or equivalently, preserved the sense but complex conjugated all the phase factors) we would have a distinct and different topological state.  Consequently, these states cannot be realized without breaking time-reversal symmetry.  Moreover, and less obviously, these states {\em always} must have a gapless boundary.  The gapless modes comprise the famous gapless edge states in the FQH effect.  This is not obvious but can be established entirely from the statistics of the anyons\cite{PhysRevX.3.021009}, and, like the anyonic statistics, the gapless edge modes exist for the Laughlin states even if all symmetries of the system are broken.

\subsubsection{Three-dimension toric code}
\label{sec:three-dimension-z_2}

A perhaps full classification of topological phases exists in two dimensions.  The problem in three dimensions is much less explored, and offers qualitative differences.  An interesting example which has the most likelihood of experimental relevance is realized by the three dimensional toric code model.  This is described exactly by Eq.~(\ref{eq:1}) on for example the cubic lattice, where the stars $S_s$ become three-dimensional clusters of six bonds emerging from a site, and the plaquettes $S_p$ remain as they are but include over all three orientations (xy, yz, and xz planes).  It is solvable in the same way as the two dimensional toric code, and realizes a three dimensional topological phase.  For example, with periodic boundary conditions, there are $2^3=8$ degenerate ground states, and no local operator can distinguish these states.  

\begin{figure}[htbp]
  \centering
  \includegraphics[width=3.3in]{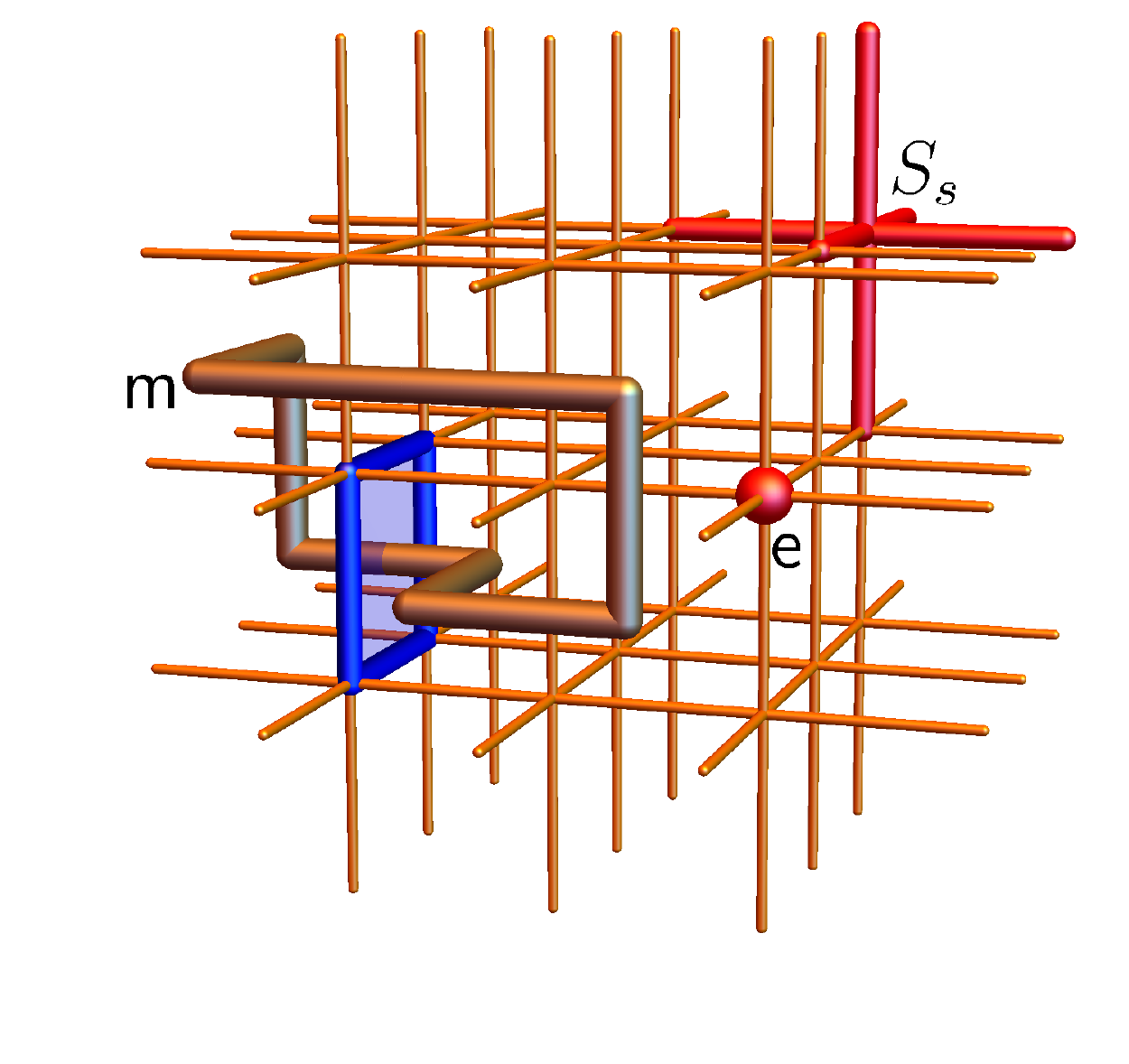}
  \caption{Three dimensional toric code.  A star operator is shown in red.  An {\sf e} excitation, which is particle-like, is shown as a red sphere.  The {\sf m} excitation becomes a loop, as indicated.  The loop resides on the dual lattice, and is defined by the plaquettes it pierces, which have negative flux, $P_p=-1$ -- one of these plaquettes is shown in the figure.}
  \label{fig:toric3d}
\end{figure}

Significant differences from two dimensions appear when we consider the {\em excitations}.  We can consider again defect sites at which $S_s=-1$.  These are point defects in three dimensions which may be called {\sf e} particles.  The magnetic excitations are, however, different.  This is because one cannot create a single plaquette with $P_p=-1$.  By construction, the product of $\prod_{p\in \mathcal{S}} P_p$ over all plaquettes forming a closed surface equals 1.  This is the equivalent of a Gauss' law integral, and implies that the magnetic excitations form {\em lines}, which can be seen as piercing these plaquettes with $P_p=-1$ and which cannot terminate in the sample.  Thus the magnetic {\sf m} defects form either closed loops or extended lines that cross the sample.  With a finite amount of energy, only small loops can be created, which are topologically trivial from far away.  Thus one does not really have a non-trivial magnetic {\sf m} {\em particle} in the spectrum of the three-dimensional toric code phase.  Instead of mutual statistics of particles, one has mutual statistics between the point {\sf e} particles and the line {\sf m} defects: if an {\sf e} particle is transported around a closed loop which {\em links} with an {\sf m} defect, a $\pi$ phase shift is acquired.

Because of the line-like nature of the {\sf m} defects, the three-dimensional toric code phase is more stable than the two-dimensional one.  Long defect lines cost an energy proportional to their length, which even at low non-zero temperature cannot be compensated by entropy.  Hence there is actually a finite temperature phase transition from a low temperature ``deconfined'' phase to a high temperature trivial one.

\subsection{How much entanglement?}
\label{sec:entangl-based-meth}

We have insisted, throughout this section, that a key, and even defining, feature of quantum spin liquid states is large and long-range entanglement. Here we want to refine this notion. If one uses entanglement entropy as quantitative measure of entanglement,  quantum spin liquid states, including topologically-ordered phases, are typically no more entangled than any other ground states.  That is apparent from Eq.~(\ref{eq:10}): the dominant term in the entanglement entropy is the area law, and the physical of topological order appears only as a correction. The cartoon picture for the area law in QSL states is that of short-range valence bonds. In forming a partition of the system, only those valence bonds astride the cut contribute to the entanglement entropy, and their number scales like the boundary size.  Indeed, adherence to the area law is believed to be a generic property of ground states of most local Hamiltonians.   Violations of the {\em leading} area law behavior occur only in situations with a very large number of gapless degrees of freedom, and even then the generic (i.e.\ stable) deviation from the area law is only by a multiplicative logarithmic factor.  For example, one dimensional quantum critical states with conformal invariance have a universal logarithmic term in the entanglement entropy, and free fermions with a non-vanishing Fermi surface have a multiplicative logarithmic correction in any dimension.  Trial wavefunctions for QSL states have been constructed (see Sec.~\ref{sec:gapless-matter-field}), which also have such a logarithmic correction.  Apart from these special examples, however, most QSL states still obey the strict leading area law.  Presumably the need for entanglement in QSLs means that the coefficient of the area law cannot be too small, but this is a non-universal quantity, which cannot be readily used.

This shortness of the range of the entanglement actually can be {\em used} for various types of analysis.  This is because area law states are actually very special, and form a highly restricted subspace in the entirety of the Hilbert space.  A randomly chosen state, and moreover even a Hamiltonian eigenstate with finite energy density, actually has much larger entanglement and obeys a volume law.  Hence one expects that area law wavefunctions are highly constrained and can be written in a special form.  The general belief is that an excellent approximation to an almost arbitrary area law state is provided by so-called Matrix Product States (MPS) in 1d, and their generalization, Tensor Network States (TNS) in higher dimensions\cite{verstraete2008matrix}.  A TNS can be considered as a quantum many body state constructed from a rule, in which the data for the state is specified by a set of finite local tensors which are contracted in some particular way.  A simple example of a TNS is a Projected Entangled Pair State\cite{verstraete2006criticality}, or PEPS, which, for concreteness written on a square lattice, takes the form 
\begin{equation}
  \label{eq:75}
   |\psi\rangle = \sum_{\{\sigma_i\} } \sum_{abc\cdots} \left[ T_{abcd}(\sigma_1) T_{aefg}(\sigma_2)\cdots \right] |\sigma_1\cdots\sigma_N\rangle,
\end{equation}
where the $\sigma_i$ are the physical ``spin'' variables defining states in some basis, that reside on sites $i$ of the lattice, and the roman letters $a,b,c,\cdots$ are ``internal'' indices which are placed on the links between lattice sites, and range over $D$ discrete values.  Each site with its surrounding links is endowed with a tensor $T_{abcd}(\sigma)$, whose values define the state.  This representation naturally encodes translationally invariant states which obey the area law.  It can represent both trivial and topological states, and for example one can easily encode the toric code ground state in this way.  So one can see explicitly from this type of construction that quantitatively, topological ordered phases and trivial ones have the same order of magnitude of entanglement.  

If one accepts that TNS can describe {\em all} fully gapped ground states, then one can classify these phases by classifying TNS, which is much simpler as one is basically classifying finite tensors $T_{abcd}(\sigma)$.  This is believed to be correct in one dimension\cite{chen2011classification}, where a TNS reduces to a MPS, since the tensor then has two internal indices and can be written as a matrix.  In two or more dimensions, fully gapped chiral topological states such as those discussed in Sec.~\ref{sec:other-topol-phas} so far have resisted representation as TNS\cite{wahl2013projected}, so completeness seems unlikely. However, the method does provide an understanding of a very large class of states\cite{chen2012symmetry}, and has led to many discoveries.  

Also important are numerical applications of TNS and MPS.  The assumption of this form of a quantum state greatly reduces the size of the Hilbert space which must be searched to find a ground state of a many body quantum system.  Indeed, it turns out that one of the most successful methods for finding quantum ground states, the Density Matrix Renormalization Group (DMRG), can be formulated as an algorithm to find an optimal MPS\cite{schollwock2011density}.  This understanding has led to various improvements of the DMRG, as well as a variety of new approaches that directly use higher dimensional TNS to seek low energy states\cite{jordan2008classical}.  These ``low entanglement'' approaches comprise a rapidly developing and very promising field of computational physics, which is particularly well suited to the study of QSLs. 

\subsection{Pr\'ecis of Section~\ref{sec:highly-entangl-quant}}
\label{sec:general-lessons}

From the toric code example, we saw that highly entangled phases can support emergent excitations with exotic properties that cannot be built from a finite number of fundamental constituents of the system.  These excitations have non-local properties (their mutual statistics) which derive from the fundamental non-local entanglement of the ground state.    The toric code is a representative of the subclass of highly entangled states which are fully gapped, which also includes fractional quantum Hall states.   These are called topological phases.  

Some general features of highly entangled phases are:
\begin{itemize}
\item The ground state cannot be smoothly deformed into, or approximated by, a product state over any finite blocks
\item The entanglement entropy shows deviations from the strict area law
\item Some elementary excitations are ``non-local'', in that they cannot be created individually by any set of local operators
\item These ``quasiparticles'' exhibit some form of long-range interactions with one another, for example anyonic mutual statistics
\end{itemize}

\section{Gauge theory}
\label{sec:gauge-theory}

The toric code model of the previous section provides a paradigmatic example of a highly entangled state of matter, and in particular of topological order.  Since Kitaev introduced it, the model has been extremely influential and led to many insights.  However, very closely related models and much of the physics was known in a different context, that of lattice gauge theory.  It turns out that gauge theory plays a central role in the current understanding of QSLs.  In this section we introduce some of the basic concepts of gauge theory and how they apply to QSLs.  

\subsection{U(1) gauge theory}
\label{sec:u1-gauge-theory}

A simple and (somewhat!) familiar place to start this discussion is the case of U(1) gauge theory.  This is nothing but a quantized lattice version of the familiar theory of electromagnetism.  The simplest model can be written on a $d$-dimensional cubic lattice, whose sites are labeled by roman letters $a,b,\cdots$.  The ``field'' variables form a single canonically conjugate pair on each bond, where we define $A_{ab} = -A_{ba}$, $E_{ab}=-E_{ba}$,
\begin{equation}
  \label{eq:18}
  \left[ A_{ab}, E_{ab}\right] = i ,
\end{equation}
and fields on different links commute.  We take the Hamiltonian to be
\begin{equation}
  \label{eq:19}
  H =  - K \sum_p \cos (\nabla \times A) + K' \sum_a \left( {\rm div}\, E \right)_a^2 + \frac{U}{2} \sum_{\langle ab\rangle}E_{ab}^2.
\end{equation}
Here we defined the lattice divergence
\begin{equation}
  \label{eq:24}
  \left({\rm div}\, E\right)_a = \sum_{b \in {\rm nn}(a)} E_{ab},
\end{equation}
as well as the curl $\nabla \times A$, defined on a plaquette by the discrete analog of a line integral,
\begin{equation}
  \label{eq:20}
  B_p=\left(\nabla \times A\right)_p = \sum_{a \in p} A_{a,a+1},
\end{equation}
with $a=1,2,3,4$ labeled sequentially clockwise around the plaquette, and $4+1=5=1$.  We can think of $A_{ab}$ as the vector potential, like in the usual electromagnetism, $E_{ab}$ as the electric field, and $B_p$ as the magnetic flux/field through a plaquette.   

The Hamiltonian as defined in Eq.~\eqref{eq:19} is periodic in $A_{ab}$ with period $2\pi$.  So we may require the wavefunction is also periodic: $\psi(A_{ab}+2\pi, \cdots) = \psi(A_{ab},\cdots)$.  This defines a ``compact'' U(1) gauge theory.  From Eq.~\eqref{eq:18}, $e^{i\theta E_{ab}}$ acts as a unitary operator to shift $A_{ab}$ by $\theta$, and so for the wavefunction to satisfy these periodic boundary conditions $E_{ab}$ must be an integer (or rather have integer eigenvalues).  

The expression of the Hamiltonian in terms of vector potential introduces, as usual, gauge invariance.  That is, $A_{ab}$ can be shifted by a gradient of some scalar function,
\begin{equation}
  \label{eq:22}
  A_{ab} \rightarrow A_{ab} + \chi_a - \chi_b,
\end{equation}
which leaves $B_\square$ invariant.  This transformation is generated by the unitary operator
\begin{eqnarray}
  \label{eq:23}
  U[\{\chi\}] &  = &  \exp\left[ i\sum_{\langle ab\rangle} E_{ab}(\chi_a-\chi_b)\right] \nonumber \\
  & = &  \exp\left[ i \sum_a \chi_a \left( {\rm div}\, E \right)_a \right],
\end{eqnarray}
The fact that this is a symmetry means that $({\rm div}\, E)_a$ commutes with $H$ for all $a$.  This is not surprising from the point of view of electromagnetism as this divergence, according to Maxwell, is just the charge density, and since we have not included any charged particles in Eq.~\eqref{eq:19}, it should be conserved.  Hence we can choose eigenstates of the Hamiltonian to also be eigenstates of charge, i.e.
\begin{equation}
  \label{eq:25}
  ({\rm div}\, E)_a |\psi\rangle = q_a |\psi\rangle,
\end{equation}
where $q_a$ are the fixed charges.  The model includes all possible integer values of these electric charges, a priori.  In the context of lattice gauge theory, it is normal to focus on the ``vacuum'' sector in which $q_a=0$.  These states are gauge invariant, since, using $q_a=0$ and Eq.~\eqref{eq:23}, $U|\psi\rangle =|\psi\rangle$.  But states with non-zero charges are just as physical, and simply transform by a phase factor $U|\psi\rangle = e^{i\sum_a q_a \chi_a}|\psi\rangle$.  

A very close analogy with the toric code model in the limit $U\rightarrow 0$.  In this limit, {\em both} $({\rm div}\, E)_a$ {\em and} $({\rm curl} \,A)_p$ commute with $H$.  So the ground state should be found just by choosing $q_a=({\rm div}\, E)_a=0$ on all sites and $B_p=({\rm curl} A)_p =0$ on all plaquettes.  We see that $q_a$ and $B_p$ are analogous to $S_a$ and $P_p$ of the toric code, respectively.  The situation is just somewhat more subtle than in the toric code case because since $B_p$ is a continuous variable, states with very small but non-zero $B_p$ (which are like magnetic charges in the toric code but with very small charge) have very low energy, and so that stability of this na\"ive ground state is less obvious.  The electric charges have a well-defined gap $K'$, however.  

\subsubsection{Coulomb or deconfined phase}
\label{sec:coul-or-deconf}

A na\"ive analysis consists of taking very small but non-zero $U$, which we expect to introduce {\em small} fluctuations of $B_p$.  Assuming they are small, we may then Taylor expand the cosine term to obtain, up to a constant
\begin{equation}
  \label{eq:26}
  H \approx K' \sum_a q_a^2 + \frac{U}{2} \sum_{\langle ab\rangle}E_{ab}^2 + \frac{K}{2} \sum_p B_p^2.
\end{equation}
This is just the usual electromagnetic energy in terms of field strengths squared.   When this expansion is valid, we say that the system is in the {\em Coulomb phase}, sometimes also known as the {\em deconfined phase}.  It turns out that the na\"ive analysis is correct, and the expansion is valid, in the limit described ($U \ll K$) in three dimensions {\em but not in two dimensions}.  We will return to the two-dimensional case later in Sec.~\ref{sec:stabiluty-u1-gauge}.

For now we concentrate on three dimensions.  Since the expanded $H$ is quadratic in $E_{ab}$ and $A_{ab}$, it can be readily diagonalized by normal modes.  We will not go into the details here, which are just the usual treatment of quantized electromagnetism but on a lattice\cite{hermele2004pyrochlore}.  Since the normal modes with different momentum do not interact, the low energy, long wavelength physics is captured just by the continuum limit of Eq.~(\ref{eq:26}):
\begin{equation}
  \label{eq:78}
  H \approx \int \! d^3x\, \left[ \frac{\epsilon}{2} |\vec{E}|^2 + \frac{1}{2\mu} |\vec{B}|^2\right].
\end{equation}
This is just standard vacuum electromagnetism, with some effective dielectric constant $\epsilon$ and magnetic permeability $\mu$.   As a consequence, there are propagating linearly dispersing gapless ``photon'' modes, with velocity $c_{\rm eff} = \sqrt{\epsilon\mu}$, which exist with two transverse polarizations, just as in ordinary electromagnetism. Second, above an energy gap, non-zero charges can be present.  For these charges, there is an additional contribution to the ground state energy in the corresponding sector, which is {\em finite} but long-ranged: beyond the finite $K'$ energy explicit in Eq.~\eqref{eq:26}, one finds the usual long-range Coulomb potential, proportional to $q_a q_b/r_{ab}$, where $r_{ab}$ is the distance between sites $a$ and $b$.  

A third property of the three-dimensional Coulomb phase is not apparent from the na\"ive approximate Hamiltonian in Eq.~\eqref{eq:26}.  In addition to the electric charges which are explicit here, implicitly in the original model of Eq.~\eqref{eq:19}, {\em magnetic} charges are also possible.  These are {\em point defects} in three dimensions, near which $B_p$ is not small and so the expansion of the cosine is not possible, but where this is possible far away.  Owing to the periodicity of the cosine, such a defect can act as a source of magnetic flux, so that if one sums $B_p$ over a set of plaquettes surrounding the defect (the discrete analog of the surface integral), the result is a non-zero integer multiple of $2\pi$.  This is possible because in fact $B_p$ is defined only modulo $2\pi$.  This excitation is, for obvious reasons, called a magnetic monopole.  Within the Coulomb phase, the magnetic monopoles behave similarly to the electric charges: they are separated by a non-zero but finite energy gap (of order $K$ rather than $K'$) from the ground state, and experience a $1/r$ Coulomb potential with one another.  
 
\subsubsection{Non-locality of excitations}
\label{sec:non-local-excit}

The Coulomb phase is {\em not} a topological phase, which must have a gap to all excitations.  There is no ground state degeneracy, nor is there an exponentially approximate one, on a torus or other multiply connected manifolds (there are low energy states on a torus whose energy decays as a power law of system size associated with flux threading the torus\cite{hermele2004pyrochlore}).  The system does not support anyons, which anyway are not well-defined in three dimensions.  However, the Coulomb phase does share the essential attribute that it supports excitations which are non-local, and cannot be created singly by any local operator.  

The electric charge is the simplest such excitation.  An isolated electric charge is defined, in the Kitaev-like limit where $U\rightarrow 0$, by a single site with non-zero $q_a$.  Consider a general local operator, consisting of operators within some finite region $R$.  It can be expanded in a sum of terms of the form
\begin{equation}
  \label{eq:21}
  \mathcal{O} = F(\{ E_{ab}\}_{\langle ab\rangle \in R}) \times e^{i \sum_{\langle ab\rangle \in R} m_{ab} A_{ab}},
\end{equation}
where $m_{ab}$ must be integers to maintain periodicity.  Acting on any given state, this operator shifts $E_{ab} \rightarrow E_{ab} + m_{ab}$.  Notably $m_{ab}=0$ for bonds $ab$ outside of $R$.  Now we can consider the total charge in a region $\hat{R}$ larger than $R$, 
\begin{equation}
  \label{eq:27}
  Q_{\hat{R}} = \sum_{a \in \hat{R}} ({\rm div}\, E)_a = \sum_{\langle ab\rangle \in \partial \hat{R}} E_{ab},
\end{equation}
where we rewrote it as a surface ``integral'' using Gauss' law.  The change in this charge induced by $\mathcal{O}$ is then
\begin{equation}
  \label{eq:28}
  \Delta Q_{\hat{R}} = \sum_{\langle ab\rangle \in \partial \hat{R}} m_{ab} = 0.
\end{equation}
This vanishes because the bonds $m_{ab}$ in the final sum are outside the domain $R$ acted on by $\mathcal{O}$.  So local operators cannot create single charges, only neutral combinations such as dipoles.  Nevertheless, in the Coulomb phase, these charges are elementary excitations.  A different way to see this is that an electric charge can be sensed infinitely far away by its electric Coulomb force, or, the electric field lines emanating from it.  Since a local operator cannot generate field lines far from the region it acts upon, it cannot create a non-neutral charge configuration.  Likewise, magnetic monopoles cannot be created singly by local operators, since they too experience long-range Coulomb forces and are associated with magnetic field lines which extend to infinity in the Coulomb phase.  The existence of such non-local but finite energy excitations is characteristic of a highly entangled phase.  

\subsection{Toric code as a gauge theory}
\label{sec:toric-code-as}

In the prior subsection, we introduced U(1) gauge theory, which has an advantage in terms of understanding that we were all trained from undergraduate days in electromagnetism.  We saw that the simple compact U(1) gauge theory in fact is quite analogous to Kitaev's toric code.  Like the toric code, at least in three dimensions, it supports non-local excitations, and we can understand the non-locality of those excitations as due to the field lines emanating from them.  

We now revisit the toric code model and discuss its representation as a gauge theory.  In fact, the toric code model itself is nothing but a particular recast 
version of the lattice Ising gauge theory introduced by Wegner and others. Indeed, the correspondence between the $\sigma$, $P$ and $S$ variables of the toric code with the lattice electric $E$ and vector potential fields $A$ of Wegner goes as follows.   We can think of the $\sigma_i^z$ fields which live on the links $i$ connecting sites $a$ and $b$ of the lattice as analogous to the exponential of the vector potential $A_{ab}$, living on that link, $\sigma_i^z \sim e^{i A_{ab}}$, but with $A_{ab}=0,\pi$ only.  Then the plaquette operator $P$ is the lattice curl of the vector potential, i.e.\ the magnetic flux, on the plaquette in question, $P \sim \cos (\nabla \times A)$.  In this picture, the $\sigma_i^x$ fields become the conjugate ``electric fields'' $\sigma_i^x = e^{i \pi E_{ab}}$ living on the links, with $E_{ab}=0,1$.  The star operator is $S_a = e^{i \pi ({\rm div}\, E)_a}$, the exponential of the divergence of the electric field.  By Gauss' law we identify $S_a=+1$ with a neutral site, and $S_a=-1$ with a unit electric charge on site $a$.   This explains the use of ``electric'' and ``magnetic'' terminology for the toric code example.   The sense in which this is a gauge theory is that the Hamiltonian enjoys a local symmetry.  Since $S_a$ commutes with the Hamiltonian, we can consider it as a generator of a local symmetry.  Generally we can consider the unitary and hermitian operator
\begin{equation}
  \label{eq:12}
  U[\{ s\}]\equiv U[s_1,\cdots,s_N] = \prod_a S_a^{s_a},
\end{equation}
where $s_a=0,1$.  Under this symmetry, we have
\begin{equation}
  \label{eq:11}
 \sigma_{ab}^z \rightarrow  U[\{ s \}] \sigma_{ab}^z U[\{ s\}]  = (-1)^{s_a} (-1)^{s_b} \sigma_{ab}^z.
\end{equation}
Here we replace the bond label $i$ by the two sites $a$ and $b$ which it connects.  In this way $\sigma_{ab}^z$ appears like a ``link'' variable.  We can freely choose the site variables $s_c$ and under this transformation $\sigma_{ab}^z$ transforms like a gauge connection, and $H$ is invariant.

\subsection{Stability of the toric code: {\em emergent} gauge structure}
\label{sec:stab-toric-code}

A key attribute of both the Coulomb phase of the three-dimensional U(1) gauge theory and the topological phase of the $Z_2$ gauge theory is their stability.  They are robust to {\em all} weak perturbations, even those which explicitly break the gauge symmetry.  This is even possible to prove rather rigorously\cite{hastings2005quasiadiabatic}.  Here we discuss this at a physicist's level of rigor, using the toric code model, where it is simplest, for illustration. The stability may seem strange from the gauge theory point of view, since the fact that $S_a$ commutes with the Hamiltonian is a special feature of the toric code model.  Let us consider a generalization of the toric code which breaks these symmetries:
\begin{equation}
  \label{eq:13}
  H = H_{\rm tc} - h_z \sum_i  \sigma^z_i - h_x \sum_i \sigma^x_i.
\end{equation}
How do we discuss this in the gauge theory language?  Well, we can rewrite the problem by introducing some redundant degrees of freedom.  Let us introduce some new variables $\tau_a^x = \pm 1$ on the original sites, and $\mu_p^z = \pm 1$ on the original plaquettes.  We define an enlarged Hilbert space $\mathfrak{H}_{big}$ which is the direct product of the original Hilbert space $\mathfrak{H}_{0}$ and those of these two new variables, $\mathfrak{H}_{big}=\mathfrak{H}_{0}\times\mathfrak{H}_\tau\times\mathfrak{H}_\mu$.  Now for each physical state of the original system, we define a single state in the enlarged space, by taking $\tau_a^x = S_a$ and $\mu_p^x = P_p$ (in practice, we can consider a complete basis of $\sigma_i^z$ product states, and define $\tau^x_a$ for each one, and then expand each such basis in the dual $\sigma_i^x$ basis, and then define $\mu_p^x$ for each expanded component).  So we map in this way each state $|\psi\rangle$ in the original Hilbert space to a unique state $|\Psi(\psi)\rangle$ in the enlarged space. Now we would like to write a new Hamiltonian $\mathcal{H}$ which operates in the enlarged space and in that space reproduces the original Hamiltonian matrix elements:
\begin{equation}
  \label{eq:14}
  \langle \Psi(\psi)| \mathcal{H}|\Psi(\psi)\rangle = \langle \psi|H|\psi\rangle. 
\end{equation}
This is achieved by writing
\begin{equation}
  \label{eq:15}
  \mathcal{H} = H_{\rm tc} - h_z \sum_{\langle ab\rangle} \tau_a^z \sigma_{ab}^z \tau_b^z - h_x \sum_{\langle pp'\rangle} \mu_p^z \sigma_{pp'}^x \mu_{p'}^z ,
\end{equation}
where $\sigma_{pp'}^x$ is defined on the link of the original lattice which is shared by neighboring plaquettes $p$ and $p'$.
This must be supplemented by the constraints which impose the physical subspace of the larger Hilbert space,
\begin{equation}
  \label{eq:16}
  \tau_a^x = S_a, \qquad \mu_p^x = P_p.
\end{equation}
Using those, we are free to rewrite the toric code Hamiltonian in terms of the new variables to obtain
\begin{eqnarray}
  \label{eq:17}
  \mathcal{H} & = & - K \sum_p \mu_p^x -K' \sum_s \tau_s^x \\
  && - h_z \sum_{\langle ab\rangle} \tau_a^z \sigma_{ab}^z \tau_b^z - h_x \sum_{\langle pp'\rangle} \mu_p^z \sigma_{pp'}^x \mu_{p'}^z.\nonumber
\end{eqnarray}
This Hamiltonian (Eq.~\eqref{eq:17} or Eq.~\eqref{eq:15}) has gauge symmetry by construction, which is generated by the relations in Eq.~\eqref{eq:16}, i.e.\ the operators $S_s \tau_s^x$ and $P_p \mu_p^x$ commute with ${\mathcal H}$.  It can be considered as a lattice Ising gauge theory with ``matter'' fields, with $\tau_a^z$ realizing the {\sf e} particles and $\mu_p^z$ realizing the {\sf m} particles.  By construction, all microscopic observables are gauge invariant, and hence cannot create either particle singly, since operators which create single anyons are not gauge invariant.  

More importantly, the ground state for small $h_z, h_x$ has an {\em emergent} gauge symmetry, which is descended from the gauge symmetry of the ideal toric code.  When $h_z=h_x=0$, then $S_s$ and $P_p$ {\em separately} (or equivalently $\tau_s^x$ and $\mu_p^x$) commute with $\mathcal{H}$.  This is clearly not an exact gauge symmetry of the full Hamiltonian.  However, since the ground state for $h_z=h_x=0$ is unique up to global degeneracies which depend upon the geometry, and separated by a gap from the excited states, it retains this structure to all orders in perturbation theory.  It is physically obvious from Eq.~\eqref{eq:17}, since the $\tau$ and $\mu$ spins are obviously polarized and spin flips have a large gap, and so transverse fluctuations can be systematically integrated out order by order in $h_x,h_z$.  Physically, when the matter fields are gapped, they can be integrated out to obtain an effective pure gauge theory.  In doing so, one obtains virtual {\em pairs} of electric and magnetic particles in the ground state.  Crucially, however, members of a pair are always nearby, and so do not disturb the physics on long distances.

\subsection{Confinement}
\label{sec:confinement}

While the topological phase of the toric code is stable to arbitrary small perturbations, large perturbations can destroy it.  Consider the Hamiltonian in Eq.~(\ref{eq:13}) with very large $h_x\gg h_z, K, K'$.  In this case, the ground state is, to a good approximation, uniquely determined and equal to that state $|+\rangle = \otimes_i |\sigma_i^x=+1\rangle$ on  all links.  Clearly this is a completely unentangled state, and by the arguments of Sec.~\ref{sec:entanglement} cannot possibly support non-local anyonic excitations.  To see this explicitly, consider trying to create an {\sf e} or {\sf m} anyon.  For the {\sf e} anyon, we must create a site around which the product of $\sigma_i^x$ is minus one.  This obviously requires introducing some bonds with $\sigma_i^x=-1$.  In fact, to introduce an {\em isolated} {\sf e} particle, we must introduce a semi-infinite string of these negative bonds.  Each such bond costs an energy of $2h_x$, so that the energy of the putative {\sf e} particle is linear in the length of the string, which is the system size, or the distance to the nearest other {\sf e} particle.   Hence, if we only supply a finite amount of energy to the system, we can never create an isolated {\sf e} particle: it remains {\em confined} to within some short distance of a partner.  The linear growth of energy with separation of two gauge charges is a characteristic behavior of confinement, which occurs in many gauge theories.  

What happens to the {\sf m} particle?  Well, recall that it corresponds to a plaquette in which the flux $P_a=-1$.  To make one, one needs to create a semi-infinite row of bonds with $\sigma^z_i=-1$.  This is created by an operator which is a product of $\sigma^x_i$ along these same bonds.  Now, let us imagine acting with this string upon the $|+\rangle$ state.  It is an eigenstate of $\sigma_i^x$, so this operator does not create any excitation at all!  So the {\sf m} particle has ceased to exist.  Using a bit of poetic license, it is tempting to say that the ground state is a ``condensate'' of {\sf m} particles, since the operator which creates {\sf m} actually gives back the ground state -- which implies this operator has off-diagonal long-range order.  This is actually a valid viewpoint!  Indeed, one can even develop a theory of the transition from the topological phase to the confined phase based on condensate of the {\sf m} particle.  From the point of view of the {\sf m} excitations, the trivial state $|+\rangle$ is what is called a {\em Higgs phase}.  Since it is both a confined phase for {\sf e} and a Higgs phase for {\sf m}, we have seen that, in this case at least, a Higgs phase (gauge particle condensate) and confined phase are the same thing.  This is true in many gauge theories.  

\subsection{Stability of the U(1) Coulomb phase}
\label{sec:stabiluty-u1-gauge}

Because it is gapless, the question of stability of the Coulomb phase of the U(1) gauge theory is more subtle than its counterpart for a topological phase. It turns out that the na\"ive procedure is correct in three dimensions but not in two.  The subtlety is that large $K$ implies that ${\rm curl} \, A$ is {\em typically} small but it can experience large but short-lived, short-distance fluctuations.  These have a qualitative effect.  If we assume, as we do with the na\"ive expansion of the cosine, that $B = {\rm curl}\, A$ is {\em always} small, then we conclude that there is a topological conservation law.  In three dimensions, since ${\rm div}\, B = 0$, the flux through (sum of $B$ over plaquettes) through a closed surface is always zero.  In two dimensions, the flux through a large area in the plane can change only by flux passing in or out of the boundary.  However, this flux conservation is violated if we allow for variations of $A_{ij}$ or $B$ on the scale of $2\pi$.  

One can show that the violation of flux conservation takes the form of {\em monopoles}.  In three dimensions, these are point defects in which a net flux of $\pm 2\pi$ exits a microscopic volume of the lattice.  They really behave like particles with magnetic charge.  Due to the large flux (e.g.\ of order $2\pi/6$) through plaquettes near the center of the monopole, they cost a large energy of order $K$.  However, the total energy of a monopole is finite.  In three dimensions, this makes the monopole a gapped excitation, and this gap is sufficient to prevent these excitations from entering the ground state and disrupting the Coulomb phase.  Thus the Coulomb phase is stable in three dimensions.  

In two dimensions, the monopole is not a particle but an {\em instanton}.  In a path-integral formulation, it is a configuration in space-time in which the flux through an area evolves and changes by $\pm 2\pi$.  In the Hamiltonian formalism, operators must be added to the Hamiltonian which insert such fluxes locally with some amplitude, known as a fugacity.  One should imagine that in the middle of the event, there is a large flux concentrated through some set of plaquettes at the location of the instanton.  During this temporary period, there is a large energy cost.  However, the cost is short-lived, and so the resulting {\em action} is finite.  Consequently, these instantons appear with a finite density in space-time, and the conservation of flux is spoiled.  In a renormalization group or scaling view, we say that the monopole fugacity is a relevant perturbation to the Coulomb ``fixed point''.  As shown first by Polyakov, this leads to destruction of the Coulomb phase, a gap for the photon, and confinement of gauge charged excitations: {\em the Coulomb phase of the pure gauge theory is unstable in two dimensions}.  That is, there can be no Coulomb or U(1) phase whose only low energy mode is the photon in two dimensions.  
Instead, the gauge theory in Eq.~\eqref{eq:19} in two dimensions is actually in a {\em confined} phase, which is smoothly connected to the limit of large $U$.  

The above discussion describes the stability of the Coulomb phase in the model of Eq.~\eqref{eq:19}, which has U(1) gauge invariance.  One may wonder if the phase remains stable if this invariance is broken explicitly, which is certainly possible since it is not a microscopic symmetry.  For example, a term such as $-t \sum_{\langle ab\rangle} \cos A_{ab}$ may be added to the Hamiltonian.  One might expect that, since this term breaks gauge symmetry, it will immediate generate a mass for the photon, in any dimension.  Indeed one obtains a mass if one na\"ively expands the cosine.  

However, in fact the Coulomb phase in three dimensions is also stable to such perturbations: no gap obtains for small but non-zero $t$.  First, it is clear that expanding the cosine is not justified, since there is no reason for fluctuations of $A_{ab}$ to be small (large $K$ only controls fluctuations of $B= {\rm curl} \,A$).  Indeed, fluctuations of $A_{ab}$ are enormous.  In a perturbative expansion of the partition function in powers of $t$, one can actually see that only terms which combine to form gauge invariant combinations of $A_{ab}$ -- i.e.\ fluxes $B_p$ -- yield non-zero contributions.  This is in fact a manifestation of Elitzur's theorem.  More physically, the term $\cos A_{ab}$ acting on a state creates two opposite electric charges at sites $a$ and $b$.  An isolated charge has a gap of order $U$, and hence these charges cannot unbind and remain trapped as dipoles.  Perturbation in $t$ simply produces virtual dipole fluctuations in the ground state, renormalizing the dielectric constant and magnetic permeability of the effective Maxwell electrodynamics.  We see that the effect of ``breaking'' gauge invariance in the pure gauge theory is simply to introduce electric excitations {\em above a gap} into the theory.  As we did in Sec.~\ref{sec:stab-toric-code} for the toric code case, one can make this explicit by introduce some slave matter fields analogous to the $\tau_a^z$ variables in Eq.~\eqref{eq:15}, which we leave as an exercise to the reader. So long as the gap remains non-zero, the Coulomb phase is stable.  We see that U(1) gauge invariance can ``emerge'' even if it is broken explicitly.  This was first noted in Ref.\cite{foerster1980dynamical}.

\subsection{Universal properties of U(1) Coulomb phases in three dimensions}
\label{sec:univ-prop-u1}

The arguments of the previous subsection clarify that the Coulomb phase of the U(1) gauge theory is stable in three dimensions.  What are its universal properties?  We may try to develop a picture which is as close as possible to that for the $Z_2$ topological or toric code phase in two dimensions, i.e.\ to ask what is the ``minimal data'' needed to characterize such a U(1) QSL?  

We have already noted that at low energy, the Coulomb phase is simple (though not as simple as a topological phase, which has {\em no} low energy excitations!), and is described just by Maxwell electromagnetism.  This implies the existence of a propagating photon mode, with two transverse polarization states, just as in for real electromagnetic radiation.  However, this ``light'' is artificial, and couples to emergent electric charge rather than real charge.  The emergent photon of a spin system would be expected to have a slow propagation speed, related to the strength of magnetic exchange.  

Above an energy gap, the Coulomb phase possesses excitations that can be labeled by their emergent electric and magnetic charges.  For the standard compact U(1) gauge theory presented in previous sections, these two charges are quantized in integer multiples of those of the primitive electric and magnetic charges.  So we can view all possible charges as superpositions of primitive {\sf e} and {\sf m} particles (and their antiparticles $\overline{\sf e}$ and $\overline{\sf m}$, which are distinct since the charges in U(1) gauge theory are integers rather than $Z_2$ values as in the toric code), where we have borrowed notation in an obvious way from the toric code case.   In this standard theory, the {\sf e} and {\sf m} particles are bosons.  As in the toric code, one can consider a composite particle $\varepsilon$ which is the fusion of {\sf e} and {\sf m} and is here called a ``dyon''.  By considering the Aharonov-Bohm Berry phase associated with moving the flux around the charge, one can derive that the dyon $\varepsilon$ is actually a fermion in this case: interchanging two dyons results in a net $\pi$ phase rotation.  In general the excitations can be labeled by two integers $n_1$ and $n_2$, corresponding to the number of {\sf e} and {\sf m} particles fused (these integers can take any sign or vanish), and the result has electric and magnetic charge $(Q,M)=(n_1,n_2)$ in fundamental units.  The particle is a boson or a fermion if $n_1 n_2$ is even or odd, respectively.

Do {\em all} U(1) Coulomb phases behave in this way?  In general they do not.  From quantum field theory, it has been known for some time that an interesting modification of Maxwell theory is possible, known as an axion electrodynamics.  This consists of adding to the action the term
\begin{equation}
  \label{eq:73}
  S_{\theta} = \frac{\theta}{8\pi^2} \int dt d^3x \epsilon_{\mu\nu\gamma\lambda} \partial_\mu A_\nu \partial_\gamma A_\lambda.
\end{equation}
This has the same number of derivatives as the usual Maxwell term, and so is on the same footing.  Indeed it can be rewritten as just a multiple of $\vec{E}\cdot \vec{B}$, so can be thought of as a kind of mixing of diamagnetic and dielectric response.  It has the unusual feature of being a total derivative, and hence can be rewritten as a pure boundary term.  Consequently it has no effect on the local dynamics of the gauge field, e.g.\ the photon mode.   In this sense it does not actually have any low energy effect.  However, it does modify the behavior of gapped excitations, and in particular magnetic monopoles.  Specifically, the $\theta$ term couples charge and spin, and imbues magnetic monopoles with electric charge, with a fundamental monopole acquiring an electric charge equal to $\theta/(2\pi)$ times the fundamental electric charge.

Because the axion term does not modify the low energy physics, it appears one can understand this modified U(1) Coulomb phase completely without invoking Eq.~(\ref{eq:73}), by instead replacing it by a statement about the quasiparticle content of the gauge theory.  This is like viewing the toric code state not as a gauge theory but as a topological phase whose minimal data is simply the set of anyons it supports and their mutual and self statistics.  The analogous statement here is that the Coulomb phase is described by the vacuum Maxwell theory at low energies, supplemented by a set of gapped quasiparticles carrying electric and magnetic charges.  In general, the allowed set of these two charges which are carried by quasiparticles does not need to coincide with that of the simple lattice compact U(1) gauge theory, but can be modified in a way that corresponds to a non-trivial $\theta$ angle.  

To arrive at this constructively, let us first suppose the existence of an electrically charged particle, whose magnetic charge is zero.  So this {\sf e} particle has $(Q,M) = (1,0)$.  Now we ask if we can allow a particle which has magnetic charge?  Dirac answer this question partially in his Dirac quantization condition, which tells that for the consistency of quantum mechanics, which involves the vector potential and not the magnetic field, the magnetic charge of a monopole must be quantized to unit values (in the units we take!).   The condition arises because the vector potential of a monopole is necessarily singular, but we require that this singularity not have physical effects.  
However,  Dirac assumed this monopole did not have electric charge.  In general, Dirac's argument says that two particles with electric and magnetic charges $(Q_1,M_1)$ and $(Q_2,M_2)$ can be consistently treated in quantum mechanics if $Q_1M_2 - Q_2 M_1$ is an integer.  Taking $(Q_1,M_1) = (1,0)$, we find that the requirement on the second particle is only that $M_2$ be integer, and $Q_2$ is arbitrary.  We take the minimal magnetic charge with $M_2=1$, and then we have $(Q_2,M_2) = (\frac{\theta}{2\pi}, 1)$.  This can be considered to {\em define} $\theta$.  The full set of allowed charges is established by superposition of these two ``elementary'' particles, so we obtain
\begin{equation}
  \label{eq:74}
  (Q,M) = (n_1 + \frac{\theta}{2\pi} n_2, n_2),
\end{equation}
where again $n_1,n_2$ are integers, which just count the number of the two particles we posited above which are fused together.  

Eq.~(\ref{eq:74}) both can be viewed as a {\em definition} of the $\theta$ angle and as a specification of the quasiparticle charges.  It remains to specify the statistics of these particles.  In three dimensions there are no mutual statistics.  The single-valuedness of the wavefunction under taking one particle fully around another is guaranteed in fact by the Dirac quantization condition.  However, we can ask about the self-statistics, which must be bosonic or fermionic (there are no anyons in three dimensions).  It has been shown that if our two ``base'' particles $(1,0)$ and $(\frac{\theta}{2\pi},1)$ are bosons, then their bound state -- the ``dyon'' -- is a fermion.  Under these assumptions, the particles with $n_1n_2$ even are bosons and $n_1 n_2$ odd are fermions.  Note that if we take $\theta=2\pi$, we arrive at the situation where the particle with charge $(1,1)$ is a boson, and then the particle with charges $(0,1)$ is a fermion.  Compare this to $\theta=0$ where the statistics of these two particles is reversed.  If we further increase $\theta=4\pi$, the statistics of all charges are restored.  The conclusion, reached in Ref.\cite{metlitski2013bosonic}, is that the physics of this theory is $4\pi$ periodic in $\theta$.   

Let us also comment on the assumption that $(1,0)$ and $(\frac{\theta}{2\pi},1)$ are bosons.  If we instead assume $(\frac{\theta}{2\pi},1)$ is a fermion, then this is simply equivalent to the previous assumption but with $\theta$ shifted by $2\pi$.  The situation with $(1,0)$ a fermion and $(\frac{\theta}{2\pi},1)$ a boson is equivalent to this one by electromagnetic duality: there is no fundamental distinction between magnetic and electric fields in the vacuum Maxwell theory, and in 3+1 dimensions they can be exchanged, which relabels the charges, and changes $\theta$ (see Ref.~\cite{cardy1982duality}).  So the only remaining distinct case is that both $(1,0)$ and $(\frac{\theta}{2\pi},1)$ are fermions, but this has been argued to be impossible\cite{wang2014classification,kravec2014all} to realize in any three dimensional model.  

With these conventions, the Maxwell Hamiltonian in Eq.~(\ref{eq:78}), containing the constants $\epsilon$ and $\mu$ which determine the strengths of electric and magnetic Coulomb forces,  and the value of $\theta$, defined by  Eq.~(\ref{eq:74}), gives an apparently complete specification of the U(1) QSL, {\em neglecting symmetry}.   With symmetry, the situation becomes considerably more complex, as we briefly discuss in Sec.~\ref{sec:other-qsls}.

\subsection{Chern-Simons theory}
\label{sec:other-emergent-gauge}

We spent quite a bit of time describing $Z_2$ gauge theory and U(1) gauge theory.  We emphasized that both possess, in appropriate dimensions, deconfined phases in which the gauge charged particles ``emerge'' as finite energy, but non-local, excitations.   But there is another famous gauge theory which appears heavily in the condensed matter literature: Chern-Simons theory in two dimensions.  Here the Maxwell term is replaced or supplemented by a term with one fewer derivative.  The U(1) Chern-Simons action is
\begin{equation}
  \label{eq:72}
  S_{CS} = \int dt d^2x \, \frac{k}{4\pi} \epsilon_{\mu\nu\lambda} A_\mu \partial_\nu A_\lambda.
\end{equation}
Here we use the standard notation that lumps time in as a zeroth space-time component and index.  The integer $k$ is called the ``level'' of the Chern-Simons theory, and the theory is often denoted by the symbol U(1)$_k$.  Recall that the action for Maxwell electromagnetism involves two derivatives of the gauge field, so that the Chern-Simons action has indeed fewer derivatives and hence totally changes the low energy properties of the gauge theory, dominating any Maxwell term if it is present.  

Indeed, the presence of the Chern-Simons term opens a gap in the spectrum, which is of the order of the ratio of the Cherm-Simons coefficient to the coefficient of the Maxwell term.  In the pure Chern-Simons theory, the gap is infinite, and the theory describes {\em only} global topological excitations.  Consequently, the Chern-Simons gauge theory is known as a Topological Quantum Field Theory (TQFT).   In fact, the U(1)$_k$ theory actually describes the Laughlin series of topologically ordered phases, discussed in Sec.~\ref{sec:other-topol-phas}.  The level of the Chern-Simons theory corresponds precisely to the notation used there.  

Chern-Simons theory is a very well-developed subject.  We will not discuss it further here.  The interested reader can find many reviews on the topic.  We would like to emphasize here only that the U(1)$_k$ Chern-Simons action contains no more and no less information than the mutual statistics of anyons in the Laughlin states as discussed in Sec.~\ref{sec:other-topol-phas}.  Moreover, there are generalization of Eq.~(\ref{eq:72}) to multi-component Chern-Simons fields (so-called ``K matrix'' theories) and to non-abelian gauge fields, which can describe other topological phases.  In our view Chern-Simons theory is just one way to represent the universal properties of some particular topological phases.  It is convenient for some purposes, but certainly not essential.  This is quite different from usual U(1) Maxwell theory of Sec.~\ref{sec:u1-gauge-theory},  where the electric field fluctuations and the photon mode of the gauge theory itself are observable by local measurements.

\subsection{Phase transitions}
\label{sec:phase-transitions}

``Conventional'' phases are typically characterized by a broken symmetry, and continuous phase transitions to and from such states can be understood in terms of symmetry breaking (or restoration) of the set of symmetries under which the Hamiltonian is invariant. Without spontaneous symmetry breaking as a defining characteristic of the phase, how can one describe continuous phase transitions out of quantum spin liquid states, which need not break any symmetries?   We have already discussed, in Sec.~\ref{sec:confinement}, the need for a phase transition between a QSL phase and a confined one.  Moreover, in the case of a $Z_2$ topological phase, we suggested that the confinement transition could be understood as condensation of the {\sf m} particle.  

This viewpoint is actually quite general, and indeed a rather general way to contemplate transitions {\em out} of QSLs.  Whenever we can define a quasiparticle with an energy gap in a QSL, we can imagine tuning a parameter so that this energy is brought to zero.  At that point, we reach an instability of the QSL state.  In the simplest case, when this particle is a boson, the transition can be viewed as a condensation of this particle.  In the toric code case, this could be either an {\sf m} or an {\sf e} particle.  In the $Z_2$ gauge theory language, these two cases would be called the ``confinement'' or ``Higgs'' transitions.  The language is based on focusing on the {\sf e} excitations.  The {\sf m} particle, carrying $\pi$ magnetic flux $P_p$, when it condenses in the ground state, induces large magnetic fluctuations, i.e.\ fluctuations of $\sigma_i^z$.  Consequently, the conjugate variable $\sigma_i^x$ which describes electric field lines becomes very ``stiff'', and cannot fluctuate.  This freezes string of electric field lines emanating from an {\sf e} particle, so that it develops a non-zero line tension.  Hence the {\sf e} particle is ``confined''.  If instead the {\sf e} particle condenses this also removes it from the spectrum, but in the opposite way by bringing its energy to zero rather than infinity.  Of course, these two scenarios are equivalent if we interchange {\sf m} and {\sf e}.  

In this condensation scenario, we can associate a bose field $\phi$ with the particle that condenses, e.g.\ {\sf m}, and develop a field theory description of this transition.  A priori this is slightly delicate since $\phi$ is not gauge-invariant, as the {\sf m} carries the magnetic gauge charge, and a single {\sf m} particle cannot exist in full isolation.  However, since we approach the transition from the topologically ordered phase, the fluctuations of the gauge fields are weak, and can be neglected.  This is equivalent to the statement that {\sf m} is a good quasiparticle in the QSL phase.    From this reasoning, in the simplest model exhibiting a confinement transition, for example the toric code with an applied field $h_x$ (Eq.~(\ref{eq:13}), the critical properties are described by a simple scalar field theory, implying a 2+1-dimensional Ising universality class.  This is a specific example, but for transitions out of topological phases, the anyon condensation picture has been developed in some detail\cite{PhysRevB.79.045316}.

In a similar way, we can consider phase transitions from U(1) QSLs in three dimensions as condensations of electric or magnetic charges, if these are bosonic.  Indeed, the transition associated with the condensation of an elementary bosonic electric charge out of the U(1) Coulomb phase is {\em the} original Higgs transition.  Likewise, the condensation of a magnetic monopole corresponds to the confinement transition in the U(1) gauge theory.  From this reasoning, we can guess that this transition in simplest U(1) gauge theory is described by the theory of a complex scalar field (associated to the monopole and the antimonopole) coupled to a U(1) gauge field, i.e.\ the U(1) abelian Higgs theory, in 3+1 dimensions.  In fact, this guess can be confirmed explicitly by a duality transformation, which directly map the lattice compact U(1) gauge theory to the abelian Higgs model\cite{banks1977phase,peskin1978mandelstam}.  This leads to the prediction that this transition is generically first order, using a famous analysis of the abelian Higgs model\cite{halperin1974first}.

Because the charges are integers in the U(1) case rather than just parities, there are further possibilities.  For example, a bound state of two electric charges could form, and this object could condense.  This is a ``partial'' Higgs transition, and the result is {\em not} a fully confined phase, because in the condensate the parity of the electric charge remains a good quantum number.  In fact, the charge two condensate is actually at $Z_2$ topological phase: the remnant odd charge state becomes the {\sf e} excitation of the topological phase.  Similarly, the charge two condensate supports vortex excitations, which are essentially the {\sf m} particles (though in 3 spatial dimensions they are line defects).  So we see that the condensation paradigm can also describe transitions between different QSL phases.  In the U(1) case, in developing a theory of the transition itself, it is necessary to include the gauge fields explicitly, since these do fluctuate at low energy within the Coulomb phase.  

Such a condensation mechanism, in both the $Z_2$ and U(1) cases, can also describe symmetry breaking.  Though we have not yet discussed this in detail (it is postponed to Sec.~\ref{sec:symm-fract}), when we take into account symmetries of the system, the quasiparticles are generally imbued with conserved quantum numbers.  Thus when such a particle condenses, it brings with it symmetry breaking.  Interestingly, the field representing a non-local quasiparticle itself is {\em not} the same as a symmetry breaking order parameter.  Since a physical order parameter is a local observable, it must be a composite object built of a product of the quasiparticle fields.  This allows for a variety of interesting phenomena at such phase transitions. For example, several order parameters which are not na\"ively related by symmetry can be built from the same quasiparticle field, hence ``unifying'' multiple competing orders.  The composite nature of the order parameter(s) also means that the critical exponents associated with them will be anomalous, and very different from those associated with conventional phase transitions in which the order parameter itself is the field which condenses.  

These ideas can be extended in various ways.  We may contemplate a transition which occurs when a quasiparticle which is not a boson is brought to zero energy.  Then obviously it cannot condense.  However, the crossing with the ground state is still a phase transition.  The simplest example is again in the toric code, if the $\varepsilon$ particle is brought to zero energy.  Since this is a fermion, the problem is perfectly well-behaved even if the energy is made slightly negative.  The system is stabilized by the Pauli exclusion principle, and a Fermi surface or other gapless state can form.  The precise nature of this state depends upon the symmetries of the problem.  For example, if the $\varepsilon$ particle carries some conserved U(1) quantum number (say a spin component), then a Fermi sea may form.  Otherwise, since only the number of $\varepsilon$ particles is a priori conserved only modulo two, one might have a gapless nodal state more like that of an unconventional superconductor.  These are not confined states, but rather gapless QSLs.  

Similar things can be considered even if a sharp quasiparticle does not exist.  We will encounter gauge theories with gapless gauge-charged fermionic quasiparticles in Sec.~\ref{sec:partons}.  One may still consider collective modes which are quasi-bound ``resonances'' of pairs of these fermions.  While such a collective object may not be completely sharp, it may still serve as a field with which to describe a phase transition out of the given QSL phase.  For example, we may view a possible transition from a gapless U(1) QSL to a $Z_2$ phase in such a way.  The presence of gapless fermions and low energy gauge fluctuations makes such a transitions highly non-trivial and theoretically challenging, however, so this is a rather speculative application.

We finish by remarking that the quasi-particle condensation mechanism is one way to understand phase transitions out of QSLs.  It need not be the only way.  Theoretically this problem is very much open.

\subsection{Pr\'ecis of Section~\ref{sec:gauge-theory}}
\label{sec:precis-section}

Gauge theory is pervasive in the description of QSLs.  We reviewed the paradigmatic examples of compact U(1) and Ising ($Z_2$) gauge theories.  Basic ingredients of these theories are ``electric'' and ``magnetic'' fields, and particles which may carry electric or magnetic charges, or both.  There is a natural correspondence between the non-local quasiparticles of highly entangled phases and these charges, which are intrinsically non-local because they are inevitably accompanied by field lines emanating to infinity.  Non-local quasiparticles can propagate over long distances in a {\em deconfined} phase.  Conversely, in a {\em confined} or {\em Higgs} phase, non-local quasiparticles cannot be created and separated from one another with finite energy cost.   Thus the latter phases can be short-range entangled.

Here are a few key facts:
\begin{itemize}
\item When the charged particles are gapped, the compact U(1) gauge theory does not have a stable deconfined phase in two dimensions.  This is due to the proliferation of instantons.
\item The same theory possesses a stable deconfined phase in three dimensions.  In addition to gapped particles with magnetic and electric charges, it generically contains a gapless linearly dispersing emergent photon mode.  This phase is consequently known as a Coulomb phase.  
\item The $Z_2$ gauge theory has a stable deconfined phase in both two and three dimensions.  This is in the same universality class as the toric code.  
\item These deconfined phases are stable to {\em all} weak perturbations, even those which microscopically break the gauge symmetry.   When such perturbations are present, but the deconfined phase still persists, we may say that the system possesses an {\em emergent gauge structure}.
\item U(1) Coulomb phases in three dimensions may also be characterized by a $\theta$ parameter, which describes ``mixing'' of magnetic and electric charges of gapped excitations.  The full set of charged particles in these phases generically contains both bosons and fermions.
\item In two dimensions, chiral topological phases can be described by Chern-Simons gauge theory.  
\item Gauge theories contain transitions between deconfined and confined/Higgs phases, which are generically outside the Landau paradigm of symmetry breaking order parameters.  However, in many cases, these transitions can be described by condensation of some non-local excitation, which allows the development of a field theory description.
\end{itemize}

\section{Partons}
\label{sec:partons}

The ``parton'' construction is a powerful approach to QSLs.  The basic idea is, similarly to the manipulations in Sec.~\ref{sec:stab-toric-code}, to represent the spins of a quantum magnet in terms of canonical bosons or fermions subject to some constraints, which restrict their allowed states to a subspace isomorphic to the spin Hilbert space.  Such methods have a long history, beginning just as a calculational tool for working with angular momentum algebra.  For example, a spin-S operator can be represented by Schwinger bosons according to
\begin{equation}
  \label{eq:29}
  \vec{S} = \frac{1}{2} b_\alpha^\dagger \vec\sigma_{\alpha\beta} b_\beta^{\vphantom\dagger},
\end{equation}
where $b_\alpha^\dagger b_\alpha^{\vphantom\dagger} = 2S$ (sums over the spin-1/2 indices $\alpha,\beta$ are implied), and $\vec\sigma$ is the vector of Pauli matrices.  In the following we will specialize to $S=1/2$, as the most important case.  A spin-1/2 operator is also conveniently represented by Abrikosov fermions,
\begin{equation}
  \label{eq:30}
  \vec{S} = \frac{1}{2} f_\alpha^\dagger \vec\sigma_{\alpha\beta} f_\beta^{\vphantom\dagger},
\end{equation}
with $f_\alpha^\dagger f_\alpha^{\vphantom\dagger} = 1$ (generalizations to higher spin with fermions are possible but more complex).  The canonical fermions or bosons are ``parts'' of the original spins, hence partons.  Many other parton constructions are possible, and useful in different contexts -- c.f. Secs.(\ref{sec:kita-honeyc-model},\ref{sec:ring-exch-hubb},\ref{sec:rare-earth-pyrochl}).  The hope of the parton construction is that, though these representations are introduced purely formally, if we guess correctly, the partons may take on a life of their own as true quasiparticles of the system.  This seems very bold, but there is at least one example where it definitely works -- see Sec.~\ref{sec:kita-honeyc-model}!  However, we should in general be quite careful about this na\"ive notion, and recognize that this need not be the case.

\subsection{Mean field theory}
\label{sec:mean-field-theory}

One of the first applications of partons historically was in ``slave particle mean field theory''.  The basic idea is to insert Eq.~\eqref{eq:29} or Eq.~\eqref{eq:30} into a spin Hamiltonian, and decouple the resulting four-boson or four-fermion interaction in a mean field fashion to produce an effective quadratic Hamiltonian, which can be easily solved.  For example, for an antiferromagnetic Heisenberg interaction, we may write, using bosonic partons:
\begin{eqnarray}
  \label{eq:31}
  \vec{S}_i \cdot \vec{S}_j & = & \frac{1}{4} b_{i\alpha}^\dagger \vec\sigma_{\alpha\beta} b_{i\beta}^{\vphantom\dagger} \cdot b_{j\gamma}^\dagger \vec\sigma_{\gamma\delta} b_{j\delta}^{\vphantom\dagger} \\
  & = & \frac{1}{2} b_{i\alpha}^\dagger b_{i\beta}^{\vphantom\dagger} b_{j\beta}^\dagger b_{j\alpha}^{\vphantom\dagger} - \frac{1}{4} b_{i\alpha}^\dagger b_{i\alpha}^{\vphantom\dagger} b_{i\beta}^\dagger b_{i\beta}^{\vphantom\dagger} \nonumber \\
  & = & \frac{1}{2} b_{i\alpha}^\dagger b_{j\beta}^\dagger b_{j\alpha}^{\vphantom\dagger}b_{i\beta}^{\vphantom\dagger} - \frac{1}{4} \nonumber \\
& \rightarrow & \frac{1}{2} \left\langle b_{i\alpha}^\dagger b_{j\beta}^\dagger \right\rangle b_{j\alpha}^{\vphantom\dagger}b_{i\beta}^{\vphantom\dagger}+ \frac{1}{2} b_{i\alpha}^\dagger b_{j\beta}^\dagger \left\langle b_{j\alpha}^{\vphantom\dagger}b_{i\beta}^{\vphantom\dagger}\right\rangle + {\rm const.}, \nonumber 
\end{eqnarray}
where in the third line we used the constraint $b_{i\alpha}^\dagger b_{i\alpha}^{\vphantom\dagger}=1$.  Up to that point, the expression is exact.  In the final line we made the mean-field approximation.  Note that, we rather arbitrarily chose to group the creation and annihilation terms together.  This can be justified in a certain large-$N$ generalization of the microscopic SU(2) spins (to SP(2N)), but we will not follow this route, as we are interested directly in the case of physical $S=1/2$ spins.  The interested reader may find the large-$N$ mean field theories described in great detail in many theoretical papers, and in the books of Auerbach and Sachdev.  

It is clear that, for $S=1/2$, at the mean-field level, there is an enormous amount of choice.  We might equally well decouple according to
\begin{eqnarray}
  \label{eq:32}
   \vec{S}_i \cdot \vec{S}_j & = & \frac{1}{2} b_{i\alpha}^\dagger b_{j\alpha}^{\vphantom\dagger}  b_{j\beta}^\dagger b_{i\beta}^{\vphantom\dagger}  + {\rm const.} \\
  & \rightarrow & \frac{1}{2} \left\langle b_{i\alpha}^\dagger b_{j\alpha}^{\vphantom\dagger}  \right\rangle b_{j\beta}^\dagger b_{i\beta}^{\vphantom\dagger}+ \frac{1}{2} b_{i\alpha}^\dagger b_{j\alpha}^{\vphantom\dagger}  \left\langle b_{j\beta}^\dagger b_{i\beta}^{\vphantom\dagger}\right\rangle + {\rm const.}\nonumber
\end{eqnarray}
We are still free to choose the form of the expectation values in Eq.~\eqref{eq:31} or Eq.~\eqref{eq:32}, which could be taken differently for each bond $i,j$ in principle, or we could use {\em both} decouplings simultaneously, and we might equally well have used fermionic partons.  We could also have decoupled just the first line in Eq.~\eqref{eq:31}, separating terms on different sites, which is just the usual Curie-Weiss mean field decoupling.  

Within a purely mean field (or large $N$) approach, one would simply decide between all these choices a priori by energetics.  One must (i) solve the resulting quadratic fermion or boson Hamiltonian, including in addition an on-site ``chemical potential'' term $\sum_i \mu_i (b_{i\alpha}^\dagger b_{i\alpha}^{\vphantom\dagger}-1)$ or $\sum_i \mu_i (f_{i\alpha}^\dagger f_{i\alpha}^{\vphantom\dagger}-1)$, with $\mu_i$ chosen so that $\langle b_{i\alpha}^\dagger b_{i\alpha}^{\vphantom\dagger}\rangle = 1$ or $\langle f_{i\alpha}^\dagger f_{i\alpha}^{\vphantom\dagger}\rangle = 1$ make the expectation values self-consistent with the solution of the quadratic problem, and (ii) then calculate the total mean-field energy, being careful to properly account for constants added or subtracted when carrying out the mean-field decoupling in Eq.~\eqref{eq:31} or Eq.~\eqref{eq:32}.  In practice it is not possible to search for all possible self-consistent mean field solutions, but calculations are usually carried out by assuming a particular decoupling scheme, and assuming some high degree of symmetry, for example that both spin-rotation and translational symmetries are unbroken.  In the bosonic parton mean field, there is also the possibility that bosons condense: this occurs when the constraint $\langle b_{i\alpha}^\dagger b_{i\alpha}^{\vphantom\dagger}\rangle = 1$ cannot be satisfied with a positive definite single-boson spectrum.  Such condensed states correspond to magnetic ordering.  

\subsection{Quasiparticle picture}
\label{sec:quas-pict}

The parton mean field approximation has many issues.  This stems from the extreme degree of approximation employed: the representation of the physical spins requires the {\em local} constraint of fixed boson or fermion number on each site.  This means an infinite number of constraints must be applied to respect the original Hilbert space of the problem. At the mean field level every one of those constraints is approximated.  Hence, it is not surprising that the energies obtained by parton mean field theory are not accurate.  This makes it unreliable for differentiating between distinct possible ground states.  Energetics of partons is better attacked from the wavefunction approach, discussed in Sec.~\ref{sec:wavefunctions} and in specific applications in Sec.~\ref{sec:models-methods}.

Instead, one may adopt a more phenomenological approach.  We try to construct a theory of partons as quasiparticles, without direct reference to which microscopic Hamiltonian the quasiparticle description applies.  This is similar in spirit to Fermi liquid theory, in which, when interactions between electrons are strong, we cannot necessarily derive the quasiparticle dispersion, Fermi surface, and Landau parameters, but we can use them to describe the physics of the system.  Here the situation is a bit more delicate, as the partons are not adiabatically connected to excitations of any weakly coupled microscopic system like a free Fermi gas.  It is more like studying the Bogoliubov quasiparticles of a superconductor, with some further subtleties we will discuss.

For concreteness we will focus on the fermionic parton formalism, which has the advantage that it lacks an instability to Bose condensation.  We suppose that a first approximation to an effective Hamiltonian for the system is a quadratic fermion one:
\begin{equation}
  \label{eq:33}
  H_{0} = \sum_{ij} \left[ t_{ij}^{\alpha\beta} f_{i\alpha}^\dagger f_{j\beta}^{\vphantom\dagger} + \Delta_{ij}^{\alpha\beta} f_{i\alpha}^\dagger f_{j\beta}^\dagger + {\rm h.c.}\right].
\end{equation}
We are free to impose $\Delta_{ji}^{\beta\alpha} = - \Delta_{ij}^{\alpha\beta}$ by fermion anticommutation relations, and we should require $t_{ji}^{\beta\alpha} = \left( t_{ij}^{\alpha\beta}\right)^*$ for hermiticity.  This Hamiltonian is mathematically the same as the Bogoliubov-de Gennes Hamiltonian for quasiparticles in a superconductor, when $\Delta \neq 0$, or, when $\Delta=0$, matches that of a metal, insulator, or semiconductor.  

\subsection{$Z_2$ states}
\label{sec:z_2-states}

One general feature of this Hamiltonian is that it is even in fermion operators, and hence has a $Z_2$ symmetry under $f_{i\alpha}\rightarrow - f_{i\alpha}$, $f_{i\alpha}^\dagger \rightarrow - f_{i\alpha}^\dagger$, that is, fermion parity is conserved.  For the moment we assume there are no other ``emergent'' global symmetries of $H_0$, i.e.\ the only other symmetries of Eq.~\eqref{eq:33} are microscopic symmetries of the underlying spin system, for example spin-rotation symmetry and space group symmetries.  We return to discuss the important case in which Eq.~\eqref{eq:33} has an emergent U(1) symmetry later.  

We may wonder about the physical meaning of the apparent $Z_2$ symmetry.  If we think of the fermions as introduced via Eq.~\eqref{eq:30}, then this symmetry is in fact just a consequence of the ansatz.  It does not correspond to any physical symmetry of the problem.  Moreover, in fact, the physical spin operators are invariant under this transformation {\em locally}: we may independently change the sign of $f_{i\alpha}$ on each site, without changing the spin operators.  This is a gauge invariance induced by the ansatz.  The {\em global} $Z_2$ symmetry of Eq.~\eqref{eq:33} is thus an artifact of treating the $f_{i\alpha}$ as quasiparticles.  It is in reality a remainder of the gauge symmetry inherent in Eq.~\eqref{eq:30}.  So to improve upon Eq.~\eqref{eq:33}, we introduce a lattice gauge field that restores this residual gauge symmetry, letting $H_0 \rightarrow H_1$, with
\begin{equation}
  \label{eq:34}
  H_{1} = \sum_{ij} \left[ t_{ij}^{\alpha\beta} \sigma_{ij}^z f_{i\alpha}^\dagger f_{j\beta}^{\vphantom\dagger} + \Delta_{ij}^{\alpha\beta} \sigma_{ij}^z f_{i\alpha}^\dagger f_{j\beta}^\dagger + {\rm h.c.}\right] + H_g,
\end{equation}
where $\sigma_{ij}^z$ is a $Z_2$ gauge field, and we demand that the theory is invariant under $f_{i\alpha}\rightarrow s_i f_{i\alpha}$, $f_{i\alpha}^\dagger \rightarrow s_i f_{i\alpha}^\dagger$ {\em and} $\sigma_{ij}^z \rightarrow s_i s_j \sigma_{ij}^z$.  We can define $\sigma^z_{ii}=1$ to include on-site terms in Eq.~\eqref{eq:34}.  We include in general terms in $H_g$ which depend upon the gauge fields alone and respect this invariance, such as
\begin{equation}
  \label{eq:35}
  H_g = - K \sum_p P_p -K' \sum_i S_i - h \sum_{\langle ab\rangle} \sigma_{ab}^x + \cdots,
\end{equation}
where $S_i$ and $P_p$ are defined as in the toric code.  The gauge invariance is then just the condition that $U_i=S_i(-1)^{f_{i\alpha}^\dagger f_{i\alpha}^{\vphantom\dagger}}$ (the generator of gauge transformations) commutes with $H$.  We can think of Eqs.~(\ref{eq:34},\ref{eq:35}) as providing an interpolation.  For large $K$, the flux is forced to zero, i.e.\ $P_p=1$ everywhere, in the ground state, and the Ising gauge theory (like the toric code model) is in its deconfined phase.  This corresponds to the QSL phase. Due to the absence of flux, we can choose a gauge in Eq.~\eqref{eq:34} for which $\sigma_{ij}^z=1$, and we return to Eq.~\eqref{eq:33}.  That is, in the deconfined phase, for excitations of fermions only, $H_0$ is an adequate description.  Conversely, if we take $h$ large, then the Ising variables become polarized along $x$, and the $\sigma^z$ variables are strongly fluctuating.  Then $H_{1}$ can act only at second order in perturbation theory, and to $O(t^2/h, \Delta^2/h)$, one recovers an Heisenberg interaction between spins defined as in Eq.~\eqref{eq:30}.  In this limit, we should expect that $f_{i\alpha}^\dagger f_{i\alpha}^{\vphantom\dagger}=1$, and since $\sigma_{ab}^x=1$, we have $S_i=1$.  Since the gauge generator is a constant of the motion, we expect it to be independent of the parameters $h,K,K'$ etc.  So we come to the conclusion that
\begin{equation}
  \label{eq:36}
  U_i=S_i (-1)^{f_{i\alpha}^\dagger f_{i\alpha}^{\vphantom\dagger}} = -1.
\end{equation}
The minus sign in Eq.~\eqref{eq:36} is very important.  It is associated with the presence of half-integral spin on a site, which has important implications for the nature and degeneracy of the ground state of spin models.  It is expected to be valid for all $Z_2$ spin liquids in this parton construction.  We will return to this in Sec.~\ref{sec:z2-qsls}.  

We have arrived at a consistent and a priori complete prescription to translate the mean-field Hamiltonian, Eq.~\eqref{eq:33} into a full-fledged effective gauge theory that describes the ground state and excitations.  This consists of the Hamiltonian $H=H_1+H_g$, Eqs.~(\ref{eq:34},\ref{eq:35}), and the constraint in Eq.~\eqref{eq:36} which should be imposed upon all physical states.  This means that a physical operator---by definition one that acts on a physical state to give another physical state---must leave $U_i$ unchanged, so that 
\begin{equation}
  \label{eq:37}
  U_i \mathcal{O} U_i = \mathcal{O},
\end{equation}
if $\mathcal{O}$ is physical (note $U_i^\dagger = U_i$).  

Independently of the sign of the right hand side of Eq.~\eqref{eq:36}, the constancy of $U_i$ implies that the fermions carry the electric gauge charge {\sf e} (we can equivalently say that the fermions {\em are} the {\sf e} particles).  We note that Eq.~\eqref{eq:36} is {\em not} the same  as the hard constraint $f_{i\alpha}^\dagger f_{i\alpha}=1$ we would impose for Abrikosov fermions.  Clearly we cannot {\em strictly} identify the $f_{i\alpha}$ in the effective gauge theory with Abrikosov fermions.  It is apparent that the effective gauge theory, when it applies, describes instead true physical {\em quasiparticles}, rather than just formally defined Abrikosov fermions.  

Returning to the deconfined phase, we can consider magnetic {\sf m} excitations.  In the toric code limit ($h=0$), we can think of these as states in which there is a non-zero flux $P_p=-1$ on a single plaquette.  The $h$ term simply gives some dynamics to the {\sf m} excitation, but does not change its topological properties.  Note that, if the Hamiltonian in Eq.~\eqref{eq:33} does not imbue the fermions with a gap, then this {\sf m} ``excitation'' may well {\em not} be a quasiparticle in the sharp sense.  Its presence is described by a string of negative $\sigma_{ij}^z$ bonds in Eq.~\eqref{eq:34}, which modifies the boundary conditions for every single-particle fermion state (they become antiperiodic on encircling the {\sf m} particle, when compared to the states without the {\sf m} particle, or equivalently, there is a semi-infinite row of negative bonds).  Since there are an infinite number of fermions in occupied states, this can have drastic effects upon the many-body state.  

\subsection{$U(1)$ states}
\label{sec:u1-states}

\subsubsection{Formulation of U(1) gauge theory}
\label{sec:formulation-u1-gauge}

It may happen, for specific choices of $t_{ij}^{\alpha\beta}$ and $\Delta_{ij}^{\alpha\beta}$, that Eq.~\eqref{eq:33} has more than just the $Z_2$ parity symmetry.  For example, if $\Delta=0$, $H_0$ is invariant under a general phase rotation of the fermions, $f_{j\alpha}\rightarrow e^{i\chi} f_{j\alpha}$.  This contains the parity transformation and enlarges it, but like the parity symmetry of the previous subsection, it is unphysical, and just part of the gauge invariance introduced by the ansatz of Eq.~\eqref{eq:30}.  As the phase $\chi$ is arbitrary, this is called a U(1) state. Thus, following the same reasoning, we should improve the mean field theory by promoting this to a gauge symmetry, and requiring physical states to be invariant under the associated gauge transformation.  We let $H_0\rightarrow H_1$, with in this case
\begin{equation}
  \label{eq:38}
  H_1 = \sum_{ij} t_{ij}^{\alpha\beta}e^{iA_{ij}} f_{i\alpha}^\dagger f_{j\beta}^{\vphantom\dagger} + H_g,
\end{equation}
where $A_{ij} = - A_{ji}$ is a lattice gauge field, analogous to the vector potential.  We require that the full theory is invariant under the gauge transformation $f_{j\alpha}\rightarrow e^{i\chi_j} f_{j\alpha}$, $f^\dagger_{j\alpha}\rightarrow e^{-i\chi_j} f^\dagger_{j\alpha}$, and $A_{ij} \rightarrow A_{ij} + \chi_i-\chi_j$.  The gauge part of the Hamiltonian has the same form introduced in Sec.~\ref{sec:u1-gauge-theory}, Eq.~\eqref{eq:19}, and involves the ``electric'' field $E_{ij}= -E_{ji}$ conjugate to the vector potential (c.f. Eq.~\eqref{eq:18}).  Because $A_{ij}$ is a periodic variable (see Eq.~\eqref{eq:38}),  the eigenvalues of $E_{ij}$ are quantized to be integers (well, generally, separated by integers).  In the literature, the periodicity of $A_{ij}$ is denoted by calling the latter a {\em compact} U(1) gauge field.

We must again express the condition that physical states are gauge invariant.  This means that the gauge generator $Q_i$, where the gauge transformation is enacted by $U_i = e^{i\chi_i Q_i}$, commutes with $H$, i.e.\ is a constant in the physical manifold.  The generator is
\begin{equation}
  \label{eq:42}
  Q_i = ({\rm div}\, E)_i - f_{i\alpha}^\dagger f_{i\alpha}^{\vphantom\dagger}=-1.
\end{equation}
where the particular constant of $-1$ is chosen so that in the limit in which the electric field vanishes, the microscopic constraint is satisfied.  This is the analog of the sign choice in Eq.~\eqref{eq:36}, and corresponds to the condition of half-integral spin per site.  

Eqs.~(\ref{eq:38},\ref{eq:19},\ref{eq:42}) give a prescription to translate the mean field solution to a physical description of low energy states in the U(1) case.  This full Hamiltonian is what is called a {\em compact U(1) gauge theory} with (fermionic) matter fields.  It corresponds most directly to the mean field approximation in the limit of large $K$, in which case fluctuations of $A_{ij}$ are suppressed.  However, because $A_{ij}$ is a continuous variable, we must be cautious: there is no gap for excitations which induce fluctuations of $A_{ij}$.  Hence it is not obvious that the singular case $K=\infty$, which in fact corresponds to the mean field state, describes qualitatively the same physics as the generic situation $K<\infty$, when fluctuations are allowed.  The question is if the mean-field U(1) state is {\em stable} to gauge fluctuations.

\subsubsection{Pure gauge theory}
\label{sec:pure-gauge-theory}

We discuss first a ``simple'' case.  For large $K$ we may certainly make a first approximation to the problem by taking $A_{ij}=0$, which is a particular gauge satisfying ${\rm curl} A=0$.  Then we can diagonalize $H_1$ in Eq.~\eqref{eq:38}, to obtain a mean-field fermion spectrum.  In the simplest cases, this is {\em gapped}, and so it appears safe to integrate out the fermions.  We expect that, in this case, we will obtain a pure U(1) gauge theory, i.e.\ a Hamiltonian of the form of Eq.~\eqref{eq:19}, but with renormalized couplings, and the constraint in Eq.~\eqref{eq:42} replaced by one no longer involving fermions, which has the form given in Eq.~\eqref{eq:25}.  The fixed ``background'' charges $q_i$ might depend upon the site $i$, and should be determined from a more careful study.  Technically we can derive Eq.~\eqref{eq:25} in particular cases systematically, if we can tune the mean-field band structure ($t_{ij}^{\alpha\beta}$) so that the fermion gap becomes very large.  For example, on a cubic lattice, one may take a mean-field state of the form \begin{equation}
  \label{eq:43}
  t_{ij}^{\alpha\beta} = \delta^{\alpha\beta}\left[ u \delta_{ij} (-1)^{\sum_{\mu=1}^d x_\mu} - t \delta_{|i-j|,1}\right],
\end{equation}
which just describes fermions with nearest-neighbor spin-preserving hopping $t$ and a staggered potential $u$.  When $u$ is large, the ground state simply has a two-sublattice structure with 2 fermions on each site of one sublattice, and 0 fermions on each site of the other sublattice.  Replacing $f_{i\alpha}^\dagger f_{i\alpha}^{\vphantom\dagger} = 1 - \epsilon_i$, where $\epsilon_i= -1$ or $+1$, respectively, on these two sublattices, we then obtain Eq.~\eqref{eq:25} with $q_i = \epsilon_i$.  This specific case is illustrative, but the specific form of the $q_i$ is not particularly important for the stability arguments summarized below.

For the case of gapped spinons, we arrive therefore at an effective pure compact U(1) gauge theory on the lattice, without fluctuating matter fields. This is just the problem formulated in Sec.~\ref{sec:u1-gauge-theory}, Eqs.~(\ref{eq:19},\ref{eq:18},\ref{eq:25}).  As discussed there, the na\"ive large $K$ limit of this theory is just Maxwell electromagnetism, which describes the familiar electromagnetic photon modes (with two polarizations) in free space in three dimensions.  In two dimensions, it describes a similar photon with a single polarization (as the magnetic field can only be directed out of the plane).  Because of the similarity to standard electromagnetism, the state described in this way is known as a {\em Coulomb phase}, or sometimes a {\em photon phase}.  This Coulomb phase is, as discussed in Sec.~\ref{sec:stabiluty-u1-gauge}, stable in three dimensions but not in two.  This means that for this class of mean field states in two dimensions, the parton mean field is qualitatively wrong at low energies and long distances.  

\subsubsection{Gapless matter fields}
\label{sec:gapless-matter-field}

An interesting twist on the stability issue in two dimensions is the situation in which the spinons are gapless.  Integrating out these gapless modes is a dangerous and uncontrolled procedure, and so the analysis of the pure gauge theory is not applicable.  In general, there are two conflicting tendencies.  First, the spinons themselves (if they are bosons), or some collection of them (e.g.\ a spinon-anti-spinon pair) might condense, disrupting the state.  This seems more likely in the gapless case than the gapped one since the spinons or collections of them can be generated with very small energy.  Conversely, the monopole events, which involve fluctuations of magnetic flux, should be {\em suppressed} by the zero point fluctuations of the spinons, which carry the dual electric gauge charge: due to the commutation relation in Eq.~\eqref{eq:18}, fluctuations of electric fields suppress fluctuations of $A_{ij}$ and hence magnetic flux.  

The theoretical challenge is that the theory even without monopoles (i.e.\ the {\em compact} theory) of gauge fields coupled to gapless spinons is generally complicated and not easy to solve, and there are many possible cases involving different ``band structures'' of the spinons.  Consequently, this problem is still largely unresolved.  We summarize the major known results here:
\begin{itemize}
\item {\bf Dirac fermions:} Neglecting monopoles, theorists have heavily studied the problem of quantum electrodynamics in three space-time dimensions, or QED$_3$, described by the low energy action,
  \begin{eqnarray}
    \label{eq:44}
    && S = \\
    && \nonumber \int \! d\tau\, d^2x \Big[\sum_{i=1}^N \overline\psi_{i} \left( \partial_\tau - i a_0 - i (\vec\nabla - i \vec{a})\cdot\vec\sigma \right)\psi_{i}\\
    && \nonumber + \frac{1}{4g} (\partial_\mu a_\nu - \partial_\nu a_\mu)^2 \Big],
  \end{eqnarray}
  where $\psi_i$ is an $N$-component vector of 2-component spinors (spinor indices suppressed), and $a_0,\vec{a}$ are the time and spatial components of a U(1) gauge field.  It is known that, for sufficiently large $N$, i.e.\ $N>N_c$ with some finite $N_c$, this theory is in a conformally invariant phase with power-law correlations, described by some non-trivial RG fixed point.  This is known as an ``algebraic spin liquid''\cite{rantner2001electron} or  ``Dirac spin liquid''\cite{ran2007projected} phase.  For sufficiently large $N$, moreover, the critical phase is stable to monopoles\cite{hermele2004stability}.  It is believed that, for small $N$, the critical phase of QED$_3$ becomes unstable.  The precise nature of this instability, the critical $N$ for which this occurs, and the dependence of this instability upon microscopic details, are not well understood.  We do not know if a critical Dirac spin liquid state is stable in any physically relevant situation (for example, $N=4$ for the proposed Dirac spin liquid on the kagom\'e lattice\cite{ran2007projected}).
\item {\bf Fermi surface:} Another situation of interest is the case in which the spinons form a partially filled band with a Fermi surface (or multiple such surfaces).  This is known as a ``spinon Fermi surface state''\cite{ioffe1989gapless,nagaosa1990normal,PhysRevB.46.5621}.  Following the heuristic arguments above, since this case has {\em more} low energy ``electric'' excitations (spinons) than the Dirac fermion case, it is expected to be more stable to monopoles than the latter situation.  However, many other instabilities are possible involving pairing, particle-hole condensates, etc.  Many theories postulate that, like the Dirac case, this situation supports a gapless power-law regime, known as a ``strange metal''.  The properties of such a strange metal itself are not well understood, and the subject of much current study\cite{metlitski2010quantum}.  As for the above case, we do not know definitively if a spinon Fermi surface state is stable in any physically relevant context, but such states have been suggested to be relevant to organic triangular lattice materials -- see Sec.~\ref{sec:triangular-organics}.
\item {\bf Critical bosons:} Unlike fermions, bosons are never generically gapless, because negative energy single-particle modes are inherently unstable to condensation.  Thus for generic spin liquid states described by the bosonic parton construction, one obtains only gapped spinons.  However, one may contemplate the situation in which a system is tuned to a quantum phase transition point at which the gap of bosonic spinons vanishes.  It is possible for a U(1) gauge theory to be stabilized in two dimensions at this point.  This phenomena is known as ``deconfined quantum criticality''\cite{senthil2004deconfined}, and has been argued to describe the phase transition between N\'eel and Valence Bond Solid orders on the square lattice.
\end{itemize}

\subsection{Wavefunctions}
\label{sec:wavefunctions}

The effective field theory or quasiparticle approach of Sec.~\ref{sec:quas-pict} allows the parton mean field approach to be interpreted as a consistent phenomenology of different phases.  However, a quasiparticle approach, like Landau's successful Fermi liquid theory of metals, largely dodges the question of describing the ground state itself for a microscopic model.  We do not expect the na\"ive mean field theory to be energetically accurate, or even predictive of {\em which} QSL state might apply to a given Hamiltonian.  Variational wavefunctions inspired from the parton mean field theory give one general approach to the latter problem.

Instead of elevating the mean-field free parton Hamiltonian $H_0$ in  Eq.~\eqref{eq:33}, to an effective gauge theory, as we did in Secs.~\ref{sec:z_2-states}, and \ref{sec:u1-states}, we construct a physical wavefunction from it.  This is done in two steps.  First we write down the literal ground state wavefunction $|\Psi_0\rangle$ for $H_0$.  This wavefunction resides in the enlarged Hilbert space $\mathcal{H}_>$ which defines the Fock space for the partons.  For either Schwinger bosons or Abrikosov fermions for spin $S=1/2$, $\mathcal{H}_>$ is a $4^N$-dimensional space of states for $N$ sites, with vacuum, spin up/down, and doubly occupied states possible on every site. However, $|\Psi_0\rangle$ is not a legitimate wavefunction for the original spin system: it contains components outside the physical Hilbert space $\mathcal{H}$ of the original spins, which is only $2^N$-dimensional, and contains only up and down states on every site.  To obtain a wavefunction which is at least in the correct Hilbert space, we must project to the latter Hilbert space, so we define the (unnormalized) variational state by
\begin{equation}
  \label{eq:45}
  |\Psi_v\rangle = \hat{P}_G |\Psi_0\rangle,
\end{equation}
where $\hat{P}_G$ annihilates all the non-physical states within $\mathcal{H}_>$.  It is called a Gutzwiller projector, and can be written as
\begin{equation}
  \label{eq:46}
  \hat{P}_G = \prod_i n_i (2-n_i),
\end{equation}
where $n_i$ is the number operator for partons on site $i$.  More explicitly, one can imagine evaluating Eq.~\eqref{eq:45} by expanding $|\Psi_0\rangle$ in a Fock space basis and erasing all components which do not have one parton per site.  The result is a sum of terms in a site-diagonal basis, which might be written as
\begin{equation}
  \label{eq:47}
  |\Psi_v\rangle = \sum_{\sigma_i\cdots\sigma_N=\pm 1} \Psi_{\sigma_1\cdots \sigma_N} |\sigma_1\cdots \sigma_N\rangle,
\end{equation}
where $\sigma_i=\pm 1$ is the eigenvalue of $\sigma_i^z = 2 S_i^z$.  The state is described by the explicit wavefunction 
\begin{equation}
  \label{eq:48}
  \Psi_{\sigma_1\cdots \sigma_N} = \langle 0| f_{1\sigma_1}\cdots f_{N\sigma_N} |\Psi_0\rangle,
\end{equation}
where $\langle 0|$ is the vacuum bra, and for concreteness we wrote fermionic partons, but an identical formula applies for bosonic partons with $f\rightarrow b$.   It is possible in many cases to express $\Psi_{\sigma_1\cdots\sigma_N}$ as a product of determinants, which allows for relatively fast numerical evaluations, and in particular, a Monte Carlo procedure exists for calculating expectation values of observables in the $|\Psi_v\rangle$ state.  We do not discuss details of this algorithm here, but merely remark that it can be used both to calculate a variational energy as $\langle \Psi_v|H|\Psi_v\rangle$ as well as to measure other observables.  
\begin{figure*}[htbp]
  \centering
  \includegraphics[width=\textwidth]{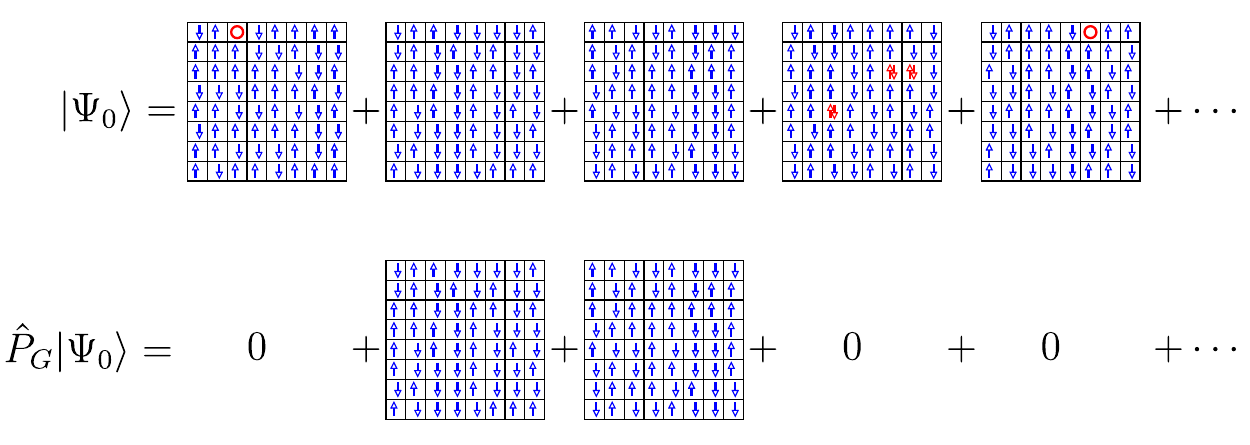}
  \caption{Illustration of Gutzwiller projection.  The unprojected wavefunction is expanded in a basis of fermion number eigenstates at each site.  Then all components which involve empty or doubly occupied sites are ``erased'' to form the variational state.   }
  \label{fig:gutz}
\end{figure*}
The wavefunction constructed in this way can be considered representative of a QSL state. It certainly generates highly entangled wavefunctions.  The initial state $|\psi_0\rangle$ is created as a product state in momentum space, but it is highly entangled in real space.  The Gutzwiller projection selects some subset of real space Fock states, but the massive superposition is preserved.  See Fig.~\ref{fig:gutz} for an illustration.

Na\"ively, the state built from a particular Hamiltonian $H_0$ is describing the same phase as the phenomenological gauge theory constructed from $H_0$ in Sec.~\ref{sec:quas-pict} by gauging its unphysical global symmetries.   How can we understand whether this is actually correct?  For the case of gapped topological phases, i.e.\ cases belonging to the $Z_2$ example of Sec.~\ref{sec:z_2-states}, some rigorous comparison can be directly made, as we understand the universal properties of $Z_2$ topological phases very well.  Most directly, one may consider the topological entanglement entropy $\gamma$ introduced in Sec.~\ref{sec:entanglement}.  It can be calculated for {\em any} wavefunction, and its values are quantized in fully gapped states. Importantly, any non-zero value immediately implies intrinsic topological order. The $Z_2$ state has the value of $\ln 2$, and observation of this value leaves only a few alternative options\cite{rowell2009classification,jiang2012identifying}: a doubled semion state, or a pair of $\nu=1/2$ chiral FQHE states, which can be then distinguished from the $Z_2$ state by a small number of other comparisons.  Furthermore, one can construct trial wavefunctions representing the distinct topological sectors on a torus, and check that these are both degenerate and orthogonal in the thermodynamic limit\cite{ivanov2002projected}.

If we accept that the Gutzwiller wavefunctions are characteristic of a true QSL phase, which seems correct for example for $Z_2$ states, one can make profitable use of them.  For example, one can compare wavefunctions for $Z_2$ states obtained using bosonic and fermionic spinons.  Prior to projection these wavefunctions look extremely different, and it might appear that these are necessarily different phases.  However, the discussion in Sec.~\ref{sec:example:-toric-code} shows that all $Z_2$ QSLs contain both bosonic and fermionic excitations, so there is hope.  Indeed, it has been shown that in some particular limits identical wavefunctions may be obtained, after projection, from bosonic and fermionic partons.  Then the states represented by smoothly varying the wavefunctions within each parton construction are expected to remain in the same phase.

However, it is not clear that the validity of projected wavefunctions as representatives of a true QSL phase is generally true. Indeed, there are clear counter-examples.   For instance, the projected wavefunction for a gapped U(1) state in two dimensions cannot represent a U(1) QSL phase, as these are {\em unstable} and do not exist generically!  Similarly, one may project gapless U(1) Dirac states, and this may lead to an apparently stable QSL wavefunction.  However, from the field theory point of view -- c.f. Sec.~\ref{sec:gapless-matter-field} -- such a state, if it exists at all, is expected to be described by a strongly coupled conformally invariant $2+1$-dimensional theory with non-trivial exponents.  The projected wavefunction's properties can be obtained by an effective classical Monte Carlo procedure.  The most non-trivial result which might conceivably arise from such a classical Monte Carlo calculation is a $2+0$-dimensional critical theory, so we cannot expect this to capture properly the universal properties of a putative U(1) Dirac state.  

Despite these issues, projected wavefunctions are a powerful tool for exploring QSL ground states of physical models.  They can be constructed quite systematically (see also the next Section), and different states can be compared energetically by variational Monte Carlo.  Refinements of these wavefunctions are also possible, starting from the simple projected states.  These techniques lead to very good energies for many problems of interest, and have suggested interesting QSL states as phenomenological candidates to explain experiments.  Notable proposals (which realize some of the states introduced in Sec.~\ref{sec:gapless-matter-field}, within the context of lattice models to be discussed in Sec.~\ref{sec:models-methods}) are a gapless U(1) Dirac state and a gapless U(1) Fermi surface state for the nearest-neighbor S=1/2 Heisenberg model on the kagom\'e lattice, and the S=1/2 Heisenberg plus ring exchange model on the triangular lattice, respectively.  

\subsection{Pr\'ecis of Section~\ref{sec:partons}}
\label{sec:precis-section-4}

Parton approaches consist of embedding microscopic states into a larger, fictitious Hilbert space, in which physical operators are broken into several ``parts'' -- hence partons.  This can be carried out exactly microscopically on the lattice in many ways, standard representations being Schwinger bosons and Abrikosov fermions.  This is the starting point for various parton mean field approximations.  Beyond mean field theory, one can use partons to formulate effective theories of QSLs, which consist of gauge fields coupled to the partons.  One may also write down variational wavefunctions for QSL states using partons, by applying to a mean field wavefunction a Gutzwiller projection back to the physical Hilbert space.  A loose connection exists between the effective theory, which is a complete description of a phase of matter, a specific parton mean field solution, and a Gutzwiller projected wavefunction.

Using partons, one arrives not only at the ``pure'' gauge theories (with gapped charged excitations) discussed in the previous section, but also at gauge theories with gapless matter.  For example:
\begin{itemize}
\item Fermionic partons may be gapless and exhibit Dirac points or a Fermi surface.  When this occurs, these gapless fermions {\em may} stabilize a U(1) deconfined phase even in two dimensions.  
\item Bosonic partons may also be gapless at ``deconfined'' quantum critical points.
\item At least for some examples of $Z_2$ QSLs, the same phase may be represented by either bosonic or fermionic partons.
\end{itemize}

\section{Symmetry and fractionalization}
\label{sec:symm-fract}

Up to now we have focused on the role of entanglement and emergent non-locality in QSLs.  In general, we are also interested in symmetry considerations.  Many interesting models, and to a first approximation many experimental systems, have spin-rotationally invariant Hamiltonians.  Ideal materials also have space group symmetries.  We may be interested in whether these symmetries are preserved by the ground state, or some subset may be spontaneously broken.  Those symmetries which are preserved may be used to label quantum numbers of the excitations.  For many years, QSL states were considered to be {\em defined} by the absence of symmetry breaking.  We hope that the prior sections strongly rebut this notion, favoring instead entanglement as the essential feature of QSLs.  However, in some cases these two attributes are linked: states with certain symmetries {\em must} be non-trivially entangled.  This connection, and roles and uses of symmetry in QSLs, will be discussed in this section.

\subsection{Lieb-Schultz-Mattis theorem and related constraints}
\label{sec:lieb-schultz-mattis}

While we have emphasized that the absence of symmetry breaking does not constitute a good definition of a QSL, it can in many instances be a sufficient indication for a QSL.  Probably the simplest intuition for this comes from the band theory of solids.  We learn as undergraduates that each filled band comprises one single electron state per unit cell, and so that, if spin degeneracy is unbroken, and one has an insulator, so that all bands are full or empty, there must be an even number of electrons per unit cell.  Therefore in band theory, any system with an odd number of electrons per unit cell must be a metal, which is of course gapless.  This very simple argument furthermore extends perturbatively to interacting systems, for which Luttinger's theorem guarantees that the volume inside the Fermi surface, which is equal to half the electron density times $(2\pi)^d$, is preserved to all orders by interactions.

For spin systems, similar results were obtained first by Lieb, Schultz, and Mattis\cite{lieb2004two}, and have been extended in various ways by Oshikawa\cite{PhysRevLett.84.1535} and, more rigorously, by Hastings\cite{PhysRevB.69.104431}.   We will loosely refer to these results as ``LSM'' constraints. In rough terms, these results prove that a local SU(2) invariant spin Hamiltonian on a lattice system with a total {\em half}-integer spin per unit cell and periodic boundary conditions {\em must} have, above its absolute ground state, at least one low-lying state whose energy scales to zero with system size (length $L$), at least as rapidly as $1/L$.  The low energy state can be achieved in various ways.  For example, if the system is ordered, then symmetry breaking implies a quasi-degeneracy in the large system limit, between different linear combinations of the broken symmetry ground states.  Note that either magnetically ordered states or states with lattice symmetry breaking but which preserve SU(2) can satisfy the requirements.  The latter case includes what is typically called ``valence bond solid'', or VBS, order, in which spins arrange into a static pattern of singlets on the lattice.  

If the system does not break symmetry, then there is either a bulk gapless excitation or a ground state degeneracy, both of which are, by assumption, {\em not} related to symmetry breaking.  Neither of these is compatible with a trivial, fully gapped, quantum paramagnet without ground state degeneracy.  So far as we know, a fully gapped state with robust ground state degeneracy on the torus and no symmetry breaking is always topologically ordered.  The gapless situation is even more interesting. This mode, since it is not protected by Goldstone's theorem, must be protected by non-trivial entanglement of the ground state.  So we conclude that, at least for systems with SU(2) symmetry and half-integer spin per unit cell, the absence of order is enough to imply an interesting, highly entangled, QSL state.  

These results extend or generalize in various ways.  Oshikawa's arguments \cite{PhysRevLett.84.1535}\ require only U(1) symmetry ($S^z$ conservation) and not SU(2).  It may also be true that for certain lattices, the existence of space group operations beyond translations implies the absence of a gap even for some integral spin values per unit cell.  A recent preprint uses entanglement based ideas to argue that a trivial gapped state is ruled out using time-reversal symmetry alone, assuming its action satisfies $T^2=-1$ on any odd number of unit cells\cite{watanabe2015constraints}.

\subsection{Partons and PSG}
\label{sec:partons-psg}

It is interesting to talk about symmetry within the parton scheme of the previous Section.  There, QSL states were constructed by elevating a fiduciary free fermion or boson system to a gauge theory, or in a more ad-hoc way, Gutzwiller projecting the associated wavefunction.  How do we assess the symmetries of these states?  A natural starting point is to consider the symmetries of the mean-field free parton Hamiltonian, Eq.~\eqref{eq:33}.  However, this is deceptive since the partons are not themselves necessarily physical.  In particular, the parton Hamiltonian is invariant under a global (i.e. constant) gauge transformation, but this is not a physical symmetry.  Indeed, in the more correct associated gauge theory, e.g.\ Eqs.~(\ref{eq:34},\ref{eq:35},\ref{eq:36}), it is clear that physical states (and operators) do not transform under this operation.  So the invariances of the parton Hamiltonian form some larger object than the physical symmetry group of the ground state.

%  As an illustrative example, we give the case of the ``$\mathbb{Z}_2[0,\pi]\beta$'' state, a $Z_2$ spin liquid constructed from fermionic partons on the kagom\'e lattice.  Its mean field Hamiltonian is\cite{PhysRevB.83.224413}
% \begin{eqnarray}
%   \label{eq:58}
%   H_0 & = & \sum_i \left( \lambda_3 f_{i\alpha}^\dagger f_{i\alpha}^{\vphantom\dagger} + \frac{\lambda_1}{2} \epsilon_{\alpha\beta} f_{i\alpha}^\dagger f_{i\beta}^\dagger + {\rm h.c.}\right) \\
% && + \chi_1 \sum_{\langle ij\rangle} \nu_{ij} \left( f_{i\alpha}^\dagger f_{j\alpha}^{\vphantom\dagger}+ {\rm h.c.}\right) \nonumber \\
%   & & + \sum_{\langle\langle ij\rangle\rangle} \nu_{ij}\left( \chi_2 f_{i\alpha}^\dagger f_{j\alpha}^{\vphantom\dagger} + \Delta_2 \epsilon_{\alpha\beta}f_{i\alpha}^\dagger f_{j\beta}^\dagger + {\rm h.c.}\right),\nonumber
% \end{eqnarray}
% where $\nu_{ij}=\pm 1$ for each bond, forming a pattern which {\em doubles} the size of the unit cell of the kagom\'e lattice (see Fig.1 of Ref.\onlinecite{PhysRevB.84.020407} for a clear figure showing the pattern).  Hence $H_0$ appears non-translationally invariant.  However, Eq.~\eqref{eq:58} is invariant under a modified ``hidden'' translation operation, obtained by combining the translation with a sign change:
% \begin{equation}
%   \label{eq:59}
%   T_\mu: \qquad f_{i\alpha} \rightarrow \zeta_{i,\mu} f_{i+\mu,\alpha},
% \end{equation}
% where $\zeta_{i,\mu}=\pm 1$ is a sign that depends upon the site, and the direction of the translation ($\mu$ represents a translation by a primitive vector of the kagom\'e lattice).  

In general, moreover, the symmetries of the ground state may not be apparent from the gauge theory.  Consider the example given in Eq.~\eqref{eq:43} of Sec.~\ref{sec:pure-gauge-theory}.  In three dimensions, this represents a stable U(1) QSL phase.  The Hamiltonian, Eq.~\eqref{eq:38} is \begin{equation}
  \label{eq:56}
  H_1 = -t \sum_{\langle ij\rangle} e^{iA_{ij}}f_{i\alpha}^\dagger f_{j\alpha}^{\vphantom\dagger} + u \sum_i \epsilon_i f_{i\alpha}^\dagger f_{i\alpha}^{\vphantom\dagger} + H_g,
\end{equation}
with $\epsilon_i = +1 (-1)$ when $i$ is on the even (odd) rock-salt sublattice of the cubic lattice.  This $\epsilon_i$ factor makes $H_1$ appear non-translationally invariant.  However, there is a hidden translational invariance.  For example, under a translation by a unit in the x direction, we may take
\begin{eqnarray}
  \label{eq:57}
  f^{\vphantom\dagger}_{i,\alpha} & \rightarrow & 
\epsilon_{\alpha\beta}f^\dagger_{i+\hat{x},\beta}, \nonumber \\
  f^{\dagger}_{i,\alpha} & \rightarrow & \epsilon_{\alpha\beta} f^{\vphantom\dagger}_{i+\hat{x},\beta}, \nonumber \\
  A_{ij} & \rightarrow & \pi - A_{i+\hat{x},j+\hat{x}}, \nonumber \\
  E_{ij} & \rightarrow & -E_{i+\hat{x},j+\hat{x}}.
\end{eqnarray}
Here we have combined a particle-hole transformation of $f$ with the translation, which compensates for the staggered potential term $u$, leaving $H_1$ invariant.  This seems different from ``ordinary'' translational symmetry.  However, one can easily see that on the spin operator, Eq.~\eqref{eq:30}, this operation acts just like a the usual translation: $\vec{S}_i \rightarrow \vec{S}_{i+\hat{x}}$.  Since any physical operator -- i.e.\ in the original spin Hilbert space -- can be constructed from products of these spins, {\em all} physical operators behave properly under this translation.  Hence, the strange-looking transformation in Eq.~\eqref{eq:57} actually realizes proper translation symmetry of the physical states, and the ground state described by it {\em is} translationally invariant (unless the symmetry is spontaneously broken -- this is actually the case in two dimensions, due to the instability of the U(1) gauge theory).  One can also see this in the projected wavefunction formulation.  There the latter two lines of Eq.~\eqref{eq:57} are replaced by the Gutzwiller projection. One can show that, after projection to singly occupied states, the particle-hole transformation has no effect, and so the above operation really just translates any projected state.  

In fact, the transformation in Eq.~\eqref{eq:57} is not unique.  We can compose it with any other element of the U(1) gauge group.  The latter has no physical significance, so all such choices represent the same physical translation.  Xiao-Gang Wen introduced a nomenclature for the full set of transformations under which the mean field parton Hamiltonian is invariant, which includes the physical ones and global gauge transformations, calling this the Projective Symmetry Group, or PSG\cite{wen2002quantum}.  It is projective because a whole set of transformations corresponds to each physical symmetry.  In the present context, the PSG is just a mathematical description of the apparent symmetries of the mean field Hamiltonian. From the PSG, by ``modding out'' by the gauge symmetries, one obtains the physical symmetries of the effective Hamiltonian and in particular of the QSL ground state, if a stable one exists.  Wen also provided some systematic methods to construct PSGs and associated ans\"atze for the spinon hopping and pairing, in order to realize symmetric QSL states, i.e.\ effective Hamiltonians which retain all the {\em physical} symmetries of the underlying lattice.  This has been heavily used to simplify the task of searching for different parton constructions, e.g.\ in variational wavefunction calculations.  

\subsection{Quasiparticle PSG and quantum numbers}
\label{sec:quas-psg-quant}

\subsubsection{General considerations}
\label{sec:gener-cons}

If one is interested in universal properties of the QSL phase, the gauge theory, e.g.\ Eq.~\eqref{eq:56}, contains a great deal of extraneous information and far more degrees of freedom than necessary.  It is also tied to a particular representation, i.e.\ the parton construction.  A more minimal and also more physical question related to symmetry is to ask about the quantum numbers of the excitations.  For example, if the QSL preserved spin-rotation symmetry, we make ask about the SU(2) spin of the spinons.  Likewise, we can ask how the excitations transform under space-group operations.  

We make first a general remark about quantum numbers of excitations in general.  In a many body system in the thermodynamic limit, excitations are expected to usually be local.  If the system has a gap, then they can be considered as ``particles'', and even in the gapless case often a quasiparticle description holds.  When we talk about the quantum numbers of the excitation we really mean this relative to the ground state. What does ``relative'' mean? We should think of a quasiparticle as created by some operator.  When we talk about the quantum numbers of the quasiparticle, we really refer to the transformation properties of this operator.  So it is better to think of quasiparticle quantum numbers as referring to transformation properties of operators rather than of states.  This distinction arises already in very simple non-entangled systems.  Consider for example an ordered ferromagnet, composed of spin-1/2 spins.  The ground state itself has some large spin, which might be half-integer if the total number of spins is odd.  Due to the broken symmetry, we can only discuss the quantum number of total spin along the magnetization axis -- call it $S^z$ -- since rotation symmetry about this axis is unbroken.  The operator which creates such an excitation is $\mathcal{O}=S_i^-$, if the ground state has $S^z=N/2$.  Note that $\mathcal{O}$ creates {\em integer} spin $\Delta S^z=1$, that is, under a U(1) transformation, $U=e^{i\theta S^z}$, $U^\dagger\mathcal{O}U^{\vphantom\dagger} = e^{i\theta}$, with a unit coefficient in the exponential.  Even though the spins themselves are half-integer, the excitations have integer spin.  More formally, extending to full spin-rotation symmetry, the spin operators themselves are vectors, and transform under SO(3) rather than SU(2).  From any finite combination of them, we can build only representations of SO(3).  So in conventional, short-range entangled phases, a spin-1/2 quasiparticle, which is {\em not} a single-valued representation of SO(3), cannot arise.  

In this sense, the spin-1/2 spinons of a QSL are ``fractional'' quasiparticles.  Crudely speaking, they are created by spinors $f_{i\alpha}$ or $f_{i\alpha}^\dagger$.  This is allowed because physical operators always involve an even number of fermion operators.  More generally, the non-local quasiparticles of a QSL may form multivalued representations of physical symmetries.  They must, of course, satisfy a constraint that physical states, which are ``neutral'' collections of such quasiparticles, must decompose into ordinary representations.  

All this assumes that we can define quasiparticles as sharp excitations, which is the case in gapped, topological QSLs, but may not be the case in other QSLs.   Anyway we proceed with this assumption for now.  The representations of physical symmetries acting on individual quasiparticles may be ``projective''.  This means that if there are two symmetry operations $g$ and $h$, and their composition $g h$, then the actions $U_a$ of these operations on a quasiparticle state $a$ obeys
\begin{equation}
  \label{eq:39}
  U_a(g) U_a(h) = \omega(g,h) U_a(g h),
\end{equation}
where $\omega(g,h)$ can be a non-trivial unimodular number (exponential of a phase).  The algebra of the $U_a$ can be considered a projective representation of the symmetry group.  This is another PSG, associated now to a quasiparticle rather than a parton mean field theory. A priori, yet more complex behavior might occur, e.g.\ if there are some internal degrees of freedom of the quasiparticle, then a priori these ``flavors'' may change under a transformation, and $U_a(g)$ might be a matrix in this space.  More generally, one type of quasiparticle could transform into another under a symmetry.  When the group operations are violated by the phases $\omega \neq 1$, then we might say that the symmetry is fractionalized.  For the case of a spin-1/2 spinon, this is the case because in SO(3), we can consider for example $g=h$ to be a spin rotation by $\pi$ about some axis, so that $gh$ is the identity.  However, for a spin-1/2 particle, $\omega=-1$ for a $2\pi$ spin rotation, so SU(2) is indeed realized projectively for spinons.

Other symmetries may also be realized projectively for quasiparticles of QSLs.  For example, in the next subsection, we will discuss an instance of projectively realized translational symmetry.

\subsection{$Z_2$ QSLs}
\label{sec:z2-qsls}

For the case of a $Z_2$ QSL, we know from Sec.~\ref{sec:anyons} that there are three non-trivial anyon quasiparticles, {\sf e}, {\sf m}, $\varepsilon$, and the former are bosons while the latter is a fermion.  For a complete description of symmetry action in this QSL, we should understand the quantum numbers of all three types of anyons.  

First let's discuss spin.  We can call a spinon of a QSL a quasiparticle which carries S=1/2, and we assume a spinon exists.  Since in a topologically trivial phase, such excitations cannot exist, we know that on making a {\em transition} from the QSL to a trivial phase, the spinons must be removed from the spectrum.  This argues that S=1/2 should be bound to non-trivial anyons, since these are removed from the spectrum at this point.  Since {\sf e} and {\sf m} are equivalent (their label is a choice of definition), we may try either {\sf e} and {\sf m} to have S=1/2, or both of them to have S=1/2, or finally neither of them to have S=1/2.  If neither have S=1/2, then since we can fuse {\sf e} and {\sf m} to form $\varepsilon$, then the latter also has integer S.  In that case there are no spinons, so we can discard it.  If both {\sf e} and {\sf m} have S=1/2, then $\varepsilon$ has integer spin.  This situation is strange insofar as both bosons carry spin, and so there is a priori no way to form a topologically trivial state by anyon condensation which does not break spin symmetry.  Thus the natural situation is one in which one of {\sf e} or {\sf m} has S=1/2 and the other does not.  Then by fusion, $\varepsilon$ also has S=1/2.

Thus a $Z_2$ QSL generically has both bosonic and fermionic spinons, which we may choose, by convention, to be called the {\sf e} and $\varepsilon$ particles, respectively.  This strongly suggests that the choice of fermionic or bosonic partons to describe $Z_2$ QSLs in two dimensions is purely one of convenience.  Indeed, it has been established that at least some QSL phases can be described by both formalisms, that is there are bosonic and fermion parton constructions which describe the same phase of matter.

It is beyond the scope of this paper, and probably not really of general interest, to give a complete description of symmetry fractionalization of anyons in $Z_2$ QSLs. Moreover, even this is still a research-active area and it is not obvious to the authors that the last word has been written on this subject.  However, we will discuss some aspects of translational symmetry.  Consider a system on a lattice, whose Bravais lattice is generated by two primitive translations $\hat{T}_\mu$, with $\mu=1,\cdots,d$ in d dimensions.  For example, $\hat{T}_\mu^{-1} \vec{S}_i \hat{T}_\mu =  \vec{S}_{i'}$, where ${\bf r}_{i'} = {\bf r}_i + {\bf a}_\mu$, and ${\bf a}_\mu$ is the $\mu^{\rm th}$ translation vector.  The lattice translations commute, i.e.\ $\hat{T}_\mu \hat{T}_\nu = \hat{T}_\nu \hat{T}_\mu$.  However, we may find that, acting on an anyon, they do not:
\begin{equation}
  \label{eq:40}
  \hat{T}^{(a)}_\mu \hat{T}^{(a)}_\nu = \hat{T}^{(a)}_\nu \hat{T}^{(a)}_\mu \eta^{(a)}_{\mu\nu},
\end{equation}
where $\hat{T}^{(a)}_\mu = U_a(\hat{T}_\mu)$ indicates the effective translation operator acting on anyon $a$.  The meaning of $\eta^{(a)}_{\mu\nu}$ is a phase accumulated by anyon $a$ upon moving through a loop composed of a step in the $\mu$ direction, then one in the $\nu$ direction, then one backward in the $\mu$ direction, and finally one backward in the $\nu$ direction, to return to the original location.  In a $Z_2$ topological phase, the factor $\eta^{(a)}_{\mu\nu}$ must equal $\pm 1$.

Indeed, Eq.~\eqref{eq:40} is called the magnetic translation algebra, and is well known in the description of electrons in magnetic fields on the lattice.  It can occur however for anyons in QSLs even in zero field, and is a component of the non-trivial PSG.  Note that the phases/fluxes $\eta^{(a)}_{\mu\nu}$ are gauge invariant, and when non-zero imply some ``fractionalization'' of momentum.  For a complete description of the translational symmetry of a $Z_2$ QSL, we need these fluxes for all three non-trivial anyons.  However, any two suffice, since one can build the third by fusing the two others.  

It can be instructive to consider the example of a modified toric code model, for which we can make this construction very explicitly.  Let us consider the model of Eq.~\eqref{eq:1} by with $K'<0$.  In this model the ground state is very similar to the usual toric code, except that we require $S_s=-1$ on all sites. This corresponds in fact to the situation realized for partons in Sec.~\ref{sec:z_2-states}, c.f. Eq.~\eqref{eq:36}, if we consider a state with a large gap for spinons (hence $f_{i\alpha}^\dagger f_{i\alpha}=0$ (mod 2)).  We can construct it in the same fashion, by starting with a reference state diagonal in $\sigma^x_i$ satisfying this, and then projecting to $P_p=+1$.  Now let us consider an {\sf m} excitation.  It is created by a string operator, consisting of a product of $\sigma_{i}^z$ along a semi-infinite line.  We have freedom to choose exactly which line, but we need to make some specific choice and stick with it, which establishes a phase convention.  For example, we may take
\begin{equation}
  \label{eq:41}
  \mathcal{M}_{x,y} = \prod_{y'<y} \sigma^z_{x,y;x,y+1}
\end{equation}
to be the operator which creates an {\sf m} particle on the plaquette whose lower-left corner is $(x,y)$. Now we require
\begin{equation}
  \label{eq:60}
  \left(\hat{T}_{x}\right)^{-1} \mathcal{M}_{x,y} \hat{T}_{x} \equiv \hat{T}^{\sf m}_{x} \mathcal{M}_{x,y} = \mathcal{M}_{x+1,y}
\end{equation}
and similarly for the other translations.  Now consider the sequence of translations around a single plaquette:
\begin{equation}
  \label{eq:61}
  \hat{T}_p = \hat{T}_{-y} \hat{T}_{-x} \hat{T}_{y} \hat{T}_{x} .
\end{equation}
By acting sequentially, we obtain
\begin{eqnarray}
  \label{eq:62}
  &&  \left(\hat{T}_{p}\right)^{-1} \mathcal{M}_{x,y}\, \hat{T}_{p}  = S_{x+1,y+1} \,\mathcal{M}_{x,y}.
\end{eqnarray}
Physically, this expresses the mutual statistics of the {\sf m} and {\sf e} anyons in the usual toric code: the {\sf m} anyon acquires a phase factor of $-1$ on encircling a site (here $(x+1,y+1)$) with an {\sf e} anyon.  However, for our modified toric code, there is, relative to the usual toric code, a static background {\sf e} anyon at every site. This then tells us that
\begin{equation}
  \label{eq:63}
   \hat{T}^{\sf m}_p = \hat{T}_{-y}^{\sf m} \hat{T}_{-x}^{\sf m} \hat{T}_{y}^{\sf m} \hat{T}_{x}^{\sf m} = -1.
\end{equation}
In terms of Eq.~\eqref{eq:40}, we have $\eta_{xy}^{\sf m}=-1$.

Note that the semi-infinite nature of the string was crucial here.   If we considered a finite periodic system, we would need to form a finite string with two ends, creating two {\sf m} particles.  For such a pair excitation, upon translation, one star operator appears at each end.  However, the product of these two stars is positive in the ground state.  This is in accord with the fact that for any operator composed of a finite number of physical operators, translations must commute.  Yet we see that individual anyons can transform projectively.

The non-commuting of translations is actually a general feature of the PSG for {\sf m} particles, in any $Z_2$ QSL satisfying the LSM-like condition of Sec.~\ref{sec:lieb-schultz-mattis}, i.e.\ with a half-integer spin per unit cell.  One can crudely think that this spin is represented by a background {\sf e} charge per unit cell.  More rigorously, the constraints of Sec.~\ref{sec:lieb-schultz-mattis} imply that it is not possible to find a featureless gapped, non-degenerate ground state of the system.  In general, the {\sf m} particle can be condensed (by tuning some parameter to bring its energy to zero) to drive a transition out of the topological phase into a trivial one.  If the {\sf m} particle obeys the trivial commutative translation algebra, a uniform condensation is possible and results in a trivial translationally invariant, gapped state, which contradicts the results of the LSM theorem.  The non-trivial PSG for {\sf m} instead guarantees that an {\sf m} condensate breaks translational symmetry, which satisfies the LSM constraints by trading the topological degeneracy of the $Z_2$ QSL for a symmetry-breaking one. 

\subsection{Other QSLs}
\label{sec:other-qsls}

One can also attempt to fix quantum numbers of excitations in other QSLs.  The generalization of the above ideas for $Z_2$ QSLs to other topological phases is probably the simplest extension, since these phases have completely sharp (though non-local) quasiparticles, due to the presence of a gap. A topological phase with symmetry labels has been dubbed a ``Symmetry Enhanced Topological'' (SET) phase.  Classification schemes including various symmetries have appeared in the literature.

It is worthwhile to note that some QSL phases, including topological ones, are fundamentally incompatible with certain symmetries.  For example, the chiral fractional quantum Hall states of Sec.~\ref{sec:fract-quant-hall}, which are described by a Chern-Simons theory (Sec.~\ref{sec:other-emergent-gauge}), cannot be time-reversal invariant.  One can see this already from the quasiparticle content, since the phases describing the mutual statistics of quasiparticles change sign upon time reversing the braiding of two anyons, and the reversed phases are inequivalent to the original ones.  One may also see this from the fact that these states necessarily support chiral edge states which are described by conformal field theories with unequal central charges for the left and right moving sectors.  Under time-reversal these central charges interchange, and so the state is not time-reversal invariant.  More complex constraints may also be present for non-chiral states.  For example, it has been recently shown that the ``doubled semion'' topological order (which is essentially two time-reversed copies of the $\nu=1/2$ chiral topological order) is incompatible with the combination of time-reversal symmetry {\em and} translational symmetry for systems with half-integer spin per unit cell\cite{PhysRevLett.114.077201}.

Moving beyond topological QSLs, U(1) QSLs in three dimensions with gapped charges might be attacked similarly.  The gapped excitations are still fairly sharp quasiparticles, which are protected from decay by their emergent electric/magnetic charge: they are as stable as the physical electron!  Hence we can expect to apply similar PSG reasoning to them.   The monopole PSG has been considered some time ago in a specific instance\cite{bergman2006ordering}.   Recent work considers the action of only time-reversal symmetry in some generality\cite{wang2013boson}.  

The quasiparticle picture itself may break down in QSLs in which excitation which carry emergent gauge charge are gapless.  This is the case for the gapless U(1) gauge theories discussed in Sec.~\ref{sec:gapless-matter-field}.  Still, it has been argued that an emergent enlarged physical (i.e.\ non-gauge) symmetry group appears at low energy in some such phases\cite{hermele2005algebraic}.  This is a conjectured property of the effective field theory description of these states.  It has immediate consequences such as the presence of multiple types of power-law correlations in these exotic ground states, with identical exponents for correlators of different types of operators, which are na\"ively unrelated.  This is an interesting mechanism for unifying different types of competing orders in correlated systems.  

\subsection{Pr\'ecis of Section 5}
\label{sec:precis-section-5}

Symmetry places constraints on and at the same time enriches QSLs. We review the result, due to Lieb-Schultz-Mattis (LSM) and others, that a spin system with half-integer spin per unit cell cannot form a trivial (short-range entangled in modern terms) ground state without breaking symmetries.  Conversely, a fully gapped state which retains the half-integer spin per unit cell must have topological order.  Given a QSL state, we can discuss the quantum numbers of the non-local excitations.  Some key points are:
\begin{itemize}
\item Symmetry operators may be realized projectively, i.e.\ up to a phase, on non-local quasiparticles.  This allows excitations to carry fractional quantum numbers in a QSL.  Most notably, if spin is a good quantum number, ``spinons'' may carry S=1/2.
\item The symmetry operations, combined with global gauge transformations, acting on partons or quasiparticles form a projective extension of the symmetry group.  This is the Projective Symmetry Group or PSG.
\item A classification of the symmetry assignments to anyons in two dimensional topological phases is possible: this constitutes a scheme to organize Symmetry Enriched Topological (SET) phases.
\item For $Z_2$ QSLs, a PSG can be defined for the {\sf e} and {\sf m} anyons (and the $\varepsilon$ follows from this).  We can choose the $Z_2$ vortex or vison, {\sf m}, to be invariant under spin rotation and for the systems constrained by LSM to be non-trivial, then {\sf m} must obey a non-trivial PSG under translations.  
\item The same full set of anyon PSGs can be obtained in some cases from both fermionic and bosonic parton constructions of $Z_2$ states.  This suggests that the corresponding QSLs are in the same phase.  This has been proven in some cases.  
\item Some types of QSLs are incompatible with some symmetries.  For example, chiral topological phases such as FQHE states, cannot be time-reversal invariant.
\item PSGs may be assigned to partons/quasiparticles of non-topological QSLs.  Only a limited understanding exists at present.
\end{itemize}

\section{Models and methods}
\label{sec:models-methods}

In this section, we discuss a range of models which have been heavily studied as descriptions of QSLs or of materials which are likely to host QSLs. We also discuss some of the techniques that have been used to attack such lattice models.

\subsection{Exactly soluble models}
\label{sec:exactly-soluble-mode}

Exactly solvable models play an important role as proofs of principle for existence of different QSL phases, and testbeds to understand their properties in detail.  We have already discussed Kitaev's toric code model in Sec.~\ref{sec:example:-toric-code}.  Here we discuss a number of other examples.

\subsubsection{Kitaev honeycomb model}
\label{sec:kita-honeyc-model}

\begin{figure}[htbp]
  \centering
  \includegraphics[width=3.3in]{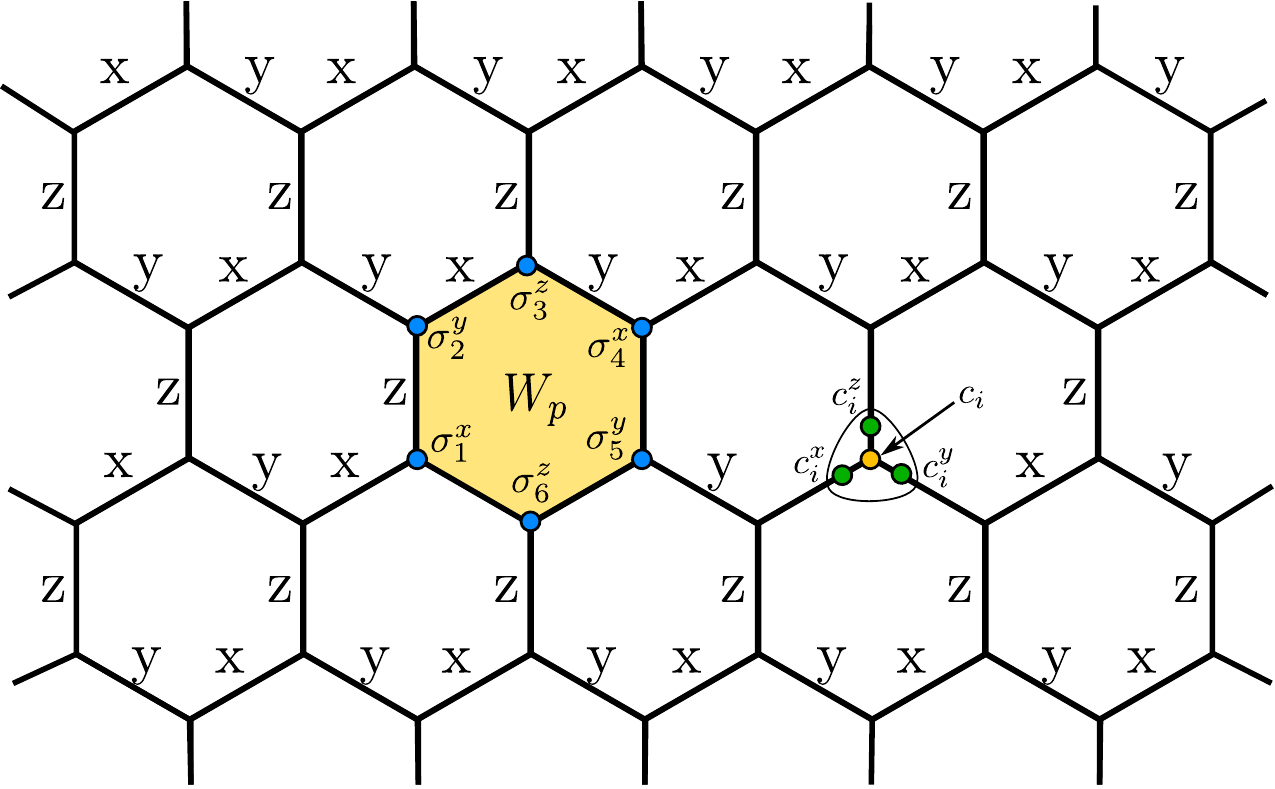}
  \caption{ The Kitaev honeycomb model, showing labeling of links, plaquette flux operator, and the representation of an individual site spin in terms of Majorana fermions.}
  \label{fig:honeykitaev}
\end{figure}

Kitaev in Ref.~\cite{kitaev2006} (see also Ref.~\cite{kitaev2009} for a summary) introduced a simple and appealing model of spin-1/2 spins with anisotropic exchange on a honeycomb lattice: 
\begin{eqnarray}
  \label{eq:50}
&&  H=J_x\sum_{\langle ij\rangle\in x}\sigma_i^x\sigma_j^x+J_y\sum_{\langle ij\rangle\in y}\sigma_i^y\sigma_j^y+J_z\sum_{\langle ij\rangle\in z}\sigma_i^z\sigma_j^z,\nonumber \\
&&
\end{eqnarray}
where the three different directions of bonds are labeled by $\mu=x,y,z$ (see e.g.\ Fig.~\ref{fig:honeykitaev}), and $\sigma_i^\mu$ are Pauli matrices as usual.  Kitaev showed that Eq.~\eqref{eq:50} is, remarkably, exactly soluble by a fermionic parton construction!  That is, a version of the parton mean field method discussed in Sec.~\ref{sec:mean-field-theory} is actually exact for this model.

% \begin{figure}[htbp]
%   \centering
%   \includegraphics[width=3.3in]{figures/honeycomb.jpg}
%   \caption{Honeycomb lattice.}
%   \label{fig:honeycomb}
% \end{figure}

Let us define the plaquette operator, $W_p=\sigma_1^x \sigma_2^y\sigma_3^z\sigma_4^x\sigma_5^y\sigma_6^z$. The latter commutes with the Hamiltonian, $[W_p,H]=0$, for all plaquettes, i.e.\ each $W_p=\pm 1$ is a constant of the motion.  This is a local $Z_2$ gauge symmetry of $H$ which is {\em physical}: states with different eigenvalues of $W_p$ are allowed and $W_p$ is measurable in principle.   Since there is one $W_p$ per unit cell, the subspace of states with all $W_p$ fixed has only one two-level system per unit cell.  We can bring this out by introducing Majorana partons. Defining $\sigma^\mu=icc_\mu$ where $c,c_\mu$ are Majorana fermion operators, $c^\dagger=c$, we obtain
\begin{equation}
  \label{eq:68}
  H=\frac{i}{4}\sum_{\langle ij\rangle}\hat{A}_{ij}c_ic_j,
\end{equation}
where $\hat{A}_{ij}=2J_{\gamma_{ij}}\hat{u}_{ij}$, with $\hat{u}_{ij}=ic_i^{\gamma_{ij}}c_j^{\gamma_{ij}}$, where $\gamma_{ij}=\mu$ if $\langle ij\rangle\in\mu$, following the convention defined below Eq.~\eqref{eq:50}. This is an exact faithful rewriting of the Hamiltonian {\em if} we keep {\em only} states such that
\begin{equation}
  \label{eq:70}
  D_i|\psi\rangle=|\psi\rangle, \quad\mbox{with}\quad D_i=c_i^xc_i^yc_i^zc_i.
\end{equation}
For example, we can enforce Eq.~(\ref{eq:70}) by acting on an arbitrary state with the projector $P=\prod_i\frac{1+D_i}{2}$.   At this point, we note that $D_i$ in Eq.~\eqref{eq:70} is the generator of a $Z_2$ gauge symmetry, which is artificial and induced by the Majorana representation.

Now, because in Eq.~\eqref{eq:68}, each $c_i^{\gamma_{ij}}$ appears only once, $[\hat{u}_{ij},\hat{u}_{kl}]=0$, and so $[\hat{u}_{ij},H]=0$, so that we may treat $H$ as a free ($c_i$) fermion operator by working in each of the eigenvalue sectors where $\hat{u}_{ij}=\pm1$. Recognizing that the plaquette operator may be rewritten in the physical subspace as:
\begin{equation}
  \label{eq:69}
  W_p=\prod_{\langle ij\rangle\in\partial p}\hat{u}_{ij},
\end{equation}
we identify $\hat{u}$ as a gauge potential, and $W_p$ as the associated $\mathbb{Z}_2$ flux operator, as seen by the $c$ fermions. 

The ground state should have no vortices\cite{lieb2004flux}, and therefore, in the ground state, $\forall p$ $W_p=1$. An acceptable choice of $\{u_{ij}\}$ satisfying this constraint is $u_{ij}=1$ if $i$ belongs to the A sublattice, and $u_{ij}=-1$ otherwise. Diagonalizing Eq.~\eqref{eq:68}, we find the following dispersion for $c$ fermion excitations: $\epsilon_\mathbf{k}=\pm2|J_xe^{i\mathbf{k}\cdot\mathbf{n}_1}+J_ye^{i\mathbf{k}\cdot\mathbf{n}_2}+J_z|$, where $\mathbf{n}_{1,2}=(\pm\sqrt{3}/2,3/2)$ (for the choice of honeycomb lattice where $z$ bonds are vertical). When $J_x\leq J_y+J_z$ and $J_y\leq J_z+J_x$ and $J_z\leq J_x+J_y$, the fermion spectrum contains two accidental zero-energy Dirac points which lie along the $k_y=0$ line. This allows to distinguish between two types of phases in the phase diagram: one gapless, and one gapped. 

Let us now connect to the discussion of the toric code and $Z_2$ gauge theory/topological order in Secs.~\ref{sec:highly-entangl-quant} and \ref{sec:toric-code-as}.  The presence of the $Z_2$-valued $W_p$ operators indicates the presence of an emergent $Z_2$ gauge structure.  A ``defect'' plaquette with $W_p=-1$ represents a non-local excitation which we would like to identify with one of the anyons of the topological phase.  Strictly speaking this need only be possible for the gapped phase, but we can, following the considerations in for example Sec.~\ref{sec:partons}, regard the gapless phase as a $Z_2$ gauge theory with gapless matter.  Following the ``flux'' notation, it is natural to regard the $W_p=-1$ particle as an either an {\sf m} or {\sf e} anyon.  If we in fact track how the fluxes should be described if we make the system anisotropic and enter a gapped phase, we see that some of the fluxes must be identified as {\sf m} and others as {\sf e} particles, in alternating rows of plaquettes.  To understand this, we can consider the loop operator which moves a flux around a closed loop.  Imagine for example taking a flux around the filled plaquette in Fig.~\ref{fig:honeykitaev}.  To move it from one plaquette to its neighbor, we should change the sign of one link $\hat{u}_{ij}$, which is shared by the two plaquettes in question.  For example, a z link with $\hat{u}_{ij} = i c_i^z c_j^z$ anticommutes with $\sigma_i^z$, so $\sigma_i^z$ performs the action of moving the flux.  To move it around the entire shaded plaquette, we must perform six such steps, with appropriate spin operators.  One can readily see that the operator which moves a flux around the central loop is exactly $W_p$!  Thus we see that the if the central plaquette contains a flux $W_p=-1$, then moving a second flux in a loop around this central one leads to a $\pi$ phase shift.  This is precisely the mutual statistics expected between {\sf e} and {\sf m} particles.  It holds already in the isotropic limit.

What is clear is the the $c$ fermion is unambiguously a fermion, so should be identified with the $\varepsilon$ particle.  As expected, the $\varepsilon$ particles sense the {\sf m} or {\sf e} particles as a $\pi$ flux.  It is interesting to then think of the decomposition of the spin operator into the fundamental anyons.  The formula $\sigma^\mu_i = i c_i c_{i\mu}$ implies that a $\varepsilon$ particle is created (by $c_i$).  The $c_{i\mu}$ operator in fact creates a pair of fluxes on the two plaquettes sharing site $i$ of separated by the type $\mu$ bond.  By the above discussion, these should be identified as a {\sf e} and an {\sf m} particle.  So we see that a spin operator creates one each of the anyons; symbolically,
\begin{equation}
  \label{eq:82}
  \sigma \sim {\sf e}\, {\sf m} \, \varepsilon. 
\end{equation}
This is rather natural since {\sf e} and {\sf m} fuse to $\varepsilon$ and two $\varepsilon$ particles fuse to the identity, i.e.\ this combination is gauge neutral, and contains zero net electric and magnetic charge.

The identification with the topological order of the toric code is complete in the gapped phase, and indeed one can actually map the honeycomb model to the toric code in the limit in which one of the couplings is large.  In the gapless state, the {\sf m}  and {\sf e} statistics are ill-defined, since moving the vortex induces dynamics in the low energy fermions: adiabatic transport is not possible.  However, interestingly, upon introducing a time-reversal breaking perturbation, the Dirac cones become gapped, and one observes that a Majorana fermion binds to each {\sf m} and {\sf e} particle.  The resulting statistics of these composite excitations is non-Abelian, and the system also displays robust chiral modes at the edge.   This can be shown explicitly by introducing, for example, time-reversal breaking second nearest neighbor interactions, but it is expected that all phases obtained by other time-reversal breaking perturbations should be smoothly connected to one another.

The gapless phase has become of particular interest because of the shown-relevance of the model in Eq.~\eqref{eq:50} with $J_x=J_y=J_z$ to honeycomb materials with strong spin orbit coupling and superexchange through 90$^\circ$-bond paths. We will return to this in detail in Sec.~\ref{sec:honeycomb-iridates}.  We would also like to point out the existence of a great many generalizations of Kitaev's soluble honeycomb model to other lattices with a similar local structure\cite{PhysRevB.89.235102,PhysRevLett.114.157202,lee2014heisenberg,PhysRevLett.114.116803}.  This includes a variety of three-dimensional models, which feature $\mathbb{Z}_2$ QSLs with an intriguing zoology of fermionic band structures.  Some three-dimensional ``hyperhoneycomb'' structured lithium iridates actually exist and have been studied experimentally\cite{takayama2015hyperhoneycomb,modic2014realization}.

\subsubsection{Projector models}
\label{sec:projector-models}

A wide set of highly entangled phases have been constructed using a method which might be considered ``reverse engineering''.  The idea is to find a Hamiltonian $H$ for which a desired state minimizes the energy.  The strategy is to construct  $H$ as a sum of terms, each of which is a positive coefficient multiplying a semidefinite local operator, i.e.\ whose eigenvalues include zero and are otherwise positive, i.e.
\begin{equation}
  \label{eq:64}
  H = \sum_a c_a P_a,
\end{equation}
where $P_a$ are the operators with semi-definite eigenvalues, and $c_a>0$.    Typically these operators are chosen as projectors, i.e.\ to have all eigenvalues zero or one.  In this situation, the energy is bounded below by zero, and so any state $|\psi\rangle$ which is annihilated by $H$, which requires $P_a|\psi\rangle=0$, is a ground state.  The uniqueness of these states, and their stability, must be established separately, and on an individual basis for each such model.  

We see that the toric code model belongs to this class.  Some other iconic models can be written in the same way---for example the Heisenberg ferromagnet has this form, if the $P_a$ are taken to be projectors onto the spin-0 sector on each bond.  A wide class of models of this type are the ``string-net'' ones of Levin and Wen, which realize a large set of highly-entangled phases.  In general, however, they are even more contrived than the toric code Hamiltonian.

One popular family of models in this class are {\em quantum dimer} Hamiltonians tuned to so-called ``Rokhsar-Kivelson points''.  In these models, the Hilbert space consists not of physical spins but of nearest-neighbor dimer coverings of some lattice. Resorting to quantum dimer models is motivated by Anderson's RVB picture of QSLs, in which the states are described as superpositions of singlet $S=0$ pairings of individual spins, e.g.\ $|\psi\rangle=\frac{1}{\sqrt{2}}(|\!\uparrow\downarrow\rangle-|\!\downarrow\uparrow\rangle)$ on each bond for $S=1/2$ spins.  Correspondingly, in the quantum dimer model, the spins are replaced by a Hilbert space spanned by a basis of states consisting of all dimer coverings: colorings of bonds, representing the singlets, in which every lattice site is covered once and only once.  The Hamiltonian contains terms which are diagonal and off-diagonal in this basis.  For example, a well studied dimer model on the triangular lattice is
\begin{eqnarray}
  \label{eq:65}
H & = & \sum_{\small \plaquette} \Big[ - J (|\plaquettev\rangle\langle\plaquetteh|+|\plaquetteh\rangle\langle\plaquettev|) \nonumber \\
  && + V(|\plaquettev\rangle\langle\plaquettev|+|\plaquetteh\rangle\langle\plaquetteh|)\Big].
\end{eqnarray}
Here the dark/thick bonds represent links which are ``occupied'' by dimers, i.e.\ have dimer number = 1, while pale/thin bonds are unoccupied links with dimer number = 0, and Hilbert space is defined as a direct product over bonds which have either 0 or 1 dimer.  We included the two standard couplings: $J$ is an ``off-diagonal'' term which flips the dimer configuration on an elementary plaquette, while $V$ is a ``diagonal'' term which penalizes plaquettes with ``flippable'' dimers.   

In general these two terms compete.  Large positive $V$ favors states with no flippable plaquettes.  Such states are annihilated by $J$, which on its own can achieve, on a single plaquette, an energy of $-J$, by forming a superposition of flippable states.  In between, a compromise can occur, as a superposition of states with both flippable and unflippable plaquettes.  Such a maximal superposition state has the best chance of realizing a QSL.  

On the triangular lattice, one indeed finds a QSL ground state for a region where $J \lesssim V$.  This can be understood from a method introduced by Rokhsar and Kivelson.  Then notes that for Hamiltonians of this type, when $J=V$ ,it becomes a sum of projectors:
\begin{eqnarray}
  \label{eq:75}
  H_{RK} & = & \sum_{\small \plaquette} 2J |\psi(\plaquette)\rangle\langle\psi(\plaquette)|, 
\end{eqnarray}
where
\begin{equation}
  \label{eq:76}
  |\psi(\plaquette)\rangle = \frac{1}{\sqrt{2}}(|\plaquettev\rangle - |\plaquetteh\rangle).
\end{equation}
The projector Hamiltonian is known as a ``Rohksar-Kivelson'' (RK) point in the dimer literature.  Similar Rokhsar-Kivelson points occur in deformations of XXZ model with Ising frustration.  These models will be discuss in Sec.~\ref{sec:frustr-xxz-models}.
 Following the general discussion above, any state which is annihilated by the projector $P_{\small\plaquette}=|\psi(\plaquette)\rangle\langle\psi(\plaquette)|$ is an exact ground state of $H_{RK}$.  This includes both unflippable states {\em and} massive superpositions which are actually $Z_2$ QSL states in the toric code universality class.  

To construct a highly entangled ground state $|\Psi\rangle$, we can do the following.  We simply add {\em all} configurations of dimers, in the occupation basis, and give them exactly equal and positive weight in the superposition.  Consider the action of the operator $P_{\small\plaquette}$ for a single plaquette on this state.  Singling out this plaquette, we can write the state as
\begin{equation}
  \label{eq:77}
  |\Psi\rangle = |\Psi_u\rangle + (|\plaquettev\rangle + |\plaquetteh\rangle)\otimes |\Phi\rangle,
\end{equation}
where $|\Psi_u\rangle$ is a state in which the given plaquette is not flippable, and $|\Phi\rangle$ is some state defined on the space of bonds outside the given plaquette.  The form of the term in parenthesis follows from the choice of equal weights which defined $|\Psi\rangle$.  Now if we act on this wavefunction with $P_{\small\plaquette}$ the first term in Eq.~(\ref{eq:77}) is annihilated because that plaquette is unflippable.  The second term is annihilated because the state in parenthesis is orthogonal to $|\psi(\plaquette)\rangle$ in Eq.~(\ref{eq:76}).  The same argument applies for all plaquettes, so the full Hamiltonian annihilates $|\Psi\rangle$.  

In general, the superposition in $|\Psi\rangle$ contains a few too many states, for example it includes non-flippable states which are zero energy eigenstates on their own.  % However, it is our expectation that, in the large system limit, the maximally flippable component vastly outnumbers the less flippable states.  
We can correct this by a simple procedure.  We start with some base state containing a maximal number of flippable plaquettes, for example with aligned columns of occupied bonds.  Then we construct a ``tree'' of additional states by successively and repeatedly acting with the flip term on each plaquette, and keeping all the states (in the occupation basis) which we accumulate in this way.  Once we have (for a finite lattice) obtained all the states we can find in this way, we just superimpose them with equal weight.  This is again an exact zero energy eigenstate at the RK point.

Does the answer depend upon the starting state?  Yes, but in an interesting way.  In particular, consider the case with periodic boundary conditions, i.e.\ a torus, with an even number of links in both directions.  Now draw a straight line cutting through a set of bonds that encircles the torus in one of the directions.  We can count the number of occupied dimers crossing this line.  The {\em parity} of this dimer number is the same for any two states connected by the flip term.  Hence we obtain two different (orthogonal) states if we superimpose configurations in which this is even or odd.  We can do the same for a line in the other direction around the torus, so the total number of ground states is $2\times 2 = 4$.  This is precisely the expected four-fold degeneracy of the toric code on a torus.  One can verify the other universal ground state properties associated with $Z_2$ topological order hold for these wavefunctions: dimer correlations are exponentially decaying, no local observables can distinguish the 4 degenerate states, and they have the proper topological entanglement entropy\cite{furukawa2007topological}.  

At the RK point, these toric code states are degenerate with other, less flippable, states.  However, if we perturb away from the RK point to $V=J-\epsilon <J$, then these less flippable states are disfavored, and the toric code states emerge as the true unique ground states.  In this way, one can argue that Eq.~(\ref{eq:65}) supports a $Z_2$ QSL state.   If we perturb in the opposite direction, to $V=J+\epsilon>J$, then the unflippable states dominate, and we lose the superposition entirely.  Thus we see that the soluble RK point lies at a phase transition between the topological phase and a trivial one.  This is not unique to the triangular lattice quantum dimer model: many projector models lie at such phase transitions.  

While the above construction of ground states works for simple quantum dimer models at RK points very generally, this does not always yield a $Z_2$ QSL.   For bipartite lattices, in fact a much larger degeneracy of RK wavefunctions on the torus exists.  It turns out that in this situation, the dimer model itself, even away from the RK point, can be mapped to a U(1) compact gauge theory without matter fields.  The idea is to define an electric field variable $E_{ab}$ on link $ab$ which is equal to the dimer number if $a$ is on sublattice A and $b$ is on sublattice B, and equal to minus the dimer number if the sublattices are reversed.  Since on a bipartite lattice, all links connect one site of each sublattice, this completely transcribes the dimer states into electric field states. The dimer constraint that one dimer overlaps each site then implies that ${\rm div}\rm E$ is fixed to $+1$ ($-1$) on the A (B) sublattice.  This is exactly the Gauss' law constraint of a U(1) gauge theory, which is compact since $E_{ab}=0,\pm 1$ is an integer.  The full quantum dimer model in this case is a compact U(1) gauge theory with some fixed background charges.  From general arguments in Sec.~\ref{sec:stabiluty-u1-gauge}, we see that this model does not support a deconfined phase in two dimensions -- the Coulomb phase is unstable.  In fact the RK point for such lattices is finely tuned to a critical point where something like the Coulomb phase exists, but it becomes unstable following the general arguments as soon as it is perturbed.  A rather general understanding of this was developed by Henley and others.  In three dimensions, since the compact U(1) gauge theory can support a Coulomb phase, one can argue that the perturbed quantum dimer model on a bipartite lattice also contains a U(1) QSL phase.   

Despite the strong RVB motivation, it is hard to quantitatively connect quantum dimer models to physical systems. Indeed, the projection of wavefunctions to the nearest-neighbor sector is ad-hoc.   Moreover, the physical nearest-neighbor singlet product basis states are not orthogonal, and orthogonalizing them is quite non-trivial\cite{poilblanc2010effective,misguich2002quantum,rokhsar1988superconductivity}.   Finally, as alluded to above, dimer models, where the Hilbert space is restricted to dimer coverings, are incomplete: they do not, for example, describe excitations with non-zero spin.  

% \begin{itemize}
% \item These models are soluble only for some states, not all of them.
% \item They are typically on the edge of a phase transition, between the most flippable and the completely unflippable states.
% \item In the ``most flippable'' sector the ground state is highly entangled.  This is a putative QSL, but it may or may not be stable.
% \item 2d bipartite lattices are critical (square, honeycomb) and QSL is unstable when perturbed.  Non-bipartite are generically $Z_2$ states.
% \item 3d bipartite are U(1) QSLs.  
% \item Despite motivation, it is hard to quantitatively connect QDMs to physical systems.  Projection to NN singlet sector is ad-hoc, and physical NN singlet basis is non-orthogonal and actually ``orthogonalizing'' these states is quite non-trivial (cite some numerical efforts).  Model itself is incomplete: does not describe spin-ful excitations.  
% \end{itemize}

\subsection{XXZ models with Ising frustration}
\label{sec:frustr-xxz-models}

Another set of highly studied models are the XXZ models with Ising frustration.  These generally take the form
\begin{equation}
  \label{eq:66}
  H = \sum_{i,j} \left[ - \frac{1}{2} J_{ij} \left( S_i^x S_j^x + S_i^y S_j^y\right) + \frac{1}{2}V_{ij} S_i^z S_j^z \right],
\end{equation}
where $J_{ij} \geq 0$ implies that the interactions of the XY components of the spins are ferromagnetic (and hence unfrustrated), while $V_{ij}$ may have arbitrary sign and can be taken as frustrated as we wish.  We assume $S=1/2$ spins in the rest of this subsection.  Due to the lack of frustration in the XY plane, these models do not suffer from a sign problem in quantum Monte Carlo, and hence are especially amenable to numerical simulation.  However, they can still host highly non-trivial states when the Ising interactions are dominant, $|V_{ij}| \gg |J_{ij}|$.  In this limit, we can think of $J_{ij}$ as inducing quantum fluctuations within the classically degenerate Ising manifold of ground states of the $V_{ij}$ terms.  Because Eq.~\eqref{eq:66} still preserves rotational symmetry about the $S^z$ axis, the conditions of LSM still apply, assuming the lattice contains an odd number of spins per unit cell.  Thus trivial disordered states are ruled out, making QSL states more likely.

\subsubsection{BFG model}
\label{sec:bfg-model}

Certain models of this type are known to host QSL states.  One example is the case of the kagom\'e lattice with equal first, second, and third neighbor Ising terms: more precisely $V_{ij}=V$ when $i$ and $j$ are on the same hexagon, while $V_{ij}=0$ otherwise.  This model was first studied by Balents, Fisher, and Girvin in Ref.\cite{balents2002fractionalization}, for the case in which all the XY exchanges on the analogous bonds are taken equal (to $J$).  It has been dubbed the BFG model, though the latter condition has been relaxed in subsequent studies.  When $V \gg J$, the low energy sector consists of all classical Ising states with a total spin $S_{hex}^z=0$ on each hexagon.  This is a highly degenerate set of states, with an extensive entropy, but which is highly constrained.  Indeed, upon including the XY exchange perturbatively using degenerate perturbation theory, the resulting effective Hamiltonian within that degenerate manifold has the structure of a $Z_2$ gauge theory\cite{balents2002fractionalization}.  Through analytical arguments and numerical simulations\cite{balents2002fractionalization,sheng2005numerical,PhysRevB.84.132409,isakov2006spin}, the ground state in this limit has been indeed shown to be in the deconfined phase, i.e.\ it realizes a $Z_2$ QSL, with the characteristic excitations like those of the toric code.  Note that, in the strong Ising limit, $V\gg J$, the sign of $J$ is actually not important, and a QSL occurs also for antiferromagnetic $J$.  However, the latter case is much less accessible to numerical approaches.

\subsubsection{Pyrochlore XXZ model}
\label{sec:pyrochlore-xxz-model}

Similar methods to those used for the BFG model above can be applied to the three-dimensional pyrochlore lattice.  There, nearest-neighbor exchange is sufficient to obtain a QSL in the large $V$ limit.  In this case, the low energy states are those with $S_{tet}^z=0$ on each tetrahedron of the pyrochlore lattice.  Because these tetrahedra themselves form a bipartite diamond lattice structure (if one connects the centers of tetrahedra), the effective model constrained to the low-energy subspace ends up having the structure of a $U(1)$ gauge theory rather than a $Z_2$ one\cite{hermele2004pyrochlore}.  The ground state of this effective Hamiltonian is believed to be in the deconfined Coulomb phase.

This model is of much more practical interest than the BFG one, in that it is (in its local axes version) potentially realizable in actual rare earth magnets, known as quantum spin ice materials---see Sec.~\ref{sec:rare-earth-pyrochl}.  In this context, it is conventional to write the Hamiltonian as
\begin{equation}
  \label{eq:52}
  H=J_{zz}\sum_{\langle ij\rangle}\mathsf{S}_i^z\mathsf{S}_j^z-J_{\pm}\sum_{\langle ij\rangle}\left(\mathsf{S}_i^+\mathsf{S}_j^-+\mathsf{S}_i^-\mathsf{S}_j^+\right),
\end{equation}
where $i,j$ are nearest-neighbor sites on the pyrochlore lattice. This is the model Eq.~\eqref{eq:66} at $J_{zz}=V_{\langle ij\rangle}$ and $J_\pm=J_{\langle ij\rangle}/2$, and all other further-neighbor couplings set to zero. This model was shown to host, for $J_{zz}>0$, $|J_\pm|/J_{zz}\ll1$ and for either sign of $J_\pm$, a $U(1)$ quantum spin liquid best described by a form of compact quantum electrodynamics on the lattice. The first term, in $J_{zz}$, is the Hamiltonian of nearest-neighbor classical spin ice. Because, up to a constant, the sum over nearest-neighbor bonds can be split into a sum over tetrahedra $t$ of $\frac{1}{2}(\sum_{i\in t}\mathsf{S}_i^z)^2$, the ground state manifold of the $J_{zz}$ term alone is extensively degenerate---it is the famous two-in/two-out manifold where two spins point ``in'' and two spins point ``out'' in {\em each} tetrahedron. Degenerate perturbation theory in $J_\pm/J_{zz}$ on this manifold has, as its first non-trivial term, the ``ring exchange'' Hamiltonian
\begin{equation}
  \label{eq:54}
  H_{\rm ring}=-K\sum_{\{i,j,k,l,m,n\}=\hexagon}\left(\mathsf{S}_i^+\mathsf{S}_j^-\mathsf{S}_k^+\mathsf{S}_l^-\mathsf{S}_m^+\mathsf{S}_n^-+\mbox{h.c.}\right),
\end{equation}
where $K=\frac{12J_\pm^3}{J_{zz}^2}$.  Next we use the fact that each pyrochlore site is shared by two tetrahedra with centers $\mathbf{r}$ and $\mathbf{r}'$, which reside on a diamond lattice.  Moreover, the sites $\mathbf{r}$ and $\mathbf{r}'$ reside, one each, from the two bipartite sublattices of this diamond lattice.  This allows one to define 
\begin{equation}
  \label{eq:71} \mathsf{S}_i^z=\epsilon_{\mathbf{r}} E_{\mathbf{r}\mathbf{r}'},\qquad\mathsf{S}_i^+=e^{i\epsilon_{\mathbf{r}}A_{\mathbf{r}\mathbf{r}'}},
\end{equation}
where we define $\epsilon_{\mathbf{r}}=+1$ on one diamond sublattice and $-1$ on the other.  This is useful to introduce because only the {\em pair} $\mathbf{r}$ and $\mathbf{r}'$ and not which diamond site is which is determined for a fixed $i$.    In particular Eq.~(\ref{eq:71}) is well-defined with this inclusion if the $E$ and $A$ variables are vector-like, such that $E_{\mathbf{rr}'}=-E_{\mathbf{r}'\mathbf{r}}$ and $A_{\mathbf{rr}'}=-A_{\mathbf{r}'\mathbf{r}}$. Now, enforcing the constraint $S_i^z=\pm1/2$ with a term $\frac{U}{2}(E_{\mathbf{r}\mathbf{r}'}^2-1/4)$, $U>0$ and large, one obtains
\begin{equation}
  \label{eq:55}
  H=\frac{U}{2}\sum_{\langle \mathbf{r}\mathbf{r}'\rangle}E_{\mathbf{r}\mathbf{r}'}^2-2K\sum_{\hexagon}\cos B_{\hexagon},
\end{equation}
i.e.\ the Hamiltonian of compact quantum electrodynamics, and exactly of the form discussed in Sec.~\ref{sec:u1-gauge-theory}. (Here, $B_{\hexagon}=(\nabla\times A)_{\hexagon}$, i.e.\ the lattice curl of $A$, as defined in Eq.~\eqref{eq:20} for $p=\hexagon$.) Not only the Hamiltonian is now of the form of a compact $U(1)$ gauge theory, but also we obtain, as in Sec.~\ref{sec:u1-gauge-theory}, Gauss' law, ${\rm div}E=0$.   

The rewriting in terms of rotors is helpful for intuition and some formal manipulations, but we emphasize that this is entirely equivalent to the ring model in Eq.~(\ref{eq:54}), which already has the compact $U(1)$ gauge invariance.  The model, in either form, is non-trivial, however, since there is no small parameter (e.g.\ in Eq.~(\ref{eq:55}) we must take $U\rightarrow \infty$ to accurately represent the original spin system -- this limit is non-trivial because $E$ is half-integer).  However, Monte Carlo calculations indicate that the small $U$ Coulomb phase of the rotor Hamiltonian of Eq.~(\ref{eq:55}) actually extends all the way to the infinite $U$ limit\cite{shannon2012quantum,banerjee2008unusual}.  Therefore we conclude that, at small transverse coupling, the XXZ model on the pyrochlore lattice realizes a three-dimensional $U(1)$ QSL.

The above treatment is based on perturbation theory in $J_\pm$.  In  Sec.~\ref{sec:rare-earth-pyrochl}, we will discuss how to extend the rewriting Eq.~\eqref{eq:71} to include matter fields and therefore obtain a more complete gauge theory valid for all values of the exchange.

\subsubsection{Other frustrated Ising models}
\label{sec:other-frustr-ising}

Many other models of ``Ising frustration'' of this type have been studied.  Moessner and Sondhi particularly focused on models in which quantum fluctuations are induced by a transverse field rather than XY exchange.  This violates the LSM conditions and thus trivial states are possible and highly entangled states are less likely.  Indeed, to our knowledge, no non-trivially entangled states have been found in such models.  XXZ models on various other lattices and with other forms of $V_{ij}$ have also been studied---for example with nearest-neighbor terms on the triangular and kagom\'e lattices. Apart from the cases discussed in the previous subsections, these models, though degenerate when $J_{ij}=0$,  {\em order} strongly at $T=0$ when an infinitesimal $J_{ij}$ is introduced, realizing conventional short-range entangled gapped states.

\subsection{Sign-free Monte Carlo}
\label{sec:sign-free-models}

For systems above one dimension, Quantum Monte Carlo (QMC) is by far the most efficient unbiased numerical method to attack spin Hamiltonians, capable of quite large size simulations.  However, it is applicable, at low temperature at least, only to problems which lack a ``sign problem'', i.e.\ which can be formulated as a stochastic statistical mechanical problem with positive definite weights.  A number of sign-free problems have been studied as models of potential QSLs and highly entangled states and critical points. 

One class of such models are the XXZ Hamiltonians of Sec.~\ref{sec:frustr-xxz-models}.  Large-scale QMC simulations have studied variants of the BFG model, obtaining the $Z_2$ QSL phase and demonstrating the universal topological entanglement entropy (Sec.~\ref{sec:entanglement}) which characterizes it\cite{isakov2011topological}.  Furthermore, the QMC method has been used to study the quantum phase transition between the QSL and a conventional XY ordered ferromagnet.  This can be thought of as condensation of a bosonic spinon  ({\sf e} particle), described by the theory of a complex order parameter.  However, because the spinon is non-local, there is an interesting twist: the physical XY order parameter behaves like the square of the spinon field.  This is readily observed from critical exponents, which are strongly modified from the conventional ones describing classical XY transitions.  Consequently, the transition is, strictly speaking, in a distinct universality class from the latter, which the authors' denote by the addition of an asterisk: XY$^*$.  

Similar studies have been carried out on the pyrochlore XXZ model.  Banerjee {\em et al} directly studied this model by QMC\cite{banerjee2008unusual}.    Their results parallel those for the BFG model.  For large $V/J$, a QSL phase was found, with characteristics consistent with a Coulomb phase.  On reducing $V/J$, a transition occurs to an XY ferromagnet for $V/J \lesssim 10$.   This transition was found to be first order, which theoretically may be attributed to the effects of fluctuations of the U(1) gauge field.  Recent works have returned to the effective model derived by Hermele {\em et al}, which also does not have a sign problem\cite{shannon2012quantum,benton2012seeing}.  This has the advantage that the Hilbert space is smaller and low energy states are more readily accessed, but it has the disadvantage that it can access only the QSL phase.  These studies show clearly the existence of the photon mode in the QSL state.  

We conclude this subsection by briefly mentioning a few more applications of QMC to QSLs and related problems.  Motome has studied the $T>0$ properties of some generalized Kitaev honeycomb-type \cite{nasu2014vaporization} and toric code\cite{kamiya2015magnetic} models.  The so-called ``J-Q model'' containing competing Heisenberg and four-spin terms on the square lattice is the focus of numerical examinations of deconfined quantum criticality\cite{kaul2013bridging,sandvik2007evidence}. Several works have studied Hubbard and related itinerant models on bipartite lattices using determinental QMC, and found some indications of possible QSL phases or interesting quantum critical points\cite{meng2010quantum,PhysRevB.71.075103,PhysRevB.88.125108,sorella2012absence}.

\subsection{Heisenberg models}
\label{sec:heisenberg-models}

SU(2)-symmetric Heisenberg models are among the most studied models of quantum antiferromagnets.  Here,
\begin{equation}
  \label{eq:67}
  H = \frac{1}{2}\sum_{i,j} J_{ij} \vec{S}_i \cdot \vec{S}_j,
\end{equation}
and $J_{ij}$ is usually taken non-zero only for a few bonds of short length, for example nearest-neighbors and perhaps second neighbors.  These models are appealing for their high symmetry, which enables a full analysis in terms of SU(2) multiplets.  In real materials, SU(2) symmetry is at best approximate, and stems from the smallness of spin-orbit coupling at the atomic level.  This is most valid for antiferromagnets composed of third row (3d) transition metal elements, in the case when an ionic crystal field multiplet is half-filled and well-separated from other multiplets.   Typically, in this best case, violations of SU(2) symmetry add terms to Eq.~(\ref{eq:67}) which are of order 5-10 percent of the symmetric Heisenberg exchanges.   For those materials at least, Heisenberg models provide a good starting point.

SU(2) symmetric models are often what people have in mind when they envision a QSL.  The most popular cartoon of a QSL is a ``Resonating Valence Bond'' (RVB) state.  Such a state is constructed by first forming a non-magnetic but short-range entangled state by grouping pairs of nearby spins, placing each such pair into a spin zero singlet, and then forming a direct product of those singlets.  This coincides with a single ``dimer covering'' of the lattice, though the dimers need not connect only nearest neighbor bonds.  Such a state is already non-magnetic, and is known as a Valence Bond Solid (VBS) or Valence Bond Crystal (VBC) state, assuming the singlets are arranged in an ordered fashion, or perhaps a Valence Bond Glass if they are not.   Typically such a state breaks symmetries of the lattice, due to the placement of the singlet pairs.  A {\em resonating} valence bond state is formed by superimposing a massive number of these singlet coverings, so that the superposition is highly entangled and breaks no symmetries.  We can imagine there are many ways to form such a superposition, by choosing different weights for different terms in the sum, and by including longer range bonds in different fashions.  For example, the Gutzwiller projected parton wavefunctions of Sec.~\ref{sec:wavefunctions}, if the parton construction respects SU(2) symmetry, can be rewritten as RVB states, with different parton mean fields yielding different weights.  So one can imagine constructing a variety of QSL states that fit this schematic RVB picture.

Going beyond cartoons, symmetric Heisenberg models are challenging theoretically because the high symmetry leaves few or no small parameters to use to control an analytic approach.  The old school analytic approach is to consider the large $S$ limit, where spin wave theory applies.  However, the expansion around $S=\infty$ is semi-classical, which strongly encourages order, so it appears impossible to obtain highly entangled states like QSLs in this limit.  A pioneering approach in the 1990s was to generalize the model from SU(2) spins to more complex objects transforming under SU(N) or Sp(N), and consider the large-$N$ limit.  When properly taken, this limit encourages quantum fluctuations and can describe highly entangled states.  This was very useful in developing a theoretical framework for QSLs, and indeed segued to the modern parton constructions and effective gauge theory descriptions.  However, this approach does not directly imply any specific type of ground state for particular models of interest with physical SU(2) spins.  

\subsubsection{Numerical methods}
\label{sec:numerical-methods}

Instead, sharp predictions of QSL phases in SU(2) symmetric Heisenberg models come from numerics.    Typically if the exchange couplings are not frustrated, an ordered phase emerges, so the focus is on highly frustrated situations.  In these cases, direct QMC is prohibited by a strong sign problem.  Instead, one turns to less scalable methods.  It is not our purpose here to describe different computational methods in any depth, but only to discuss some of the most used techniques, and their advantages and disadvantages.  

The most unbiased method is exact diagonalization, for which the only bias is the choice of finite cluster.  However, only very small systems can be solved in this way (the current record is 48 spins\cite{laeuchli}), so it may not be clear how to extrapolate to the thermodynamic limit.  Still largely unbiased is the Density Matrix Renormalization Group (DMRG), which can obtain numerically exact ground state properties (as well as properties of some low-energy states) in many cases for systems with a strip geometry, for which the DMRG is best suited by its intrinsically one-dimensional formulation.  In many cases the length of the strip can be made very large, but expanding the width is prohibitive, and limited to at best of order 10-12 sites. This is still a significant enlargement compared to exact diagonalization in terms of system size.   DMRG does have a slight bias toward low entanglement states, but since the difference between QSLs and competing states typically comes from a subdominant correction to the area law term, this is probably not a severe limitation, and it is one that can be diagnosed and overcome.  

In recent years, DMRG has come to regarded as a member of a larger family of techniques which rely upon the entanglement structure of ground states.  Quantum information theory has emphasized that the area law obeyed by nearly all ground states puts strong constraints upon the wavefunctions of these states, which can be captured well by tensor network forms as discussed in Sec.~\ref{sec:entangl-based-meth}.  The DMRG in fact can be viewed as an algorithm to find the optimal MPS for a given Hamiltonian, which has led to various improvements in the DMRG taking explicit advantage of the MPS representation.  This allows DMRG to address directly various entanglement-related properties.  For example calculation of the entanglement entropy and indeed full spectrum of the reduced density matrix is straightforward by DMRG (though extracting the {\sl universal} terms in entanglement entropy may be challenging due to finite size effects).  Furthermore, the MPS form allows a conceptual extension of a wavefunction to {\em infinite} size, by defining the wavefunction for arbitrary size through a translationally-invariant tensor.  This allows one to much more directly address topological properties, and for example construct explicitly all the degenerate states on the torus, and to calculate the full exchange statistics of anyons in a topological phase\cite{cincio2013characterizing}.  Other numerical approaches based on higher dimensional tensors, and which may thus be more intrinsically suited to two or three dimensions than DMRG, are actively being developed.   However results so far are limited, and are not yet clearly competitive with more established methods. 

There are a variety of techniques which incorporate a higher degree of bias, in exchange for various benefits.  We have already discussed the use of variational wavefunctions based on Gutzwiller projected fermionic partons.   Calculations using these wavefunctions can be carried out by a Monte Carlo procedure quite efficiently.  The results can also be improved by introducing additional variational parameters, for example by multiplying the Gutzwiller projected wavefunction by additional factors which do not introduce additional signs into the calculation.  Various improved techniques have been explored by several groups, and certainly yield, as expected from the variational principle, lowered estimates of the ground state energy, which compare well with other approaches.  It is somewhat less clear how accurate these approaches are for correlations and other physical properties, as compared to the true ground state.  In general there is potential here for variational methods to mislead, since the variational principle strictly speaking implies that the variational state is close to the ground state when its total energy, which scales with the system volume, is close to the exact ground state energy.  This means that a small error in energy density can lead to a qualitatively wrong wavefunction if the system size is large \cite{balents2014energy}.  Whether this is actually a problem in practice is unclear.  We will mention results from Gutzwiller projected variational wavefunctions for some specific models below.

Other biased techniques are designed to describe particular phases.  These methods include high temperature series expansions and series expansions around particular ordered phases, and coupled cluster methods.  These techniques require a guess or knowledge of the phase the system is in, but with that assumption can provide accurate calculations of many quantities such as ground state and excitation energies, order parameters, etc.  High temperature series is often a useful technique for quantitatively understanding thermodynamics of a model, which can be compared directly with experiment, as a means of refining the model description. Series expansions generally do not describe QSL states of Heisenberg-type models (though they have been successfully applied to special models such as the perturbed toric code).  

Having outlined the methods of analysis, we now summarize the state of understanding of QSLs in several iconic frustrated Heisenberg models. 

\subsubsection{Kagom\'{e} models}
\label{sec:kagome}

The spin-1/2 Heisenberg model on the kagom\'{e} lattice has been one of the most intensively studied problems in frustrated quantum magnetism.  Historically, most of the focus, naturally, has been on the minimal model with a single antiferromagnetic exchange $J_1$ on nearest-neighbor bonds, while recently effects of second and third neighbor exchange, $J_2$ and $J_3$, have been studied. 

We first focus on the nearest-neighbor model, which has a tumultuous past and remains controversial today.  Classically it has an extremely large ground state degeneracy, which would be expected to encourage large quantum fluctuations and invalidate spin wave theory for small spin $S$.  Very early on, this model was suggested to have a non-magnetic ground state\cite{PhysRevLett.62.2405}, which might be a QSL\cite{sachdev1992kagome}\ or a VBS.  Subsequently, exact diagonalization studies on up to 27 spins supported a gapless QSL ground state\cite{PhysRevB.56.2521}, and countless theorists studied the problem in innumerable ways.  In 2007, a Gutzwiller projected fermion variational calculation proposed a gapless U(1) Dirac QSL state\cite{ran2007projected}, while series expansions argued strongly for a VBS state\cite{singh2007ground}.  The VBS prediction stood as the leading numerical candidate for several years, until a pioneering DMRG study conclusively demonstrated that VBS order vanishes, and strongly suggested a {\em gapped} QSL state\cite{yan2011spin}.  Soon other DMRG simulations supported the gapped QSL, both by improved estimates of gaps \cite{depenbrock2012nature}\ and calculations of the topological contribution to the entanglement entropy on cylinders, which was found to be equal to $\ln2$ within small numerical errors\cite{jiang2012identifying} (note that in the latter paper, a small $J_2=0.1J_1$ was included, which further stabilized the QSL state and improved numerical convergence).  This is the expected universal value for a toric code phase, and together with the other DMRG results point strongly toward a gapped $Z_2$ QSL ground state.  Gapped $Z_2$ QSLs on the kagom\'e lattice can be described by partons. Though variational wavefunctions based on Gutzwiller-projected fermion wavefunctions\cite{PhysRevB.83.224413} seem not to yield good energies for $Z_2$ states\cite{PhysRevB.84.020407}, the energy of a Gutzwiller projected {\em bosonic} wavefunction was competitive\cite{tay2011variational}.  The the difficulty of working with such wavefunctions limited the size of the latter calculations, however.  Taken together with the DMRG, these works point toward a specific ``symmetry enriched'' $Z_2$ QSL state, i.e.\ with a specific PSG for all the toric code anyons, for the kagom\'e lattice.  Rather remarkably, the {\em same} QSL phase can be described in both the bosonic and fermionic formulations.  It is known as the $Q1=Q2$ Schwinger boson state\cite{sachdev1992kagome} or the or $Z_2[0,\pi]\beta$ fermionic mean field state\cite{PhysRevB.83.224413}.  While the matching of the PSGs for both parton constructions is strong evidence for their equivalence\cite{PhysRevB.91.100401,zaletel2015measuring,lu2014unification}, there is even stronger evidence: the states have been explicitly matched in a certain limit: i.e.\ their wavefunctions become identical\cite{PhysRevLett.109.147209}.

Though a fairly complete picture of a putative $Z_2$ QSL is emerging, improvements on the Dirac QSL wavefunction also provide competitive variational energies, so that a gapless U(1) QSL state continues to be advocated\cite{iqbal2014vanishing,PhysRevB.87.060405,PhysRevB.84.020407}.  It should be emphasized that the intrinsic stability of the U(1) Dirac QSL state has not been established, i.e.\ we do not know whether this state actually {\em exists as a phase} in any two-dimensional system without fine tuning of parameters.  Nevertheless, the variational results raise valid concerns.  Hence, the presence of a gapped $Z_2$ QSL in the nearest-neighbor kagom\'e system and even $J_1-J_2$ model remains an open issue.

Recent simulations including third-neighbor exchange (here $J_3$ is defined as the exchange between sites on opposite corners of a hexagon of the lattice) have uncovered a second and less controversial QSL on the kagom\'e lattice.  When $0.3 < J_2\approx J_3 <0.7$, DMRG finds a {\em chiral} QSL state\cite{gong2015global,PhysRevLett.112.137202,gong2014emergent}.  This realizes an old idea of Kalmeyer and Laughlin\cite{kalmeyer1987equivalence}, who suggested a QSL state could be described by the standard mapping of spin-1/2 spins to hard-core bosons (i.e.\ $S^z=-1/2$ is an empty site, and $S^z=+1/2$ is an site occupied by a boson), and constructing a bosonic fractional quantum Hall state at filling factor $\nu=1/2$.  This is a chiral topological phase, of the type described in Sec.~\ref{sec:fract-quant-hall}.  It not only has anyonic excitations and topological order, but, as discussed in Sec.~\ref{sec:other-qsls}, breaks time-reversal symmetry.  Consequently, it displays a non-zero order parameter: the scalar spin chirality, $\langle \vec{S}_i \cdot \vec{S}_j \times \vec{S}_k\rangle \neq 0$, for certain sets of three nearby spins.  In this case, the DMRG has fully verified the expected anyons in this state, and their statistics\cite{PhysRevLett.112.137202,gong2014emergent}, and there is no controversy with respect to variational wavefunctions: these calculations find the same state \cite{hu2015variational}.

\subsubsection{Square $J_1$-$J_2$}
\label{sec:square-j_1-j_2}

The frustrated $J_1-J_2$ Heisenberg model on the square lattice is another long-studied model, which was considered for possible relevance to the cuprates, and more recently for the iron pnictides and their relatives.  Classically, it undergoes a transition from the usual N\'eel state with ordering wavevector $(\pi,\pi)$ to a striped antiferromagnet with ordering at $(\pi,0)$ or $(0,\pi)$, with increasing $J_2/J_1$ (both couplings are antiferromagnetic). It is believed that near the classical degeneracy point $J_2=J_1/2$, a non-magnetic ground state occurs\cite{PhysRevB.38.9335,PhysRevB.60.7278,PhysRevLett.84.3173,PhysRevB.86.075111,richter2010spin,jiang2012spin,gong2014plaquette,PhysRevB.88.060402}.  

The nature of this ground state is controversial, with the most popular contender being a VBS with either a structure of columns of singlet bonds or an array of square singlet plaquettes.  However a QSL state has also been proposed in this region.  The most recent works seem to favor a plaquette VBS in in the region $0.5<J_2/J_1<0.6$, and leave open the possibility of an intermediate QSL phase for $0.4\gtrsim J_2/J_1<0.5$.  Alternatively, there may be a direct transition from the N\'eel to VBS phase at some point in this interval, which might realize the interesting proposal of deconfined quantum criticality\cite{senthil2004deconfined} (see also Sec.~\ref{sec:gapless-matter-field}).

\subsubsection{Triangular $J_1$-$J_2$}
\label{sec:triangular-j_1-j_2}

Anderson first proposed the RVB QSL state for the nearest-neighbor spin-1/2 Heisenberg model on the triangular lattice.  This proved to be incorrect: this model has long-range magnetic order\cite{PhysRevLett.99.127004,PhysRevLett.82.3899,PhysRevLett.69.2590}, consisting of spins oriented at 120 degrees to one another in a coplanar three-sublattice pattern.  This led the search for QSLs to move away from the triangular lattice Heisenberg model (though other types of models on the triangular lattice were heavily pursued).

Nevertheless, recent DMRG studies have revisited the problem including second neighbor exchange $J_2$.  These studies found a gapped spin liquid region lying at $0.08\leq J_2/J_1\leq0.16$ \cite{zhu2015spin,hu2015topological}.  Evidence was not provided through the entanglement entropy signature, but via the more conventional detection of exponentially-decaying spin and dimer correlations, which give evidence for the absence of such long-range order. A gap to the singlet and triplet spin excitations, as well as a two-fold ground state degeneracy for a cylinder geometry may point to a $\mathbb{Z}_2$ quantum spin liquid.  Some indication of spin chirality was also observed\cite{hu2015topological}, however, so there may be some competition between a chiral spin liquid and a $Z_2$ state.  Reminiscent of the situation in the kagom\'e Heisenberg model (Sec.~\ref{sec:kagome}), variational wavefunction calculations support instead of either of these a gapless QSL state\cite{kaneko2014gapless}.  

\subsubsection{Heisenberg pyrochlore}
\label{sec:heis-pyrochl}

The pyrochlore lattice, which is formed of corner-sharing tetrahedra, is the three-dimensional archetype of a frustrated lattice.   The nearest-neighbor model is clearly extremely frustrated, and like its two dimensional relatives, for S=1/2 Heisenberg spins lacks any small parameter with which to attack it analytically.  Being three-dimensional, it does not lend itself to numerical simulations well, so computational results are extremely minimal.  Exact diagonalization on very small clusters of up to 12 spins was performed years ago \cite{canals1998pyrochlore} pointing to a quantum spin liquid phase as evidenced by exponentially-decaying spin-spin correlation functions.   

Even biased calculations are limited.  We found only two published works proposing fermionic parton wavefunctions for the pyrochlore lattice\cite{PhysRevB.79.144432,PhysRevB.78.180410}, but these works assumed a U(1) structure and did not carry out a full analysis of even all possible symmetric QSL states, much less symmetry breaking states.  Several papers study the pyrochlore by breaking inversion symmetry and introducing two distinct exchange interactions $J$ and $J'$, within each of the two types of tetrahedra, treating $J$ exactly but $J'$ perturbatively or approximately\cite{PhysRevLett.90.147204,tsunetsugu2001antiferromagnetic}.  This approach results in a VBS state.  It is unclear whether this approach implies anything for true centrosymmetric pyrochlores, but so-called ``breathing pyrochlores'' do exist with symmetry appropriate for the $J-J'$ model.  The XXZ model, discussed in Sec.~\ref{sec:pyrochlore-xxz-model}, is tractable in the Ising limit, but it seems unlikely that the QSL found there extends to the isotropic point.

\subsection{Spin-orbital models}
\label{sec:spin-orbital-models}

Largely motivated by materials, a number of spin-orbital models have been studied as possible hosts for QSLs.  In these problems, the Hamiltonian involves not only spin operators $\vec{S}_i$ but also orbital ones, $L^\mu_i$, which represent atomic degrees of freedom other than spin.  Indeed, it is very common in magnetic ions that such additional degeneracy is present, descending from the 5-fold degeneracy of d orbitals in free space.  The operators $\L_i^\mu$ are typically two or three dimensional matrices.  When spin-orbit coupling is weak, the $L_i^\mu$ and $\vec{S}_i$ operators are independent at the single-site level, but are coupled through superexchange due to the directionality of orbitals, which controls the overlaps between nearby sites.   Typically the Hamiltonians of such models involve terms up to fourth order, schematically
\begin{equation}
  \label{eq:83}
  H =\frac{1}{2}  \sum_{ij}\left[ \vec{S}_i \cdot {\sf J}_{ij} \cdot \vec{S}_j + L_i \cdot {\sf Q}_{ij} \cdot L_j + L_i \vec{S}_i \cdot {\sf K}_{ij} \vec{S}_j L_j \right],
\end{equation}
where ${\sf J}_{ij}, {\sf Q}_{ij}$ and ${\sf K}_{ij}$ are tensors.  These are collectively called Kugel-Khomskii models\cite{kugel1973crystal,kugel1982jahn}.  

The complexity of this type of Hamiltonian means that they are considerably less understood than others we have discussed.  One reason to hope for QSL effects is a certain similarity to SU(N) spin models with more spin states and higher symmetry.  For example, if $L_i^\mu$ are two-dimensional matrices, they can be cast as pseudo-spin 1/2 operators, and if $\vec{S}_i$ are spin-1/2 operators, then Eq.~(\ref{eq:83}) contains a limit in which it describes an SU(4) Heisenberg model (with states in the fundamental representation)\cite{li19984}.  It is known that magnetic order is increasing suppressed in SU(N) Heisenberg antiferromagnets with increasing $N$.  Physically, this arises because the Hamiltonian contains many terms which generate transitions between all $N$ possible spin states -- this is very different from an SU(2) Heisenberg model with $S>1/2$, which, even though it has more spin states, generates transitions only between states with adjacent $S_i^z$ quantum numbers, because the bilinear Hamiltonian contains at most one spin raising/lowering operators on each site.  The presence of transitions between all spin states greatly increases the quantum fluctuations and reduces the tendency of of the ground state to ``localize'' in some region of the spin Hilbert space.  For the Kugel-Khomskii model, it is clear that the biquadratic (${\sf K}_{ij}$ terms in Eq.~(\ref{eq:83})) play a crucial role in inducing the necessary transitions.

Physics of this type led to early suggestions of spin-orbital liquids\cite{ishihara1997orbital,PhysRevLett.85.3950}, and QSL states have continued to be been pursued in Kugel-Khomskii models\cite{mila2007emergence,corboz2012spin,savary2015quantum}.  Closely related are recent works which combine the Kugel-Khomskii model with strong spin-orbit coupling, leading to a projection of Eq.~(\ref{eq:83}) onto an effective ``total angular momentum'' algebra of $J_i^\mu$ operators.  If $J>1/2$, one generically obtains biquadratic and higher interactions, which may induce QSLs\cite{chen2010exotic,natori2015chiral}.

\subsection{Ring exchange and Hubbard models}
\label{sec:ring-exch-hubb}

Up to now, we have restricted the discussion of QSLs to spin models with simple short-range pairwise interactions between spins.  In real solids the notion of local moments is not a completely precise one: it becomes sharp only when the Hubbard repulsion $U$ between electrons in an orbital is much larger than hopping integrals $\sim t$ between orbitals, i.e.\ when the electron wavefunction is strongly localized.  We can extend the notion of QSLs, however, to a general Mott insulator, in which the ratio $U/t$ is large enough to open a charge gap, but not necessarily so great as to strongly localize electrons to single orbitals.  Any non-trivially highly entangled state of such a Mott insulator may qualify to be called a QSL.  

In fact, it has been suggested that ``weak'' Mott insulators, in which electrons are relatively poorly localized, may preferentially nurture QSL states.  This notion is bolstered by the study of Hubbard models on frustrated lattices.  An archetypal example is the single-orbital Hubbard model on the triangular lattice:
\begin{equation}
  \label{eq:51}
  H = -t \sum_{\langle ij\rangle} \sum_\alpha c_{i\alpha}^\dagger c_{i\alpha}^{\vphantom\dagger} + U \sum_i n_i (n_i-1),
\end{equation}
where $c_{i\alpha}^\dagger$,$c_{i\alpha}^{\vphantom\dagger}$ respectively create and annihilate an electron with spin $\alpha=\uparrow,\downarrow$ at site $i$ on the triangular lattice, and $n_{i} = \sum_\alpha c_{i\alpha}^\dagger c_{i\alpha}^{\vphantom\dagger}$ is the number of electrons at site $i$ with spin $\alpha$.  Pioneering work using stochastic variational improvement of a Hartree-Fock wavefunction\cite{morita2002nonmagnetic} predicted the existence of a non-magnetic insulating state -- which, by LSM reasoning (Sec.~\ref{sec:lieb-schultz-mattis}) must be a QSL -- in a wide intermediate range of $4<U/t<10$.  

Later work has refined this notion, suggesting that this non-magnetic insulator is actually the gapless U(1) Fermi surface QSL state mentioned in Sec.~\ref{sec:gapless-matter-field}.  This has a simple parton mean field description\cite{lee2005u} using ``slave rotors''\cite{florens2004slave}.  Here one embeds the physical Hilbert space in a larger one, similarly to what one does for the Abrikosov fermions and Schwinger bosons of Sec.~\ref{sec:partons}, but different in detail.  Specifically, one introduces a rotor Hilbert space with an integer quantum number $Q_i$ on each site, along with a conjugate phase variable $\theta_i$, such that $[Q_i,\theta_j]=i \delta_{ij}$.  We use $Q_i$ to keep track of the charge on a site, i.e.\ impose the constraint $Q_i = n_i$.   So a physical state with for example one electron $c_{i\alpha}^\dagger |0\rangle$ is mapped to a state $f_{i\alpha}^\dagger |0\rangle \otimes |Q_i=1\rangle$, where $f$ can be considered a spinon variable.  One then has the relation
\begin{equation}
  \label{eq:79}
  c_{i\alpha}^\dagger = e^{i\theta_i} f_{i\alpha}^\dagger.
\end{equation}
This representation introduces a U(1) gauge redundancy, due to the constraint,  which manifests as the invariance of $c_{i\alpha}$ under $\theta_i \rightarrow \theta_i + \chi_i$, $f_{i\alpha} \rightarrow e^{i\chi_i} f_{i\alpha}$.  This representation is exact, so one can rewrite $H$ in Eq.~(\ref{eq:51}) in the rotor and $f$ fermion variables, and notably replace $n_i$ by $Q_i$, so that the spinons are free apart from the coupling to the $\theta_i$ fields which appears in the kinetic energy.  In a mean field theory, this term is decoupled, resulting in an effective spinon hopping $t_{\rm eff} = t \langle e^{i(\theta_i-\theta_j)}\rangle$, and an effective XY exchange for the rotors $J_{\rm eff} = t \langle f_{i\alpha}^\dagger f_{j\alpha}^{\vphantom\dagger}\rangle$.  This treatment leads to two phases: the Fermi surface QSL if the rotors are disordered, $\langle e^{i\theta_i}\rangle=0$, and a ``Higgs phase'' which represents a Fermi liquid when the rotors are condensed,  $\langle e^{i\theta_i}\rangle=\sqrt{z}$, where $z$ determines the spectral weight of the quasiparticle pole.  

This is purely a mean-field treatment, and can be elaborated by considering fluctuations to develop a phenomenological theory.  If small fluctuations are considered, the resulting theory is that of a Fermi surface of spinons coupled to a fluctuating U(1) gauge field governed by a Maxwell action, a problem with a long history\cite{ioffe1989gapless,nagaosa1990normal,metlitski2010quantum,mross2010controlled,senthil2008critical,lee2005u,lee2009low}.  This is a difficult problem in field theory, which remains at best partially understood.  From the microscopic point of view, the slave rotor theory is both biased and highly uncontrolled, so different derivations of the spinon Fermi surface state have been sought.

One approach is to follow the strong coupling expansion in powers of $t/U$, which to second order gives back the nearest-neighbor Heisenberg model, but include terms which appear at higher order that are not of Heisenberg type.  The leading contribution at fourth order takes the form of ring exchange around four-site plaquettes.   This motivates the study of models of Heisenberg interactions augmented by ring exchange, which might apply to weak Mott insulators, e.g.
\begin{equation}
\label{Hring-OM}
{\hat H}_{\rm ring} = 
J_2 \sum_{
\begin{picture}(17,10)(-2,-2)
	\put (0,0) {\line (1,0) {12}}
	\put (0,0) {\circle*{5}}
	\put (12,0) {\circle*{5}}
\end{picture}
} P_{12}
+ J_4 \sum_{
\begin{picture}(26,15)(-2,-2)
        \put (0,0) {\line (1,0) {12}}
        \put (6,10) {\line (1,0) {12}}
        \put (0,0) {\line (3,5) {6}}
        \put (12,0) {\line (3,5) {6}}
        \put (6,10) {\circle*{5}}
        \put (18,10) {\circle*{5}}
        \put (0,0) {\circle*{5}}
        \put (12,0) {\circle*{5}}
\end{picture}
} \left( P_{1234}+P_{1234}^\dagger \right) ~,
\end{equation}
where $P_{i_1\cdots i_n}$ is the operator which permutes the states cyclically clockwise around the sites $i_1\cdots i_n$, i.e.\ $P_{12} = P_{12}^\dagger = 1/2 + 2\vec{S}_1\cdot \vec{S}_2$ and for $n>2$,
\begin{equation}
  \label{eq:81}
  P_{i_1\cdots i_n} = \sum_{s_1 s_2 \cdots s_n} |s_2 s_3\cdots s_{n} s_1\rangle \langle s_1 s_2 \cdots s_n|.
\end{equation}
In Eq.~(\ref{Hring-OM}), the sums in $J_2$ and $J_4$ are taken over all nearest neighbor pairs and all compact 4-site rhombi, respectively, on the triangular lattice.  The advantage of such models over the Hubbard one, theoretically, is that within the effective ring exchange model the Hilbert space is just that of S=1/2 spins, so that, for example, Gutzwiller projected wavefunctions may still be used.  A study of such projected parton wavefunctions was carried out by Motrunich\cite{motrunich2005variational}, who indeed found a spinon Fermi surface state to be optimal when the ring exchange is substantial.  More recently, by combining parton wavefunctions with unbiased DMRG simulations, the precursor of such a Fermi surface state was identified in a more rigorous manner in the same model restricted to a four-leg ladder\cite{PhysRevLett.106.157202}.

\section{Materials and experiments}
\label{sec:mater-exper}

Clearly the most outstanding question, and also the most important and difficult, is the experimental realization and characterization of QSLs. This endeavor unfortunately suffers from two main impediments: {\em (i)} the scarcity of highly entangled states in materials, {\em (ii)} the subtleties of their theoretical description, which leaves only few smoking-gun experimental signatures. It is currently very difficult to probe {\em nonlocal} features or measure the degree of entanglement, which are the true hallmarks of QSLs.   Instead, traditional experimental procedures are tailored to detect order and {\em local} excitations: probes such as neutron scattering, $\mu$SR, NMR, and x-ray scattering all measure one or two point correlation functions of local operators.  Given the reliance on the latter techniques, which probe QSL physics indirectly, very detailed theoretical modeling, the outcome of which depends sensitively upon the type of QSL assumed theoretically,  is required to confront experiment. Unfortunately, this requires a good understanding of the theoretical description, which itself in principle presumes, among other things, a detailed knowledge of the material's Hamiltonian.   As a consequence, while many materials which seem not to order down to the lowest measurable temperatures are often claimed to realize quantum spin liquid ground states, {\em definitive} evidence for a single quantum spin liquid in experiment has still not been realized. The discrepancy between the number of theoretical proposals and actual realizations is quite striking. In a nutshell, theoretical models are either too complicated or too simple, in the following sense. Typically, theoretical models for which there is definitive positive evidence for spin liquid regimes will have either spin interactions of too high order, or the models will contain terms which alone would give rise to spin liquids, but are likely too simple-minded to be realized alone in materials. 

In what follows, we first review some general principles to look for and characterize quantum spin liquids in theory and in experiment. Then, we review the most prominent materials which are considered ``candidates'' to realize or nearly realize QSLs.   % where the connection to materials is recognized, namely the honeycomb iridates and the rare-earth pyrochlores, [note that it is interesting those are both very anisotropic models, quite far from what is most commonly studied theoretically] before making a list of several spin liquid candidate materials, where the connection to theory is however looser. 

\subsection{\underline{Where} to look for QSLs}
\label{sec:how-look-qsls}

\subsubsection{Frustration}
\label{sec:frustration}

Given the ubiquity of simple ordered states, it is natural to approach the problem of finding QSLs in terms of {\em destabilizing} such states. Clearly, frustration is therefore an essential feature.  In two dimensions, nearest-neighbor interactions on the kagom\'{e} (corner-sharing triangles) and triangular (``face''-sharing triangles) lattices and in three dimensions, the pyrochlore (corner-sharing tetrahedra) and FCC (edge-sharing tetrahedra) lattices are common examples of {\em geometrically} frustrated lattices. But frustration may also be introduced by further neighbor couplings (e.g., while the diamond lattice is bipartite, second neighbors form two FCC lattices), or by competing interactions, for example when interactions are anisotropic, or are higher-order in spins, or by introducing, e.g., a magnetic field.

There are many excellent reviews of the physics of frustration.  Mostly these take a classical or semi-classical perspective, which has limited impact on the QSL problem.  For our purposes frustration is merely a guide to help select systems which might be reasonable candidates for QSLs.  

\subsubsection{Small spin}
\label{sec:small-spin}

Another feature commonly advocated to be important for the realization of QSLs is a low spin.   This is based on the familiar notion that in the large $S$ limit, spin-$S$ spins become classical, which in turn is based on the fact that the commutator of two spin operators, if normalized by their ``length'' $S$, is of order $1/S$.  It is indeed true for Heisenberg models, and more generally models with bilinear exchange interactions.  However, this conclusion is {\em not} general, and must be considered a bit more carefully in practice.   A better way to think about the classicality of a spin system is in terms of localization in Hilbert space.  A classical state can be considered as the fully polarized ($S_i^z=S$) state with a spin quantization axis chosen at each site to coincide with the classical orientation of the spin vector. The single spin operator $\vec{S}$ becomes classical for large $S$ in the sense that it generates transitions only between states with $S^z$ differing by $\pm 1$, and for a Hamiltonian which involves only individual $S_i^\mu$ operators (and not products of multiple operators at a single site), the low energy states become localized near the fully polarized one, because it requires a high order action of the Hamiltonian to connect to states far from this one.   However, this conclusion rests on the bilinear exchange form of the Hamiltonian.  

Indeed, larger spin systems can be highly quantum under certain circumstances.  A simple way this occurs is if there are strong single-ion terms that split the spin degeneracy.  This can lead to a doublet ground state of the single site problem, which acts then like an {\em effective} spin-1/2, if exchange is small compared to the splitting to the next multiplet.  This is in fact typical for f electron spins in the lanthanide sequence.  A second, less common, way that larger spins can be quantum is through multipolar spin interactions.  This means couplings  involving a product of many $S_i^\mu$ operators on a single site.  Such operators do not necessarily become classical for large $S$.  Indeed, if we allow {\em all} possible operators acting on a spin-$S$ spin, it becomes simply a $N=2S+1$-level system, and we could choose a Hamiltonian which has SU(N) symmetry (see Sec.~\ref{sec:mean-field-theory}) that is {\em more} quantum than any SU(2) spin model.  Multipolar interactions escape classicality precisely by connecting all the $N$ states directly, thereby avoiding localization near any particular spin projection.  The microscopic physics of multipolar interactions is complex, but may arise out of orbitally dependent super-exchange, and spin-orbit coupling\cite{santini2009multipolar,chen2010exotic,chen2009spin}.  

% If one moves away from half-filled shells of light transition metal ions, this assumption is not necessarily good.  In fact, even the concept of a ``spin'' $S$ needs some clarification, once both spin and orbital degrees of freedom are included.  One might think of ions with both spin and orbital degeneracy, and in general exchange includes both types of operators.  In systems with strong spin-orbit coupling, it is often appropriate to combine spin and orbital angular momentum into total angular momentum $\vec{J}$, and we often write spin operators as $\vec{S}$ even though they represent total angular momentum.  However, interactions amongst these total angular momentum states may be very far from the Heisenberg type. For example, for rare earth f-electron spins, higher ``multipolar'' interactions -- involving a product of many $S_i^\mu$ operators on a single site -- are often important.   Since such multipolar operators can connect arbitrary $S^z_i$ states, they can subvert the na\"ive classicality of large spin ions.  Another complication is crystal fields, which can split the $2J+1$ or $2S+1$ states on a single atom.  This is also important for rare earth spins, often more so than exchange, and leads in many cases to a ground state doublet of the single ion problem.  In this situation, the spin is {\em effectively} spin-1/2, even when $J$ is large.  

\subsubsection{Proximity to the Mott transition}
\label{sec:prox-mott-trans}

A different strategy to find liquid ground states of spins is to look at systems close to the more common sort of quantum liquid: the metallic state.  It is not unreasonable to think that a QSL is a natural intermediate state between a metal and an ordered antiferromagnet.  This has already been discussed theoretically in Sec.~\ref{sec:ring-exch-hubb}.  Certainly proximity to a Mott metal-insulator transition leads to new sources of quantum fluctuations, e.g.\ virtual excitations across the Mott gap.  The proximity to the Mott transition is an often-cited justification for the spin liquid like behavior observed in the organics (see Sec.~\ref{sec:triangular-organics}).  

\subsection{\underline{How} to look for QSLs}
\label{sec:how-look-qsls-1}

The most popular means to identify materials which might host QSL physics experimentally is to demonstrate the lack of magnetic ordering.  For many years, the QSL state was commonly thought to be defined by the absence of order.  We have tried to emphasize that, while the lack of any broken symmetry can point to a QSL state (Sec.~\ref{sec:lieb-schultz-mattis}), this is a peripheral consideration, and not a necessary one.  QSLs in the more meaningful sense of high entanglement can occur in systems for which LSM theorems do not apply, and in systems with magnetic order.  Furthermore, the absence of symmetry breaking, even though it does sometimes imply non-trivial physics in ideal situations, may be due in real systems to complications such as disorder.  Still, the lack of magnetic ordering in a local moment material is rare enough that it can at least serve to motivate further study.  If ordering is present but occurs at anomalously low temperature, or with an anomalously small ordered moment, it may also suggest QSL physics.

Positive proof of QSL physics is unfortunately very challenging to obtain.  Hence, in practice, it is necessary to combine many different observations using multiple techniques to gradually build up as complete as possible a picture of a given material.  This is all the more necessary as the signatures of different QSL states are themselves quite distinct. In the remainder of this subsection, we will discuss various experimental probes of QSLs, and what they reveal.

\subsubsection{Neutron scattering}
\label{sec:neutron-scattering}

Neutron Scattering is often considered the method of choice to study magnetic properties. It gives access to the spin-spin correlation function $\mathcal{S}(\omega,\mathbf{k})\sim\langle S^\mu_{-\mathbf{k},-\omega}S^\nu_{\mathbf{k},\omega}\rangle$ of the system, where $\omega$ and $\mathbf{k}$ are the energy and momentum transfers, respectively.   First consider the elastic  ($\omega=0$)  signal.   Long-range magnetic order manifests as elastic scattering at magnetic Bragg peaks, whose location in momentum space gives information on the type of order.   In principle, any magnetic intensity at $\omega=0$ at $T=0$ implies a non-zero ground state expectation value of the spin operator, since using the spectral representation $\mathcal{S}(0,\mathbf{k}) \propto \langle 0 |S^\mu_{-\mathbf{k}}|0\rangle\langle 0|S^\nu_{\mathbf{k}}|0\rangle$.  This takes the form of a resolution-limited Bragg peak if the ground state moments form an ordered pattern, but may be a smooth function of momentum if the static moments are random.  A ground state with no static moments whatsoever should really have zero elastic intensity.  In practice, one cannot measure the zero of energy with absolute precision, so that one observes ``quasi-''elastic weight, i.e.\ the integral of $\mathcal{S}(\omega,\mathbf{k})$ over some narrow frequency range.  This can be non-zero without any static moments, if there are low enough energy excitations with non-zero matrix elements to the ground state.   This might arise in a QSL due to multispinon continua, collective modes like the photon of the three dimension U(1) Coulomb phase, etc.  For these types of excitations, for most QSL phases, the quasi-elastic weight would be expected to be localized in only some regions of the Brillouin zone, and show some characteristic sharp momentum dependence.   Quasi-elastic weight may also be due to quasi-static moments created by impurities, which is likely the origin if this weight depends very smoothly on momentum.

The inelastic signal, for $\omega$ on the scale of the exchange $J$, reveals excitations of the system.  The spectral representation implies that $\mathcal{S}(\omega,\mathbf{k})$ receives contributions only from excitations with energy $\omega$ and net momentum $\mathbf{k}$, for which a single spin operator has a non-zero matrix element with the ground state (at $T=0$).  The notable feature of QSLs is that their quasiparticle-like excitations are non-local, which means that individual quasiparticles do not have such non-zero matrix elements.  Rather, single spin operators acting on the ground state can create the non-local quasiparticles only in pairs.  This leads to an intrinsically smooth signal, where at each momentum transfer $\mathbf{k}$, the intensity is spread out over a continuum of frequencies $\omega$, only constrained by the condition that $\mathbf{k}_1+\mathbf{k}_2=\mathbf{k}$.  Under ideal circumstances, then, some QSLs should exhibit an {\em entirely} continuum spectrum.  This is in sharp contrast to short-range entangled states such as conventional ordered magnets, or VBS states, in which the elementary excitations are essentially spin flips, and single spin operator creates an individual quasiparticle.  In the latter case, this quasiparticle has a definite dispersion $\epsilon_{qp}(\mathbf{k})$, and contributes a resolution-limited peak to the inelastic signal.  

While an entirely continuum spectrum is possible for a QSL, it is not the only possibility.  Even if the {\em elementary} quasiparticles are non-local, a pair of them may form a bound state, which may be a local object and can contribute a sharp peak to $\mathcal{S}(\omega,\mathbf{k})$.  The presence or absence of such a bound state is a non-universal feature which can depend upon the position within a single QSL phase.   Some QSLs have, in addition to non-local quasiparticles, intrinsic emergent but local excitations.  This is the case for the U(1) QSL in a three dimensional Coulomb phase (Sec.~\ref{sec:univ-prop-u1}), which will be discussed in the quantum spin ice context in Sec.~\ref{sec:rare-earth-pyrochl}.  This is an attractive {\em feature} of three-dimensional U(1) QSLs, since it provides an unambiguous smoking gun measurement than can in principle be made using a well-established method.

% In quantum spin liquids, one expects to see either no such magnetic Bragg peaks, or only weak ones, such that most of the wavefunction contributes to the spin fluctuations. Crucial to the understanding of inelastic neutron scattering signal obtained from scattering off a spin liquid is the realization that the spin-spin correlation function provides {\em direct} access to the intensity of $S^z=0$ and $S^z=\pm1$ excitations (if $S^z$ is a good quantum number, otherwise the projection of the excitations onto $S^z$ eigenstates) only. Fractional excitations, which typically carry a fraction of $S^z=\pm1$ will therefore not appear as sharp spin-wave-like signals in the correlation function. Rather, one expects to see a diffuse signal, where at each momentum transfer $\mathbf{k}$, the intensity is spread out over a continuum of frequencies $\omega$, only constrained by the condition that $\mathbf{k}_1+\mathbf{k}_2=\mathbf{k}$. This is exactly a statement that the said excitations are non-local. A neutron interacts locally with the system's spins, and excites two particles which are essentially independent. 

\subsubsection{Resonant X-ray scattering}
\label{sec:x-ray-scattering}

While standard x-ray scattering is mainly used to measure atomic structure, resonant
x-ray scattering is being increasingly applied as a probe of magnetism, often substituting for neutron scattering's traditional role.  This is an active and rapidly changing experimental field, where capabilities have grown dramatically in the past decade.  The technique typical relies on probing a resonant atomic transition, and detecting small changes relative to the large energy of this transition.  Consequently, the energy resolution of resonant x-ray scattering is worse than neutrons: in current experiments, the energy width is of order a few tens of meV, while neutrons easily achieve $0.1$meV and a few $\mu$eV resolution is possible with specialized techniques.  Conversely, the x-rays have excellent momentum resolution, often better than neutrons.  X-rays require much smaller samples due to high resonant scattering efficiency and high intensity sources.  They are also preferable in some materials, such as iridates, which are highly absorbent to neutrons.  

Due to the complexity of resonant processes, x-ray scattering probes more than just the spin structure factor $\mathcal{S}(\omega,\mathbf{k})$.  In addition, it can measure other types of excitations which are created not by single spin operators acting on the ground state, but by more complex (but unfortunately still local) operators.  The literature on the theory of these measurements in Resonant Inelastic X-ray Scattering (RIXS) is still developing, and we refer the reader to Refs.~\cite{ament2011resonant,savary2015probing}.

\subsubsection{Specific heat}
\label{sec:specific-heat}

The most fundamental thermodynamic measurement, specific heat $c_v$ plays several roles in the study of quantum magnetism.  First, it is the most unbiased detector of phase transitions.  Any phase transition at $T>0$ is accompanied by a singularity in $c_v(T)$, which is usually detectable.  Second, it provides a means to determine entropy, using integration based on the standard thermodynamic relations.  One of the simplest tests of the validity of a model spin Hamiltonian is whether it correctly captures the spin entropy released from high to low temperature, which should equal $\ln(2S+1)$ per site for effective spin-S spins.  A classical spin liquid regime can be revealed by a plateau in the entropy, which is associated with an intermediate temperature region of low specific heat.  

The most interesting, but perhaps most subtle, use of specific heat is to gauge the number of low energy excitations in a quantum magnet.  In a gapped state, the low temperature magnetic specific heat is exponentially suppressed and obeys an Arrhenius law.  In a gapless state, it is a power of temperature which reveals the density of states of low energy excitations.  For an ordered $d$-dimensional Heisenberg antiferromagnet, for example, we expect $c_v \sim A T^{d}$. Various power laws are predicted for gapless QSL states, including three-dimensional U(1) Coulomb QSLs, and various two dimensional QSL states with gapless fermionic spinons.  

There can be complications.  In general, magnetic specific heat is not separately measurable, and some subtraction procedure is usually performed to remove the lattice contribution.  Usually this is most reliable at low temperature, where, for a gapless QSL, the magnetic contribution may be significantly larger than the lattice one.  Another difficulty is disorder, which may affect the density of states of some QSLs, changing the intrinsic behavior, or inducing entirely new low energy excitations.  

Specific heat is powerful because of its generality, but this is also its Achilles heel.  It is sensitive to {\em any} excitation, and not only excitations of spins.  In complex materials, especially at low temperatures, many other effects can contribute, such as molecular tunneling excitations and nuclear levels. So care needs to be taken to verify that a ``magnetic'' specific heat is in fact due to magnetic degrees of freedom. 

\subsubsection{Magnetic susceptibility}
\label{sec:magn-susc}

The uniform DC magnetic susceptibility $\chi$ is often the first measure of a new magnetic material.  Like specific heat, it serves several purposes.  It typically shows a peak or cusp at a phase transition, so can be a check for magnetic or other ordering. The presence of a Curie behavior $\chi(T) \sim A/T$ at temperatures high compared to the exchange interactions is the simplest indication that a material contains localized spins at all!  The coefficient $A$ determines the magnetic moment of the spins.  The {\em correction} to the Curie behavior is indicative of the strength of interactions.  Typically the susceptibility is fit to the form $\chi(T) = A/(T-\Theta_{CW})$, where $\Theta_{CW}$ is the Curie-Weiss temperature.  If this formula is applied when $T \gg T_{CW}$, it becomes $\chi(T) \sim A/T (1+ \Theta_{CW}/T + \cdots)$, so that $\Theta_{CW}$ can be viewed as the first correction to the Curie law for independent moments in a high temperature expansion.  This correction can be directly calculated for any exchange model, which allows one to extract some information on the exchange interactions.  For example for the standard Heisenberg antiferromagnetic Hamiltonian of Eq.~(\ref{eq:67}) with spin-$S$ spins, one obtains 
\begin{equation}
  \label{eq:80}
  \Theta_{CW} = - \left( \sum_j J_{ij}\right) \frac{S(S+1)}{3k_B},
\end{equation}
assuming all sites $i$ are equivalent.  Note typically $\Theta_{CW}<0$ for antiferromagnets.  From a single measurement of $\chi(T)$, one may determine $T_c$ for a magnetic ordering transition and the Curie-Weiss temperature, which allows one to construct the dimensionless ``frustration parameter'' $f=|\Theta_{CW}|/T_c$. When $f$ is large, it indicates that the system avoids ordering even when spins are highly correlated, which may suggest incipient QSL behavior, though it can have other explanations.  

The Curie-Weiss analysis is only applicable when $T \gg |\Theta_{CW}|$, a regime which may not be achievable.  Consequently, it may be better to fit experimental data to a higher order expansion or other approximate form of susceptibility determined somehow for a model Hamiltonian. Beyond just the scale of interactions, one can sometimes learn, from single crystal measurements, about more details from the full tensorial structure of the susceptibility: anisotropic g-factors, anisotropic exchange, or both.  A word of caution: fitting the susceptibility may be dangerous as it is usually relatively smooth and featureless, and so cannot reveal more than one or two combinations of parameters in the Hamiltonian.  Nevertheless, these methods are essential to get some insight into at least the scale and sometimes some details of the microscopics of a material.  

As with specific heat, susceptibility at low temperature may also be related to the low energy excitations of a material.  This seems natural since one may obtain susceptibility from thermodynamics: $\chi = - \left.\partial^2 F/\partial H^2\right|_{H=0}$.   In a more detailed sense, however, this depends upon the physics of the magnetic system.  In general the spin susceptibility arises from coupling of the external field to the magnetic moment operator.  In many models, such as Heisenberg ones, the magnetic moment is proportional to the total spin, and the total spin is conserved by Heisenberg interactions.  This implies that the field couples to a conserved quantity, and we can construct simultaneous eigenstates of the field term and the zero field Hamiltonian.  In this situation, {\em the eigenstates of the system do not evolve with field}, only their energies do.  So the susceptibility just reflects a change of occupation of eigenstates, and can be expressed in terms of spin dependent density of states.  In this situation, the low temperature susceptibility is directly related to the density of excitations: it is large when there are many gapless excitations, and small when there are few or none.  For example, within a Heisenberg description a U(1) spinon Fermi surface QSL state is expected on these grounds to have a susceptibility which tends to a constant at $T=0$ (similar to Pauli paramagnetism in a metal), while a gapped $Z_2$ QSL has an exponentially small susceptibility.  These conclusions are, however, {\em incorrect} if the Hamiltonian does not have spin rotation symmetry around the field axis.  When this symmetry is absent, the magnetic susceptibility tends to a constant at $T=0$, irrespective of the presence or absence of low energy excitations.  The non-zero susceptibility in this case arises not from a repopulation of levels, but from the evolution of the ground state wavefunction with applied field.  

\subsubsection{Nuclear magnetic resonance}
\label{sec:nucl-magn-reson}

Nuclear magnetic resonance (NMR) uses nuclear levels of atoms in a sample as a probe of their local environment.  Here we are interested mainly in how those nuclei couple to the electronic spins -- the hyperfine interaction -- and how thereby the electronic spins' behavior is reflected in that of the nucleus.  NMR is a rich and fairly complex subject, and we will only review the simplest and most common applications here.  Generally experiments study the response of the nucleus to externally applied ac fields which are resonant or nearly resonant with a nuclear transition.  One can thereby measure the energy of this transition, as well as the rates at which nuclei relax amongst and within these levels.  One identifies typically two relaxation times: $T_1$, which is the time needed to relax from one Zeeman level to another in a uniform DC field, and $T_2$, which is the time needed to {\em dephase} oscillations of the magnetization normal to the DC field.  

From the point of view of electronic magnetism, one is mostly interested in the transition frequency and the longitudinal relaxation time $T_1$.  The transition frequency is basically set by the {\em effective} Zeeman field seen by a nucleus.  This is the sum of contributions from an external applied field and internal hyperfine fields.  The latter is proportional to the magnetization of the electronic states which are coupled to that particular nucleus.  This ``shift'' of the nuclear resonance frequency is known as the Knight shift.  In the linear response regime, the magnetization of the electronic states is just proportional to their (local) susceptibility, so the Knight shift just gives a measure of the local electronic susceptibility.  This makes it a useful probe to compare with bulk uniform susceptibility.  In a magnetically ordered state, different nuclear sites may become inequivalent, resulting in splitting of a nuclear transition into multiple peaks, or even a broad shape.  Through a careful analysis NMR can give a great deal of information on the presence and arrangement of static electronic moments.

The relaxation rate $1/T_1$ is determined by the ability of the nucleus to emit energy and decay from an excited Zeeman state.  Such decay requires the energy to be deposited elsewhere into some excitation outside the nucleus.  This can be in the lattice, or, more interestingly, in the electronic spins.  The latter arises through the hyperfine coupling, and according to a Fermi's golden rule analysis, probes the density of electronic spin excitations at the transition frequency of the nucleus.  This NMR frequency is very low on magnetic scales, and can usually be considered to be just infinitesimally above zero.  Hence, formally, the relaxation obeys $1/T_1 \sim T\chi_{\rm local}''(0^+,T)$, where $\chi_{\rm local}(\omega,T)$ is the AC local susceptibility of the electronic spins.  Since $1/T_1$ probes the imaginary part of the susceptibility, which is dissipative and hence only non-zero if there are excitations, it is a true probe of the low energy density of states of the spin system, even if the uniform susceptibility and Knight shift fail due to broken spin-rotation invariance.  

We note that there can be subtleties in the NMR relaxation, since what is actually measured is the time evolution of the polarization after some initialization or pulse sequence.  The relaxation times are deduced by exponential fits of this time dependence.  This may not be a simple exponential, if there are multiple environments for the nucleus, in which case more interpretation is required.  A positive view on this is that NMR can be a good way to {\em detect} inhomogeneous or multiple environments for nuclei, which may point to a disordered or complex (e.g.\ incommensurate) magnetic state.

\subsubsection{Muon spin resonance}
\label{sec:muon-spin-resonance}

In muon spin resonance, or $\mu$SR, one implants positively charged muons into a sample, and uses the muon spin, which is like that of an electron, as a two level system to perform NMR-like experiments.  The disadvantage is the need for a facility to produce muons, but there are some advantages.  It can be used on materials for which there are no magnetic nuclear isotopes.  It can be used straightforwardly for zero field measurements: muons are inserted with a definite spin state, and then precess in whatever local field they experience after stopping in the crystal.  From measurements of the muon precession, one can determine the local field it experiences, or rather the distribution of local fields experienced by the ensemble of stopped muons.  One may also measure relaxation times, perform spin echo experiments, etc, similarly to the way one does in NMR.

An issue which sometimes arises in $\mu$SR is that, because the muon has a positive charge, it may actually perturb the sample significantly.  However, there are certainly often clear outcomes.  The observation of spontaneous oscillations of polarization in zero field $\mu$SR is incontrovertible evidence of magnetic order.  

\subsubsection{Optics}
\label{sec:optics}

Light interacts with matter through both its electric and magnetic field, and can probe magnetic systems in several ways.  The simplest effect is a one-photon absorption, which is due to the linear coupling of electromagnetic radiation to the current or charge density.  Since we are concerned with Mott insulators, the usual electron-hole excitations are not present at low energies.  Below the gap, one-photon effects are due to the linear interaction of the electric field with the electric polarization, and the magnetic field with the magnetic dipole operator.  Na\"ively the polarization operator does not involve the magnetic degrees of freedom, but due to spin-lattice interactions and spin-orbit coupling, it may in fact couple to the spins.   Various examples of this have been discussed in the literature\cite{khomskii2012electric,bulaevskii2008electronic}.  This means that magnetic excitations quite generically lead to non-zero absorption within the Mott gap\cite{potter2013mechanisms,ng2007power}.  This fact seems not to be widely appreciated, but is true not only for QSLs but also for ordinary ordered antiferromagnets.  In most cases, due to the time-reversal odd nature of the spin operator, this polarization coupling involves terms even in spin operators, which means that even in such ordered states, the sub-gap weight is multi-magnon in origin, and hence continuum-like.  However, the effect has been predicted to be significantly larger in some gapless QSLs. 

The magnetic dipole operator can be proportional to a single spin, and so in principle this can couple to the same sorts of excitations observed e.g.\ in neutron scattering.  Due to the very long wavelength of light, however, one-photon processes of this type couple only to zero momentum spin excitations.  The inability to scan momentum is to an extent made up for by excellent energy resolution.  This has been quite beautifully demonstrated in CoNb$_2$O$_6$\cite{morris2014hierarchy} and Yb$_2$Ti$_2$O$_7$\cite{pan2014low}.  

Two-photon transitions, known as Raman scattering, in which light is scattered from one energy to another, are commonly studied as probes of both lattice and spin excitations.  The latter is due to the quadratic coupling of the electromagnetic field to the spins, usually to ``bond'' operators involving pairs of spins\cite{PhysRevB.53.R14733}.  In general the signals measured by Raman are continua, and involving as they do correlations of bond operators, somewhat complex objects.  The contributions from spins must also be separated from those of phonons, etc.  Due to these complexities, applications to QSLs are limited, and open to multiple interpretations.\cite{PhysRevB.82.144412,maczka2008temperature}\  

\subsubsection{Thermal conductivity}
\label{sec:thermal-conductivity}

The thermal conductivity of a material can, like the electrical conductivity, be expressed by an Einstein relation as the product of the thermal diffusion constant and specific heat, which plays the role of the density of states.  Due to the second factor, it is closely related to the heat capacity, which we discussed in Sec.~\ref{sec:specific-heat}.  However, because of the first factor, only excitations which are mobile contribute significantly to the thermal conductivity.  Thus the thermal conductivity is useful to determine if excitations which contribute to specific heat are extended or localized.  This is important because an insulator may quite commonly have localized in-gap electronic states, which might masquerade in a purely thermodynamic measurement as spin excitations.  

In general, all propagating excitations can contribute to thermal conductivity, and nominally add in the conductivity as parallel conduction channels. A phonon contribution is always present.  Ideally, in a QSL with gapless fermion spinons, these excitations are expected to add a power-law dependence (as a function of temperature), indicative of the type of low-energy excitations present in the system.  If the contribution is sufficiently large, as in the case of a spinon Fermi surface (due to its high density of states) it would be expected to dominate over the lattice term at low enough temperatures.  A large thermal conductivity of magnetic origin in an insulator may thus be indirect evidence of fermionic quasiparticles, a surprising feature of a spin system.  However, one should be clear that thermal conductivity itself is completely agnostic as to the quantum numbers of the excitations, so that, without some corroborating evidence, one has no reason to suppose a measured thermal conductivity signal is related to spins.  Comparison of the data with that in a high magnetic field, where magnetic excitations are quenched at low energies, can help disentangle phonon contributions from magnetic ones.

The thermal Hall  effect, analogous to the electrical Hall effect,  the flow of a thermal current normal to a temperature gradient.  This requires time-reversal symmetry breaking, usually induced by a field, although it could occur spontaneously.  The difficult thermal Hall conductivity measurement has been reported in some compounds.   A priori the thermal Hall effect requires some degree of freedom besides the lattice, since it is hard to see how phonons couple significantly to an applied magnetic field.  In an insulator, there can be both orbital and Zeeman contributions that affect the spins, and, through various means, induce a thermal Hall effect.  In general we may expect that thermal Hall current can still include a lattice contribution, if the lattice degrees of freedom are scattered anisotropically by magnetic or electronic degrees of freedom which sense the field.   It may also include a direct contribution due to transverse heat current carried by spins.  

A {\em quantized} thermal Hall effect (the direct contribution) is a sharp characteristic of chiral topological phases, such as the chiral spin liquid.   It has been suggested that an unquantized direct contribution might be characteristic of some gapless QSLs.  To our knowledge no clear signature of a QSL has been observed by thermal Hall measurements to date.

\subsubsection{Other proposed experiments}
\label{sec:theor-prop}

Recognizing the lack of smoking gun signatures in standard measurements of magnetic systems, theorists have proposed unusual ways one might look for QSLs.  These include: 
\begin{itemize}
\item A state with a spinon Fermi surface differs probably most dramatically from an ordinary insulator, and various specific experiments may be envisioned to look for this Fermi surface.  For example, one might observe Friedel oscillations from an impurity,\cite{mross2011charge} or observe RKKY oscillations of exchange bias in a magnetic heterostructure\cite{norman2009measure}.  Angle-resolved photoemission might reveal the Fermi surface above the Mott gap\cite{tang2013low}. 
\item More generally, there are ideas for experiments which exploit the relations between the quasiparticles of a spin liquid and excitations of ``nearby'' phases.  For example, one may obtain a superconductor by condensing a bosonic ``holon'' (which is topologically an {\sf e} particle, but endowed with electric charge) in a $Z_2$ spin liquid.  In this picture, the {\sf m} particle is converted into a superconducting vortex.    This led to a proposal to measure persistent vorticity on quenching between a $Z_2$ QSL and a superconductor, or the possibility of observing vortices with double the minimal flux in a nearby superconductor.\cite{PhysRevB.63.134521,PhysRevLett.86.292}  Such a measurement was even attempted experimentally\cite{bonn2001limit}, with a null result. Similarly, a fermionic $\varepsilon$ particle in the QSL can be converted into a Bogoliubov-de Gennes quasiparticle in a superconductor, and one may contemplate an oscillation experiment to detect injection of this quasiparticle from a superconductor in contact with a $Z_2$ QSL\cite{barkeshli2014coherent}.
\item In ultra-cold atoms, non-local measurements might actually be possible that detect QSLs more intrinsically.  For example, using a quantum gas microscope, one can make simultaneous measurement of the state of every lattice site.  This allows one to reconstruct the amplitudes of every term in the wavefunction, and thereby reconstruct the full wavefunction if some knowledge of the phases is given.  This has been used to measure the string order in one dimensional boson Mott insulators\cite{endres2011observation}. There are also protocols to measure the 2nd Renyi entanglement entropy by producing a ``twinned'' state composed of two copies of a physical system, and then carrying out a specific quantum evolution by controlling interactions between these copies -- see Ref.~\cite{PhysRevLett.109.020505}.  Interference between such twins has been measured\cite{kaufman2014two}, and very recent the proposed experiment has actually been carried out for a four site chain of bosons\cite{greiner:_entan}.  
\end{itemize}

\subsection{Materials}
\label{sec:materials}

\subsubsection{Possible Kitaev materials}
  \label{sec:honeycomb-iridates}

The honeycomb iridate materials form a family whose simplest members are Na$_2$IrO$_3$ and Li$_2$IrO$_3$, in which Ir$^{4+}$, which is in a $d^{5}$ configuration, occupies the sites of layered honeycomb lattices (separated by a plane of sodium or lithium atoms).   These compounds rose in prominence with the suggestion\cite{jackeli2009mott,chaloupka2010kitaev} that they might be governed to a good approximation by the remarkable soluble Kitaev honeycomb Hamiltonian discussed in Sec.~\ref{sec:kita-honeyc-model}.  The basic physics descends from that of the Ir$^{4+}$ ion, which, in the octahedrally coordinated oxygen environment, can be considered to have a single hole in a 3-fold degenerate t$_{2g}$ shell.  Due to strong spin-orbit coupling, the orbital degeneracy is lifted in favor of a non-degenerate Kramer's doublet ground state: an effective S=1/2 spin (this is, in the simplest limit, called a $j=1/2$ state in the iridate literature).  Owing to the edge-sharing nature of the octahedra in these materials, there are two approximately 90 degree Ir-O-Ir bonds connecting each pair of spins, and a superexchange analysis in the strong spin-orbit limit predicts a Kitaev-type interaction between neighboring spins,\cite{jackeli2009mott} in addition to a Heisenberg term arising from direct Ir-Ir overlap.  Thus the ``minimal'' model for these materials is largely believed to be a S=1/2 Hamiltonian consisting of the sum of the Kitaev one plus Heisenberg terms.  A QSL ground state is possible in some (unfortunately narrow) range of parameters for this model.

These materials were eventually shown to order magnetically \cite{liu2011long,choi2012spin}.   Nevertheless, one may hope to observe some remnant of QSL physics either above the ordering transition, or in the higher energy excitations, or both.  There are indications of strong fluctuations at least.  Na$_2$IrO$_3$ is the best characterized, and has a Curie-Weiss temperature of $\Theta_{CW}\approx -120$K, and inelastic neutron scattering confirms that the excitation spectrum disperses on the scale of 6-10meV, which is comparable.  However, the N\'eel temperature of $T_N \approx 18K$ is significantly lower, giving it a moderate frustration parameter of $f=6.6$.  Moreover, the static moment is estimated at $0.22\mu$B from elastic neutron scattering, much reduced from the full 1$\mu$B spin-1/2 moment.  Quite recently rather direct evidence that some direction-dependent spin interactions are substantial has been uncovered by diffuse magnetic x-ray scattering, which measures equal-time spin correlations {\em above $T_N$}.  One observes that different spin components are correlated at different momenta, consistent with substantial Kitaev coupling. 

Complications arise when one looks in more detail at the ordered state, which is of ``zig-zag'' type, a collinear arrangement of spins that doubles the unit cell.  This was not expected from the simplest theoretical model, leading to two main speculations on an appropriate modified Hamiltonian.  The zigzag state can be captured by the simplest nearest-neighbor Kitaev-Heisenberg model, but only if the Kitaev term has a anti-ferromagnetic sign and the Heisenberg one is ferromagnetic, which seems not so natural.  Alternatively, it may occur with ferromagnetic Kitaev interactions and antiferromagnetic Heisenberg coupling, if second and third neighbor exchange are significant.   These are not the only possibilities, and more complicated Hamiltonians have been explored\cite{rau2014generic}; yet another potential pitfall is that (Na,Li)$_2$IrO$_3$  do not actually have exact hexagonal symmetry: the true space group is C2/m.

A yet different theoretical view is provided by {\em ab initio} calculations, which propose a ``quasi-molecular orbital'' picture of these materials as band rather than Mott insulators, and argue that spin-orbit coupling has only moderately significant effects\cite{PhysRevB.88.035107,PhysRevLett.109.197201}.    This may not be incompatible with the spin model including longer-range exchange.  However, a band picture usually fails when the gap is much larger than the magnetic exchange scale, which seems to be the case given the gap of 340meV in Na$_2$IrO$_3$ measured by ARPES.  

The lithium analog, Li$_2$IrO$_3$ actually occurs in several structures: $\alpha$-Li$_2$IrO$_3$, which is a true honeycomb like the sodium compound, and two three-dimensional polytopes, $\beta$-Li$_2$IrO$_3$ and $\gamma$-Li$_2$IrO$_3$, which locally have the coordination of a honeycomb with each IrO$_6$ octahedra sharing three edges with other octahedra, but arranged into more complex non-planar structures.  These three-dimensional structures also admit soluble Kitaev-like Hamiltonians with $Z_2$ QSL phases, that have been studied theoretically.  Experimentally, all three materials order antiferromagnetically.  The order in the three-dimensional ($\beta,\gamma$) materials is completely determined from single crystal resonant elastic x-ray scattering, while the magnetic structure of the $\alpha$ type is not yet known.  Both 3d materials exhibit incommensurate non-coplanar spiral structures with very similar wavevector, and similar ordering temperatures $\sim 40K$, while the $\alpha$ polytope orders at $T_N \sim 15K$, close to that of the sodium material.   The Curie-Weiss temperature for the $\alpha$ compound is estimated as $\Theta_{CW} \approx -33K$ and for the $\beta$ material $\Theta_{CW} \approx +40K$, neither of which would suggest significant frustration.  Single crystal measurements of the $\gamma$ compound show significant strong temperature-dependent anisotropy which cannot be fit well by a Curie-Weiss form, but does clearly indicate directional-dependent interactions.  A priori the higher ordering temperatures and lower Curie-Weiss temperatures of the lithium materials may suggest they are further away from a QSL state than is the sodium compound.  The larger ordered moment, estimated at $0.47\mu$B for the $\beta$ compound\cite{biffin2014unconventional}, seems to support this picture.

The 4d transition metal material $\alpha$-RuCl$_3$ has emerged quite recently as another possible Kitaev material, outside the iridate family.  In general, stronger correlations (i.e.\ larger Hubbard $U$) but weaker spin-orbit coupling is expected for 4d ions relative to 5d ones.  However, it can very well be sufficient to generate strong anisotropy (which even occurs for 3d ions in some situations).  Optical and x-ray studies, backed up with density functional theory, support the interpretation of this material as a spin-orbit assisted Mott insulator with Ru$^{3+}$ ions in a t$_{2g}^{(5)}$ state, just as in the iridates\cite{plumb2014alpha}.  The optical gap is $\sim 0.3eV$, also comparable to that of Na$_2$IrO$_3$.  However, if we adopt the usual superexchange approach, the octahedra in $\alpha$-RuCl$_3$ are actually less distorted than in Na$_2$IrO$_3$, and also the Ru are more separated than the Ir, both of which probably favor a more ideal realization of nearest-neighbor Heisenberg-Kitaev physics.

Unfortunately, however, like the iridates, $\alpha$-RuCl$_3$ clearly does not have a QSL ground state.  Early studies showed antiferromagnetic ordering at $15K$\cite{kobayashi1992moessbauer}, and more recently multiple transitions have been observed\cite{kubota2015successive,majumder2014anisotropic,PhysRevB.91.144420}. It is not clear if both are intrinsic or may reflect different structural domains in a sample\cite{banerjee2015proximate}.  It is plausible that this is related to stacking faults due to the weak van der Waals bonding of this compound. The susceptibility gives a ferromagnetic Curie-Weiss temperature of $\Theta_{CW} \approx $23-40K\cite{kobayashi1992moessbauer,fletcher1967x}, which is not much larger than $T_N$, but given that it is ferromagnetic and the ordering is antiferromagnetic, there may be some frustration.  Elastic neutron scattering\cite{PhysRevB.91.144420} seems to support zig-zag order as in Na$_2$IrO$_3$.  Very recent inelastic powder neutron scattering observes a high energy ($\sim 6.5$meV) feature which persists above $T_N$ and which seems similar to what is expected from the Kitaev model\cite{banerjee2015proximate}.  

In general, while none of these honeycomb-like materials seems to be an ideal QSL, they may be close enough to be interesting.  It would be interesting to see if remnants of QSL physics might be observable at high energy or high temperature (it is an interesting theoretical problem whether we would even expect this), or if any one of these materials is really close to a QSL state,  it could be somehow driven into a real QSL in some way, for example by pressure, field, or chemical modification.  

\subsubsection{Rare-earth pyrochlores}
\label{sec:rare-earth-pyrochl}

The rare earth pyrochlore oxides form a venerable class of compounds studied for their frustrated magnetism\cite{gardner2010magnetic}.  Such materials have chemical formula R$_2$M$_2$O$_7$, with R a rare-earth ion, and M a transition metal ion such as Ti or Sn.  In many of those compounds, much like in the honeycomb iridates from the previous section, very strong spin-orbit coupling and crystal fields split the single-ion spectrum into low- or zero-degeneracy manifolds. When the single-ion ground manifold is a doublet, well separated in energy from the first-excited manifold, the effective theory should be that of a spin-$1/2$.   Sometimes the interactions amongst these effective spins are extremely classical: they involve to very high accuracy only a single component of each spin, so that all terms in the Hamiltonian commute.  This is the situation in the so-called classical spin ices, such as Ho$_2$Ti$_2$O$_7$ and Dy$_2$Ti$_2$O$_7$.  However, this is not always the case.   Strong quantum effects are present in quite a few materials, e.g.\ Yb$_2$Ti$_2$O$_7$, Er$_2$Ti$_2$O$_7$, Tb$_2$Ti$_2$O$_7$, Pr$_2$Zr$_2$O$_7$.  Recognition of these materials as fully quantum effective S=1/2 systems became widespread only a few years ago, stimulated by careful investigation of Yb$_2$Ti$_2$O$_7$ by inelastic neutron scattering.  

In general, the high degree of localization of the $f$-orbitals means that exchange is very short range, and apart from a weaker dipolar interaction, can be taken to an excellent approximation to be restricted to nearest neighbors.  This restriction, and the high cubic symmetry of these compounds, allows for a rather complete understanding of the underlying spin Hamiltonian. 
For most ions (exceptions occur only if the low-lying doublet has unusual symmetry, e.g.\ Nd$^{3+}$\cite{huang2014quantum}), the most general nearest-neighbor Hamiltonian takes the form
\begin{eqnarray}
  \label{eq:53}
  H & = & \sum_{\langle ij\rangle} \Big[ J_{zz} \mathsf{S}_i^z \mathsf{S}_j^z - J_{\pm}
  (\mathsf{S}_i^+ \mathsf{S}_j^- + \mathsf{S}_i^- \mathsf{S}_j^+) \nonumber \\
  && +\, J_{\pm\pm} \left[\gamma_{ij} \mathsf{S}_i^+ \mathsf{S}_j^+ + \gamma_{ij}^*
    \mathsf{S}_i^-\mathsf{S}_j^-\right] \nonumber \\
&& +\, J_{z\pm}\left[ \mathsf{S}_i^z (\zeta_{ij} \mathsf{S}_j^+ + \zeta^*_{ij} \mathsf{S}_j^-) +
  {i\leftrightarrow j}\right]\Big],
\end{eqnarray}
There, the spin components are taken along {\em local} axes, and the coefficients $\gamma$ and $\zeta$ are constant phases related to a particular choice of local  ``transverse'' axes, which ensure the model has all the proper symmetries. All such details can be found in, e.g., Ref.~\cite{savary2012coulombic}.  Eq.~(\ref{eq:53}) should provide an excellent description of a wide set of these rare-earth pyrochlore materials.  It has three dimensionless parameters (ratios of exchanges), so the problem of finding QSLs in these materials reduces to the problem of finding them in this three-dimensional space.  

What do we know about this problem?  It is most studied (and most interesting) for the case $J_{zz}>0$, where this Ising interaction is frustrated, and this defines what one may call ``quantum spin ice''.  The first two terms of $H$ are exactly the XXZ model of Eq.~\eqref{eq:52} discussed in Sec.~\ref{sec:pyrochlore-xxz-model}. The existence of a stable spin liquid phase in the latter model was established in the $J_\pm/J_{zz}\ll1$ limit, where the model was mapped to a pure $U(1)$ compact gauge theory, as discussed in Sec.~\ref{sec:pyrochlore-xxz-model}.   This can be seen as the restriction of a full gauge theory with matter to the sector with no gauge charge, physical when the gauge charges are gapped. In fact, away from the perturbative limit the gauge charges are important, and at least virtual fluctuations must be included.  This is accomplished in Ref.~\cite{savary2012coulombic}, where a novel parton construction, different from the standard ones discussed in Sec.~\ref{sec:partons}, was introduced in a way designed to match the perturbative limit of Ref.~\cite{hermele2004pyrochlore}: 
\begin{equation}
  \label{eq:76}
\mathsf{S}^z_{\mathbf{rr}'}=E_{\mathbf{rr}'},\qquad
\mathsf{S}^+_{\mathbf{rr}'}=\Phi^\dagger_{\mathbf{r}} e^{iA_{\mathbf{rr}'}}\Phi_{\mathbf{r}'},
\end{equation}
with the constraint ${\rm div}E=Q_{\mathbf{r}}$, where $\Phi_{\mathbf{r}}=e^{-i\phi_{\mathbf{r}}}$ (and so $\Phi^\dagger_\mathbf{r}\Phi_{\mathbf{r}}=1$, a constraint in a sense ``analogous'' to the constraints  $b^\dagger b=2S$ and $f^\dagger f=1$ from the Schwinger boson and Abrikosov fermion rewritings of the spin, Eqs.~\eqref{eq:29} and \eqref{eq:30}, respectively) is a rotor variable, and $[\phi_{\mathbf{r}},Q_{\mathbf{r}}]=i$, which is invariant under the combination of $A_{\mathbf{r}}\rightarrow A_{\mathbf{rr}'}+\chi_{\mathbf{r}}-\chi_{\mathbf{r}'}$ and $\Phi_{\mathbf{r}}\rightarrow\Phi_{\mathbf{r}}e^{i\chi_{\mathbf{r}}}$. Gauss' law commutes with the Hamiltonian, a sign of the $U(1)$ gauge symmetry generated by $U(\mathbf{r})=e^{i({\rm div}E-Q_{\mathbf{r}})}$. Here the partons live on the centers of the tetrahedra, and have a very physical interpretation, in view of the physics of classical spin ice. 

In the original formulation of the theory \cite{savary2012coulombic}, the spinons were taken to be bosons. This has the advantage of allowing a simple description of the ``all-in-all-out'' tetrahedral configurations (i.e.\ with {\em two} spinons on a tetrahedron) as well as one of the transitions in terms of single-spinon condensation. The Hamiltonian we obtain through this procedure is that of spinons hopping ($J_\pm$ and $J_{z\pm}$ terms) as well as interacting ($J_{\pm\pm}$ term) in a fluctuating background. This reformulation is exact, when the constraints discussed above (rotor and Gauss' law) are enforced. This reformulation allows for the implementation of a mean field theory -- dubbed gauge Mean Field Theory (gMFT) -- which can describe quantum liquid phases. In particular, $\langle\Phi\rangle=0$ should be characteristic of a quantum liquid phase, as $\langle\Phi\rangle\neq0$ is the criterion for Higgs condensation, which induces a conventional time-reversal breaking magnetically ordered phase.  Such a mean-field approach was carried out for some large subregions of the phase space of Eq.~(\ref{eq:53}) in Refs.~\cite{savary2012coulombic,lee2012generic}.  But, we should emphasize, this is only a mean field theory, and while it agrees with solid results in some limits, is not expected to be accurate everywhere.  And indeed, the Hamiltonian includes some particularly difficult limits, such as the isotropic Heisenberg antiferromagnet, c.f.\ Sec.~\ref{sec:heis-pyrochl}.  Other parton constructions are possible, as are other mean field theories \cite{hao2014bosonic}.  In our opinion, more computational efforts, e.g.\ variational Monte Carlo calculations, are sorely needed here.

What about the experimental situation?  Among the known ``quantum'' pyrochlore materials, Yb$_2$Ti$_2$O$_7$, Tb$_2$Ti$_2$O$_7$, and Pr$_2$Zr$_2$O$_7$ are most discussed as spin liquid candidates.  Yb$_2$Ti$_2$O$_7$ is the most studied.  It clearly is a highly quantum, effective S=1/2 system, for which direct evidence is provided by the large inelastic weight and clearly defined sharp spin wave excitations in high field, with a bandwidth of order a meV \cite{ross2009two,prx}.  It seems also to be in the quantum spin ice category.  This is based in part on the exchange couplings, which were determined by the very direct technique of measuring the single magnon spectrum in a large (5T) field, for which the spins are nearly fully polarized in the ground state.  Because the ground state is rendered trivial by the strong field, one can straightforwardly and accurately calculate the magnon spectrum for arbitrary exchange couplings, and then fit the result to experiment \cite{prx,savary2012order}.  The technique is well suited to this class of materials because exchange interactions are relatively weak, hence the necessary fields are accessible and compatible with experiment, and large single crystals are available.     This experiment determined all the exchanges and in particular concluded that $J_{zz}>0$ was the strongest coupling\cite{prx}, which supports the quantum spin ice picture  (though this has recently been contested in Ref.~\cite{robert2015} and by Coldea).   Other direct evidence for spin ice physics is Ref.~\cite{chang2012higgs}, which reported the presence of ``pinch point'' like features in equal-time neutron scattering at higher temperature, consistent with the picture of a transition to a classical spin ice regime \cite{savary2013spin}.

However, it is also clear that Yb$_2$Ti$_2$O$_7$ does not realize the ideal Coulomb QSL state possible for Eq.~(\ref{eq:53}).   High quality samples show a pronounced specific heat anomaly indicating a sharp phase transition at about $T_c\approx 240$mK.    This seems largely consistent with a ``splayed'' (non-collinear) ferromagnetic ground state, which is expected according to both on the classical approximation and gMFT, given the exchanges fit in Ref.\cite{prx}.  Many experiments do indeed support ferromagnetic order \cite{yasui2003ferromagnetic,chang2012higgs,chang2014static,lhotel2014first}.  However, the transition temperature itself is suppressed, $T_{\rm MF}/T_c \approx14$, using Curie-Weiss mean field theory to determine $T_{\rm MF}$ given the fit parameters.  Moreover, spin wave excitations are strangely absent in the zero field ordered phase: inelastic neutron scattering shows instead only diffuse continuum excitations \cite{ross2009two,prx}.  Muon spin resonance also shows dynamics different from that of usual ferromagnets \cite{dortenzio2013unconventional}.

Interpretation of the zero field behavior is not settled.  We believe quantum fluctuations play an essential role in at least the unconventional excitations, though the connection to Coulombic QSLs is far from clear.  A priori, even gMFT does not place the zero field state close to a QSL phase: rather it is near a phase boundary between two ordered states.  So perhaps proximity to this boundary is important\cite{jaubert2015f-ingquestion}.  We also note that a large sample dependence due to the disorder which is known to exist in this family of compounds. In particular, it was explicitly shown that a single crystal of Yb$_2$Ti$_2$O$_7$ had approximately $2\%$ ``stuffed'' (off-stoichiometric) Yb$^{+3}$ ions, i.e.\ extra Yb ions lacunar Ti ones \cite{ross2012lightly}.  So disorder may play some role in the zero field spin dynamics, though we are inclined to think it is a secondary one.  Given continued experimental activity \cite{tokiwa2015thermal,pan2014low,pan2015measure}, we expect Yb$_2$Ti$_2$O$_7$ to be an interesting and evolving topic in the near future.

Tb$_2$Ti$_2$O$_7$ has the status of a long-standing puzzle and was the first pyrochlore to be suggested to belong to the ``quantum spin ice'' class. Theoretically, however, much less is known about this compound. This is largely due to the fact that it is unclear whether the full low-energy properties may be accounted for by a simple two-level (effective $S=1/2$) approach. Indeed, in contrast to Yb$^{3+}$ in Yb$_2$Ti$_2$O$_7$, whose lowest crystal field doublet lies 700K below the first excited one, the first two doublets in Tb$_2$Ti$_2$O$_7$ are only separated by an energy of approximately 20K. While this may seem relatively large compared to an exchange energy of a few Kelvins, inelastic neutron scattering data seems to show what can be interpreted as substantial ``mixing'' between the crystal field levels. Experimentally, Tb$_2$Ti$_2$O$_7$ has a Curie-Weiss temperature reported to be between $-19$K and $-14$K. Remarkably, for a long time, no magnetic order or signature of a magnetic phase transition had been reported down to $50$mK, with neutron scattering only displaying a diffuse signal \cite{gardner1999cooperative,gardner2001cooperative,yasui2002static,gardner2003dynamic,rule2006field,mirebeau2007magnetic}. Recently, however, thorough investigations have pointed to both the existence of more structure than initially thought in neutron scattering, like features reminiscent of ``pinch points'' (though quite different from those seen in classical spin ices) point towards power-law correlations \cite{fennell2012power} and of relatively sharp magnetic Bragg peaks in quasi-elastic scattering in field-cooled samples \cite{fennell2012power,fritsch2013antiferromagnetic,fritsch2014temperature}, as well as an anomaly in the $\mu$SR frequency shift at $T=0.15$K \cite{yaouanc2011exotic}. In fact, despite quite a large sample dependence, most crystals display a sharp upturn in the specific heat below about 400mK \cite{takatsu2012quantum}. While an upturn in the specific heat is expected because of the nuclear contributions, the minimum, which occurs in Ref.~\cite{yaouanc2011exotic} at a temperature close to where the kink in $\mu$SR is observed, could also be indicative of a transition. A magnetic transition at $150$mK would yield a frustration parameter of $f\sim125$. % In neutron scattering, the diffuse signal which was observed all the way down to $T=50$mK has been shown to, in fact, display more structure than initially thought \cite{fennell2012power,fritsch2013antiferromagnetic,fritsch2014temperature}. In particular, elastic (not equal-time) magnetic Bragg peaks are clearly seen.
Thermal conductivity -- both diagonal and Hall -- provide a new vantage point, which seems also to indicate the presence of exotic {\sl magnetic} excitations \cite{li2013phonon,hirschberger2015thermal}.  Finally, Tb$^{3+}$ contains an even number of electrons, and therefore the crystal field doublets are not Kramers doublets: their degeneracy is enforced not by time-reversal symmetry, but, rather by (possibly-approximate) crystal symmetries. As a consequence, the levels may be split by, e.g., Jahn-Teller distortions, although this latter fact is still at the center of a controversy. 

Pr$_2$Zr$_2$O$_7$ was also studied as a potential quantum spin ice candidate \cite{kimura2013quantum}. While no signs of long range magnetic order were seen in inelastic neutron scattering, pinch-point-like features were observed. The first excited crystal-field doublet lies at about 90K \cite{kimura2013quantum}, so that a spin-$1/2$ model is expected to be appropriate. This compound is, however, due to the similar sizes of the Pr$^{3+}$ and Zr$^{4+}$ ions, more prone to site inversion, and therefore to disorder, than the other, aforementioned, compounds. Moreover, like in Tb$_2$Ti$_2$O$_7$, the non-Kramers nature of the low-lying Pr$^{3+}$ doublet allows for its splitting, for example by structural disorder. Random doublet splitting has recently been advocated for experimentally. Theoretically, local symmetry breaking, even when random, rather than lead to a glassy phase, can be shown to instead give rise to the quantum spin ice phase, even in the absence of quantum spin exchanges, as well as to a ``Mott glass'' phase, depending on the strength of the disorder. Whether this is indeed the physics at play in the samples of Pr$_2$Zr$_2$O$_7$ grown so far will require more careful investigations, but this compound definitely stands as a promising candidate for the realization of a disorder-induced quantum spin liquid, a scenario possible thanks to the the fact that entanglement does not rely on the existence of any symmetries (such as translational).

% Overall, a common feature of all of the compounds mentioned above is the presence of pinch points, which can likely be quite generally taken as a signature of a large positive $J_{zz}$ interaction. Quantitatively, it might be possible to estimate 

The materials discussed here are only a few examples (the best studied for their possible quantum spin ice behavior) in the large family of rare-earth pyrochlore compounds. Intense experimental (new materials, and more detailed experiments) and theoretical investigations in this family of compounds, as well as in that of spinels--where the same physics could occur--are underway. In this context, new interesting developments are more than likely, and there is hope for a quantum spin liquid to be discovered.

%In particular, while the small scale of the exchange strength in $f$ compounds makes experiments challenging, the abundance of rare-earth pyrochlore compounds on the other hand holds good promise for finding quantum spin liquid physics, and place the pyrochlores in a good candidate material position.

% Very recent compounds: Ce$_2$Sn$_2$O$_7$ \cite{sibille2015candidate} from the Frenchies? Seems like this one is more like AIAO in fact.  Yb$_2$Sn$_2$O$_7$ \cite{yaouanc2013dynamical}?

\subsubsection{Triangular organics}
\label{sec:triangular-organics}

There are a great variety of organic molecular crystals which have been studied as nominally simple examples of correlated electron systems \cite{powell2011quantum}.  Because of the large size of the organic molecules, there are many atoms in the unit cell, and the electronic structure is in principle quite complex.  However, it is generally thought that each organic molecule, or pair of molecules, hosts a single partially occupied molecular orbital which is near the Fermi energy.  In this way, such materials are conceptually reduced to simple Hubbard-like models.  Many structures can be produced, by varying both the molecules containing the partially-filled orbitals, other molecules in the structure, end-groups on these molecules, and the pattern of the molecular  arrangements.  Diverse examples of both conductors and insulators have been studied.  

Out of this forest of materials, two have emerged as excellent empirical QSLs:  $\kappa$-(ET)$_2$Cu$_2$(CN)$_3$, EtMe$_3$Sb[Pd(dmit)$_2$]$_2$ -- we will abbreviate these by $\kappa$-ET and dmit, respectively.  These are quasi-two-dimensional nominally half-filled Mott insulators, in which molecular units host single orbitals which appear magnetic and are arranged into an approximate triangular lattice, each such layer being separated by non-magnetic ions.  Each of these two compounds are specific members of large families with the same conducting molecules, where all the differences within each family occur within the non-magnetic ions. Based on quantum chemistry calculations, and more recently density functional theory, an extended Hubbard model is typically considered an adequate theoretical description.   Unlike inorganic frustrated magnets based on 4f and 3d ions, which are ``strong'' Mott insulators well-described by entirely localized spins, here the electrons are close to a metal-insulator transition, and a spin-only description may be insufficient.  Evidence for this ``weak'' Mott insulating behavior comes from optics and transport, which observe a negligible or very small gap, and pressure studies, which are able to drive a Mott transition to a metallic state with relatively small pressures.  This suggests that a pure spin Hamiltonian with very short-range interactions is a bad approximation for these materials, and that either an extended Hubbard model or perhaps a long-range exchange model is better.

What is the evidence for QSLs in these materials?  First, they are insulating, as shown in transport and optics, so it is reasonable to regard them as Mott insulators.  Second, both magnetic susceptibility and NMR indicate the presence of local moments, with spin S=1/2 per molecular unit, and effective antiferromagnetic exchange interactions of the order $J\sim 200-250K$, in the sense the susceptibility is progressively suppressed below a Curie law as temperature is lowered.  However, despite the relatively large exchange coupling, there is no indication of magnetic ordering.  Thermodynamic quantities -- the susceptibility and heat capacity -- show no singular features which would clearly signal phase transitions.  More directly, NMR measurements, which are the most sensitive probes in these systems, show the absence of any spontaneous static magnetic fields developing down to very low temperature of 20mK in both materials.   The combination of these measurements seems to conclusively rule out long-range antiferromagnetic order at a temperature $\approx 10^4$ times lower than $J$!

However, this is negative evidence.  What do we actually know about the ground state and low energy excitations?  The main evidence for interesting physics comes from thermal measurements -- heat capacity and thermal conductivity -- at temperatures below $\sim$1K.  In these particular salts, in contrast to others in their families, a large linear term in the specific heat versus temperature, $c_v \sim \gamma T$, is observed\cite{yamashita2008thermodynamic,yamashita2011gapless}.  This is similar to the Sommerfeld contribution in a metal, and is in fact comparable to that in a metal (if normalized per spin) -- something unexpected in an insulator.  The former indicates some low energy excitations with a large density of states, and have been interpreted as evidence for fermionic quasiparticles, which can sustain a large constant density of states by the Pauli principle. In a Mott insulator, these cannot be electrons, so it is tempting to think this may be an indication of a spinon Fermi sea.  It was also noted that the uniform susceptibility $\chi$  (measured both by magnetization and NMR Knight shift) saturates to a substantial constant at low temperature, which is also typical for a metal.  Furthermore, the Wilson ratio,
\begin{equation}
  \label{eq:49}
  R_W = \frac{4\pi^2 k_B^2 \chi}{3 (g\mu_B)^2\gamma},
\end{equation}
is order one in both materials, which is expected for free fermions, and may indicate that the same degrees of freedom control both the susceptibility and the specific heat.  This seems like a non-trivial consistency check.  On the other hand, the specific heat data at least for dmit has a strange feature: the Sommerfeld coefficient is modified by a factor of two by partial isotopic substitution of deuterium for hydrogen in the layers between the dmit molecules\cite{yamashita2011gapless}, which one would think would have nothing to do with the spins.  

A different concern is whether the signals might arise from localized single-particle states, since specific heat and (to a first approximation) susceptibility are related just to density of states.  Therefore a measurement which proves the delocalized nature of the excitations is important.  This is the thermal conductivity $\kappa$.  Measurements on dmit over the range of $0.1K<T< 0.3K$ fit well to a $T^3$ phonon contribution plus a Sommerfeld-like term linear in temperature attributed to the spin subsystem.\cite{yamashita2010highly} This gives some weight to the idea that the low energy excitations are delocalized fermionic spinons, which is quite exciting.  In $\kappa$-ET, the non-phonon contribution appears to be strongly suppressed at the bottom of this temperature range, leading to the suggestion of a gap opening\cite{yamashita2009thermal}.  

Theoretically, modeling of these results has been based on the triangular lattice Hubbard model.  A non-magnetic insulating phase was first proposed based on numerical simulations\cite{kashima2001magnetic,morita2002nonmagnetic}. Later, following the specific heat measurements, theorists proposed specifically the spinon Fermi sea coupled to an emergent $U(1)$ gauge field\cite{lee2005u,motrunich2005variational} -- see Sec.~\ref{sec:ring-exch-hubb}.   As discussed, this is broadly in agreement with experiment.  Notably, this QSL state is described by a strongly interacting field theory, so there are theoretical uncertainties.  Nevertheless, theory predicts some even more dramatic phenomena than free fermionic spinons.   In fact it finds not a linear specific heat but one scaling as $T^{2/3}$\cite{motrunich2005variational}.  Given the limited range of data, and the presence of a nuclear Schottky term, it is unfortunately not possible to distinguish this from the linear Sommerfeld behavior.  

However, these measurements are quite indirect: they really indicate only that there are some excitations carrying heat, and that these are mobile.  There is no direct indication that these are actually associated with the spins.  Due to the limited range of temperature of the measurement, one cannot directly see that these excitations emerge when the spins become correlated, which is one would expect.  The restriction to very low temperature is, unfortunately, necessary, because the spin-1/2 moments are very dilute in the large organic crystal, and so their contribution to the energy is fully swamped by the lattice one above a few K.  Doubts have been raised about the specific heat measurements due to possible contamination by nuclear contributions.  Moreover, it is known that molecular rotational tunneling excitations can contribute entropy and thermal conductivity at low temperatures\cite{stephens1972thermal,sorai1983heat}.  The most direct probe which definitely measures the spins is NMR, but beyond the demonstration of the absence of ordered moments, it has not clarified the situation (there is plenty of data, but these do not seem to fit into the expectations from the $U(1)$ spinon theory).

There are many unexplained phenomena.  The NMR relaxation in both materials becomes non-exponential below a few K, which is one of several indications of inhomogeneity (possibly field-induced) -- see below.   The low frequency conductivity is quite large and roughly power-law in form\cite{PhysRevB.74.201101,elsasser2012power}, which is in fact expected from the spin fermi-surface model\cite{potter2013mechanisms,ng2007power}, but in detail the behavior does not agree with this theory.  A number of experiments point to some sort of phase transition or crossover at 6K in $\kappa$-ET\cite{PhysRevB.89.045138,manna2010lattice}.  Some transition is also indicated at 1K in dmit based on NMR\cite{itou2010instability,PhysRevB.84.094405}, and based on the field dependence of the thermal conductivity below this temperature, it was suggested that there are some spin excitations with a gap of this order, which opens at this temperature.  Motivated by these apparent transitions, theorists have proposed pairing instabilities of the spinon fermi surface state\cite{grover2010weak,lee2007amperean}.  

Most of these features are associated with quite low temperature.  So it is possible some of them are subsidiary phenomena sensitive to small details of the Hamiltonian. It would be more ideal to measure the spin state between 1K and room temperature, as the larger scales are the natural ones for spins with $J \sim$200K. For example, it would be very desirable to probe the correlations of spins directly through neutron scattering.  However, since organics contain a large concentration of hydrogen atoms, which strongly scatter neutrons, such measurements are extremely difficult and have not been successful in $\kappa$-ET or dmit so far.  In general, apart from NMR, we have no direct probes of the spins in this range.   

A basic question is that of the mechanism for the spin liquid behavior.  If these materials are truly QSLs, why only these two organics, out of the vast families?  Originally, based on chemical modeling of the hopping parameters, it was proposed that the key variable is spatial anisotropy: organics which are closest to isotropic fall into the spin liquid phase.  However, recent density functional based theory finds rather different hopping parameters, putting this mechanism into question\cite{PhysRevLett.103.067004,nakamura2009ab}. These calculations also estimate that on-site Hubbard interactions are significantly stronger than the earlier models do, and are accompanied by more significant off-site terms, which may further complicate the picture.  

One open issue is the extent of and influence of disorder and inhomogeneity.  For many years these materials have been viewed as extremely clean, with a mean free path estimated from thermal conductivity in dmit of 1$\mu$m\cite{yamashita2010highly}. Non-exponential relaxation in NMR below 6K in $\kappa$-ET and between 1-10K in dmit suggest some inhomogeneity, but it was suggested this is field-induced.  However, there are several indications of disorder or inhomogeneity in more recent experiments on $\kappa$-ET, based on e.g.\ transport and optics\cite{pinteric2014anisotropic}, electron spin resonance\cite{Padmalekha2015211}, and muSR\cite{nakajima2012microscopic}.  Whether any of these phenomena are also present in dmit, and whether they are important to the spin physics\cite{watanabe2014quantum}, remains unclear to the authors.

\subsubsection{Herbertsmithite}
\label{sec:herbertsmithite}

The kagom\'e lattice has long been thought theoretically to be sweet spot for spin liquid physics.  In the domain of the most quantum S=1/2 spins, however, materials realizations are sparse, and all suffer from various complications departing from the theoretical ideal.  They come from various copper minerals.  For example, there is volborthite, Cu$_3$V$_2$O$_7$(OH)$_2$$\cdot$2H$_2$O, which is very clean but lacks three-fold symmetry and is quite spatially anisotropic, and there is strong evidence for competing exchanges very far from the simple nearest-neighbor antiferromagnet.   Vesignieite, BaCu$_3$V$_2$O$_8$(OH)$_2$, also lacks three-fold symmetry but is much less characterized.  Kapellasite, Cu$_3$Zn(OH)$_6$Cl$_2$, has three-fold symmetry but is believed to have competing exchanges up to third neighbor, including {\em ferromagnetic} interactions between nearest neighbors\cite{PhysRevLett.109.037208}.

The simplest and most heavily studied of the quantum kagom\'e materials is herbertsmithite, with chemical formula ZnCu$_3$(OH)$_6$Cl$_2$.  Synthesized in 2005\cite{shores2005structurally}, a review as of 2010 can be found in Ref.\cite{mendels2010quantum}.  It has the ideal kagom\'e three-fold symmetry, and all indications are that the Cu$^{2+}$ ions interact predominantly by a nearest-neighbor antiferromagnetic Heisenberg exchange of order $J\sim $180K (as reinforced by first principles calculations in Ref.\cite{jeschke2013first}), yet avoids ordering and spin freezing down to temperatures as low as 50mK, as observed by NMR\cite{olariu200817}, $\mu$SR, neutron scattering\cite{helton2010dynamic,de2009scale}, and susceptibility measurements\cite{PhysRevB.76.132411}. These early experiments all observed a large density of low energy states, e.g.\ by specific heat\cite{PhysRevLett.98.107204}, indicating a gapless state.    

Structurally, herbertsmithite can be understood as a depleted pyrochlore lattice, with non-magnetic Zn atoms substituting for 1/4 of the pyrochlore sites, and forming triangular layers which then dividing the remaining Cu sites into kagom\'e planes.  However, the substitution is not perfect, and it is believed that between 5-10percent of these triangular sites are actually occupied by Cu rather than Zn.  At a first approximation, these interlayer Cu ions form quasi-free spins, which are then responsible for the low energy states.  More controversial is the extent of Zn atoms in the kagom\'e planes, and in general the feedback of both types of defects on the physics of the spins within the kagom\'e layers.  Optimistically, the kagom\'e layers may be almost unaffected by the defects, so that any measurement yields simply the sum of some contribution from the inter-layer Cu spins and a signal from ideal kagom\'e planes.  Less optimistically, the fragile nature of the ideal kagom\'e state may be altered by the disorder.  

The first single crystals were grown in 2011, allowing a next stage of studies. For example, single crystal measurements of susceptibility anisotropy suggest a 10\% easy-axis Ising exchange anisotropy\cite{han2012refining}.  Single crystal inelastic neutron scattering \cite{han2012fractionalized} observes only diffuse weight which persists to zero energy, and an absence of any sharp modes.  This is consistent with a QSL state, and was interpreted as evidence for gapless spinon excitations.  The extremely smooth spectral weight, however, which lacks any sharp features such as excitation edges, and has weight down to zero energy at all momenta, is not expected for any of the model QSL states, including both the gapped $Z_2$ QSL found by DMRG, and the U(1) Dirac favored variationally, which, though gapless, has low energy states only at isolated momentum points.  High field measurements have been carried out up to 55T, which show a rather linear magnetization versus field, and no indication of any field-induced transitions.  Moreover, the large low temperature, quasi-linear specific heat seems to persist throughout the field range, which is interesting because presumably the interlayer spins must polarize and develop a Zeeman gap in this regime, implying that these low energy excitations must come from the kagom\'e layers.  Terahertz optics also observes some in-plane ac conductivity $\sigma(\omega) \sim \omega^\beta$ with $\beta \approx 1.5$ for $30K< \hbar\omega/k_B < 67K$, which may be related to gapless electric dipole excitations of the spins\cite{potter2013mechanisms}.  

Various avenues have been explored in order to reconcile the ``gapless'' scattering with theory.  It was suggested that herbertsmithite might be perturbed sufficiently by interactions beyond nearest-neighbor, e.g.\ Dzyaloshinskii-Moriya coupling, to drive it close to a quantum critical point to a nearby ordered phase\cite{cepas2008quantum}.  Further neighbor exchange might also serve the same purpose. Another suggestion is that the ground state is a $Z_2$ QSL, but the bosonic spin-less {\sf m} excitations have anomalously low energy and a very flat dispersion, and the ``shake-off'' effects of these {\sf m} particles completely smooths out the intrinsic excitation spectrum of the spinons\cite{punk2014topological}.  A less appealing but highly plausible explanation is that the observed excitations are strongly influenced by disorder.  The very weak momentum dependence indeed suggests the excitations are rather local, and with this in mind the high degree of site disorder is hard to discount.  Numerical calculations including disorder have been carried out\cite{kawamura2014quantum}.

While all the above measurements point to a gapless state of some kind, a very recent NMR experiment, which utilizes the single crystal to select a specific advantageous field orientation, claims to observe a spin gap at low field in the kagom\'e plane\cite{fu}.  This might bring experiments on herbertsmithite more in line with DMRG calculations for the Heisenberg model, but reconciliation of this preliminary result with prior experiments would then become the issue.

\section{Other topics and Future directions}
\label{sec:future-directions}

It is impossible to be comprehensive in such a broad subject, and the vanguard is moving too quickly to keep entirely current.  Instead, we tried in this review to strike a balance between completeness and comprehensibility, and to give references to more in depth reviews of some sub-topics.  As we conclude the review, we would like to mention some of the subjects that have been given short shrift, and to comment on future directions for the field.

Obviously the subject of topological phases has strong overlap with the QSL problem.  But not all QSLs are topological, and not all topological phases are QSLs.  The field of topological phases, when divorced from applications to quantum magnetism (and for the most part any other experiments), has been studied much more deeply than discussed here.  There is an enormous amount of sophisticated mathematical structure describing topological phases, e.g.\ topological quantum field theory, modular tensor categories, quantum groups, etc.   Topological phases with non-abelian quasiparticles are of great current interest, and may have applications to quantum computing.  Readers might be interested in the problem of classifying topological phases, or even how to define topological phases in three dimensions, which is still a developing area.  We are intrigued by the ``cubic code'' and related models by Haah and others\cite{bravyi2013quantum,haah2011local,bravyi2011energy,yoshida2013exotic}, which apparently defy the conventional quasiparticle paradigm completely.  These models have phases which appear to be stable, highly entangled, yet not compatible with any emergent gauge theory description; for example, they seem to have excitations with mixed dimensionality and even fractal structure.  

Related but distinct to topological phases are ``Symmetry Protected Topological/Trivial'' (SPT) phases\cite{chen2012symmetry,levin2012braiding,chen2011classification}.  These are states which are {\em not} intrinsically entangled.  That is, if no symmetry constraint is imposed, they can be smoothly deformed into a product state.  However, if the symmetry is present, an SPT is distinct from a trivial state, and typically has the property that it supports non-trivial states at the boundary: either the surface breaks the symmetry, is gapless, or has intrinsic topological order.  There are two canonical examples of SPT states: the electronic topological insulator (TI)\cite{hasan2010colloquium}, which occurs for weakly interacting fermions, and the Haldane gap or Affleck-Kennedy-Lieb-Tasaki (AKLT) spin-1 chain\cite{affleck1987rigorous}.   The latter indicates the potential relevance of SPT phases to quantum magnets.  The AKLT state has the remarkable property that, although it consists of spin-1 spins, at each edge one finds a free spin-{\em half} moment.  This moment is fractional, and cannot be built from any finite number of spin-1 spins.  It persists unless one of the symmetries protecting the SPT state, for example time-reversal symmetry, is broken.  While the one dimensional AKLT state has been known for decades, it has only recently been appreciated that SPT states occur for bosons/spins in two and three dimensions as well.  A great variety of such states have been discovered theoretically, and a partial classification (``group cohomology'') has been suggested\cite{chen2012symmetry,chen2011classification}.  We direct interested readers to a particularly pedagogical and clear example in a two dimensional Ising system\cite{levin2012braiding}. It would be quite exciting to find an SPT phase in a two or three dimensional quantum magnet.

Topological and SPT phases are particularly simple because they exhibit a gap to all excitations, and this facilitated a great deal of theoretical progress on them.  However, a priori such fully gapped states are not the most common QSLs.  Indeed, the most celebrated QSL materials, as reviewed in Sec.~\ref{sec:mater-exper}, seem to be gapless -- though whether this is intrinsic or a result of defects is not clear.  Theoretically, within the Abrikosov fermionic parton approach, moreover, states with gapless partons quite commonly emerge as variational ground states.  Phases like this, with emergent gauge structure and gapless matter fields, are much less understood.  In particular the examples with U(1) gauge structure described in Sec.~\ref{sec:gapless-matter-field} are all central to QSL research, yet are fundamentally not understood.   We do not even know in which cases these states are actually stable phases of matter!   This clearly constitutes an important avenue for fundamental research.  There are very interesting recent developments using quantum field theory and duality.  However, we felt they were too technical to describe here, and instead direct the reader to some references.

One possible path to progress on such gapless QSL states is through entanglement. In general, the entanglement entropy of gapless phases is much less universal than of gapped ones.  We expect, for example, that even for a simple bipartition of the plane into a compact, connected region and its complement, the entropy depends in a non-trivial way upon the full shape of the boundary.  For some gapless states, even for a fixed boundary the results may be non-universal: for example, for free Fermions and presumably also for a spinon Fermi surface state, the coefficient of the leading $\sim L^{d-1}\ln L$ term depends upon the full shape of the Fermi surface\cite{eisert2010colloquium}.  However, recent results in quantum field theory show that some powerful general results are possible.   For gapless phases which are described by conformal field theories, a term analogous to the topological entanglement entropy can be defined for a specific circular/spherical bipartition, and importantly has been shown\cite{myers2010seeing,casini2012renormalization,cardy1988there,komargodski2011renormalization} to obey a monotonicity property analogous to the famous ``c theorem'' for two dimensional critical theories\cite{zamolodchikov1986irreversibility}.   Recently the 2+1-dimensional analog, known as the ``f theorem'', has been used to address the stability of U(1) Dirac QSLs\cite{grover2014entanglement}.

It is rewarding to explore connections of QSL states of spin systems to exotic phases of matter discussed in other contexts. The essential features of entanglement and non-local excitations can of course manifest in systems without localized spins.  Conceptually, indeed, there can be a smooth evolution between a Mott insulating QSL with well-defined local moments and a QSL state near the metal-insulator transition in a Hubbard model (Sec.~\ref{sec:ring-exch-hubb}).  Thus states such as topological Mott insulators\cite{pesin2010mott} and fractional topological insulators\cite{levin2009fractional,maciejko2015fractionalized}, while envisioned in distinct venues, may be considered cousins of QSLs, and many of the considerations of this review extend to them as well.   We have already mentioned parallels between two dimensional QSLs and fractional quantum Hall states, which are found in semiconductor electron gases and in graphene.    QSL states and their relatives might also be observed in ultra-cold trapped atomic gases.  Both the tunability of and certain types of measurements possible in these systems go far beyond what is possible in condensed matter.   Indeed, recently experimentalists have successfully carried out a measurement of a non-local string order parameter in a one dimensional boson chain\cite{endres2011observation}, and of the second Renyi entropy in a one dimensional bose gas\cite{islam}.  So in principle signatures of long-range entanglement might actually be {\em directly} probed in ultra-cold atoms!  

Clearly the most important direction of ongoing work is connection to experiment.  Indisputable proof of a QSL should consist of clear evidence for an emergent excitation, or of long-range entanglement.  It is challenging to envision how either of these might be accomplished even in principle for most QSL states.  A notable example is a 3d U(1) QSL, for which the emergent photon mode could be measured by inelastic neutron scattering.  Ideas for measuring fermionic spinons also seem feasible, if challenging, provided an appropriate material is found.  Additional novel experimental concepts are very much needed.  In our opinion, its success will probably also require a proven microscopic understanding of the QSL's Hamiltonian.  This in turn almost certainly necessitates moving beyond the simplest SU(2) symmetric Heisenberg models, which are at best correct up to 5-10\% corrections.   These corrections are important, since in just about all unbiased numerical studies which find QSL states, these states are only stable over a range of 5-10\% of dimensionless couplings.  Another ``correction'',  quenched disorder in the form of missing or extra spins, random exchange, etc., is doubtless influential in many QSL candidates.  It is interesting to ask if quenched randomness ever facilitates long-range entanglement, or always favors more classical, albeit glassy, states.   We should also emphasize that there are many interesting aspects to the physics of quantum frustrated magnets aside from QSLs.  A nice review of some of those topics can be found in Ref.~\cite{0034-4885-78-5-052502}.    We conclude, optimistically, that a definitive and generally accepted identification of QSLs in experiment may occur in the not too far future.  Reaching this goal will probably require the combination of innovative experimental concepts, a concerted effort at materials synthesis and refinement, and the synergy of analytical theory and intense numerical simulations.

\ack We are grateful to SungBin Lee, Gang Chen, Oleg Starykh, Cenke Xu, Matthew Fisher, Mike Hermele, Kate Ross, Bruce Gaulin, Donna Sheng, Hong-Chen Jiang, and Zheng-Han Wang for collaborations that deepened our interest and understanding of the subject.  We thank Marmy T. Chat and Veronica L. Voiture for encouragement and inspiration.  LB was supported by the NSF through grant DMR-15-06119 and the DOE basic energy sciences grant DE-FG02-08ER46524.  LS was supported by a postdoctoral fellowship from the Gordon and Betty Moore foundation through the EPiQS initiative, grant GBMF-4303.

\bibliography{qsl.bib}

\Tables

\begin{table}
  \caption{  \label{tab:QSLs} Varieties of quantum spin liquids. The acronyms used in this table are: FQH: Fractional Quantum Hall effect, CSL: Chiral Spin Liquid, RVB: Resonating Valence Bond state, QED: Quantum Electrodynamics, ASL: Algebraic Spin Liquid.}
  \begin{indented}
    \item[] \begin{tabular}{cp{1.2cm}||p{1.5cm}|p{2.5cm}|p{3cm}|p{4cm}}
    Class & & Stability & Excitations & Models/Systems & Other properties \\
\hline
   Topological& & $d\geq 2$ & gapped anyons &  & topological entanglement entropy $\gamma$ \\
    & gapped $Z_2$ & & {\sf e}, {\sf m}, $\varepsilon$ & toric code, kagom\'e Heisenberg? & $\gamma = \ln 2$  \\
              & FQH/CSL & \mbox{$d=2^{a}$}& Laughlin $1/m$ anyons & semiconductor electron gases, $J_{1,2,3}$ kagom\'e Heisenberg & broken time-reversal, gapless edge states, $\gamma = -\frac{1}{2}\ln m$ \\
Gapless $Z_2$ && $d\geq 2$ & gapless fermions & Kitaev honeycomb and generalizations, d-wave RVB & May convert to nonabelian phase with appropriate perturbation \\ 
U(1) QSL && & gapless photon, electric, magnetic, and composite charges & & \\
& ``pure'' & $d\geq 3$ & gapless photon, gapped charges & compact QED, quantum spin ice & sharp photon pole in $S(q,\omega)$ , $T^3$ specific heat \\
& ASL = U(1) Dirac & $d =2$?$^{b}$ & gapless fermions with electric charge and strongly coupled gauge field & QED$_3$, kagom\'e Heisenberg? & conformal field theory, emergent SU(N) symmetry.  $S(q,\omega=0^+)> 0$ at isolated $q$. \\
& spinon Fermi surface & $d\geq 2$? & gapless fermions with electric charge and strongly coupled gauge field & triangular Heisenberg+ ring exchange? & non-trivial power laws, emergent SU(N) symmetry.  Violates area law.  $S(q,\omega=0^+)> 0$ at all $q$\\
  \end{tabular}
  \item[] $^{a}$ Can exist in layered three-dimensional systems.
  \item[] $^{b}$ Stable with $N$ fermion flavors for sufficiently large $N$.
\end{indented}
\end{table}

\end{document}